%% file: ms.tex
\documentclass[11pt,letterpaper]{article}
\include{commands0}
\def\citep{\cite}
\usepackage[caption=false]{subfig}
\newtheorem{lemma}{Lemma}
\newtheorem{proposition}{Proposition}
\newtheorem{theorem}{Theorem}
\newtheorem{corollary}{Corollary}
\usepackage{authblk}

\begin{document}

\title{Cox Point Process Regression}

\author[1]{ \'Alvaro Gajardo}
\author[1]{Hans-Georg M\"uller}
\affil[1]{Department of Statistics, University of California, Davis, Davis, CA 95616 U.S.A.}

\maketitle

\begin{abstract}
Point processes in time have a wide range of applications that include the claims arrival process in insurance or the analysis of queues in operations research. Due to advances in technology, such samples of point processes are increasingly encountered. A key object of interest is the local intensity function. It has a straightforward interpretation that allows to understand and explore point process data. We consider functional approaches for point processes, where one has a sample of repeated realizations of the point process. This situation is inherently connected with Cox processes, where the intensity functions of the replications are modeled as random functions. Here we study a situation where one records covariates for each replication of the process, such as the daily temperature for bike rentals. For modeling point processes as responses with vector covariates as predictors we propose a novel regression approach for the intensity function that is intrinsically nonparametric. While the intensity function of a point process that is only observed once on a fixed domain cannot be identified, we show how covariates and repeated observations of the process can be utilized to make consistent estimation possible, and we also derive asymptotic rates of convergence without invoking parametric assumptions.
\end{abstract}

 \single  \no {\sc \small Cox Process,  Fr\'echet Regression, Intensity Function,  Nonparametric Regression, Wasserstein Metric.}

\double

\section*{\sf 1. Introduction}\label{s:intro}

Temporal point processes are encountered in insurance in the form of the claim arrival process, risk processes and ruin theory which targets the solvency of the insurer \citep{miko:09}; queue theory in operations research \citep{dale03}; seismology; demand patterns in bike sharing systems \citep{gerv:19}; or bid arrivals in online auctions \citep{redd:06,shmu:07}. 
Single realizations of point processes have been well studied in the literature \citep{cox:80,dale03,digg:85,digg13,reyn:03}. One important target is the intensity function, due to its straightforward interpretation as the rate of occurrence of points per unit time  \citep{dale03}. In the context of seismology, one expects the intensity function of the aftershock arrival process to depend on
the size of the earthquake that triggered the aftershocks. This exemplifies point process data for which the  intensity function depends on covariates, and provides the motivation to develop flexible nonparametric methods for such data. Specifically, we propose a nonparametric regression method for point processes as responses, coupled with Euclidean predictors in $\mathbb{R}^p$.

Poisson processes are one of the most important point processes as they have been shown to provide successful models for  a wide range of scientific applications involving random phenomena and allow the construction of more complex processes \citep{dale03} such as the Cox or doubly stochastic Poisson process \citep{cox:80}. The estimation of the intensity function of a non-homogeneous Poisson process (NHPP) has found much  interest in the literature. For a single realization of a NHPP, \cite{reyn:03} proposed an approach based on penalized projection estimators but consistency towards the intensity function can only be achieved when the expected number of observed events diverges. A standard asymptotic framework has been to assume that the intensity of the observed process $\tilde{\lambda}(t)$ can be written as a scalar multiple of an underlying intensity of interest $\lambda(t)$, where the scalar is allowed to diverge and thus enables to observe increasingly more points \citep{reyn:10,cowl:96:1,will:07}. For multiple realizations or replicated NHPP, \cite{bigo:13:2} consider the situation where the intensities are common across replications but only differ in that they are time shifted at random i.i.d. time points with a known density function.   In \cite{Leem:91} a replicated NHPP framework with a common non-random underlying intensity function is considered with results on consistent estimation of the cumulative intensity function,  while in \cite{hend:03} convergence results towards the common intensity function are obtained. These previous approaches do not incorporate covariates and thus do not study a regression framework.

In the context of stationary Cox processes \citep{cox:80}, nonparametric kernel estimation of the intensity function for just one observed point process has been proposed \citep{digg:85},  connecting this problem to kernel density estimation. However, when one has just one realization of the point process, no consistent estimator of the intensity function exists \citep{zhan:10}, due to the unavailability of a consistent estimator of the scale factor of the intensity function, i.e. $\int \lambda(t)dt$. In other work that explores the interface of spatio-temporal point processes and functional data analysis, \cite{li:14} proposed a semi-parametric generalized linear mixed model with a latent process component, and established asymptotic properties under increasing domain asymptotics, a design assumption that is commonly used for spatial processes.
We consider here a different scenario, where replications of a temporal point process are available, along with an Euclidean covariate  $X \in \mathbb{R}^p$. A previous replicated point process regression approach \citep{lawl87} also dealt with repeatedly observed non-homogeneous Poisson processes as responses; however this previous approach relied heavily on parametric specifications, for which diagnostics is very difficult, whereas we aim here at a flexible nonparametric approach that applies much more generally.

In the context of Cox processes, where the intensity function $\Lambda(t)$ is a positive locally integrable random function, a common approach when dealing with repeatedly observed point processes (but without a covariate $X$) has been to combine likelihood methods and techniques from Functional Data Analysis \citep{bouz:06:2,bouz:15,gerv:17,gerv:19}. For example,  in \cite{gerv:17}  a Karhunen-Lo\`eve expansion \citep{gren:50, klef:73}  was applied to the log intensity random functions,
\begin{equation*}
\log(\Lambda(t))=\mu(t)+\sum_{k=1}^\infty U_k \phi_k(t), \quad 0 \le t \le T,
\end{equation*}
where $0 < T < \infty$ and $\mu \in L^2([0,T])$ is the mean function, the $U_k$ are uncorrelated random variables and $\phi_1,\phi_2,\dots$ form an orthonormal basis of $L^2([0,T])$. Then, by using techniques such as Functional Principal Component Analysis (FPCA), a truncated version of the expansion is considered with only $p$ components and thus the functional problem of estimating $\mu$ and the $\phi_k$ is reduced to a multivariate approach, modeling the functions in a finite dimensional function space of basis functions like B-splines. Distributional assumptions such as Gaussianity of the $U_k$ are also introduced in order to justify a likelihood approach to obtain estimates for the basis coefficients. The previous transformation approach is extrinsic since it does not target directly intensity functions, which are subject to a positivity constraint.  This constraint  makes the intensity space convex but not linear and thus functional data analysis (FDA)  methods and especially functional principal components analysis (FPCA) directly applied to $\Lambda(t)$ are not well suited \citep{mull:16:1}.

Similarly, \cite{mull:13:1} proposed a functional approach that decomposes the intensity function into an intensity factor and a shape function and then performed a Karhunen-Lo\`eve expansion for the shape function, borrowing strength across the replications. These methods face constraints due to the  non-negative nature of the intensity function and cannot be directly extended to establish regression models for point processes. 
Since the random intensity functions are not observed, the event arrival times are used in the estimation procedures. For Cox processes, the key relation that allows this is that conditional on $n$ events occurring in an interval $[0,T]$ and $\Lambda=\lambda$, the unordered arrival times form an i.i.d. sample with density $\lambda/\int \lambda$; this  is a well known property for non-homogeneous Poisson processes \citep{dale03}. Furthermore, as noted in \cite{mull:13:1}, the intensity function $\lambda$ can be decomposed into a shape function $f=\lambda/\int \lambda$ and an intensity factor $\tau=\int \lambda$. This decomposition is the key relationship that will enable us to split the problem of estimating the intensity function conditional on covariates into two parts: Estimating the conditional shape function and estimating the conditional intensity factor of the process. While this decomposition is defining the structure of the problem, in order to achieve consistent estimation of  the shape function,  $\tau$ is assumed to diverge so that increasingly more points are available. Therefore, we observe a Cox process $(N^{(n)},\Lambda)$ such that, conditional on $\Lambda$, $N^{(n)}(T)$ has a Poisson distribution with rate $\alpha_n \tau$ for some positive sequence $\alpha_n\to\infty$.

The proposed regression approach for the intensity function of replicated temporal point processes on Euclidean predictors utilizes conditional Fr\'echet means \citep{mull:18:3} for a suitable metric on the space of intensity functions, where this metric can be decomposed into two parts, one that quantifies differences in shape and a second part that quantifies differences in the intensity factors.  As we need to estimate the density function associated with the arrival times of the point process, our asymptotic consistency results for the intensity function also utilize tools that were developed in  \cite{pana:16}. The common assumption of letting the observation window $T\to \infty$ is often not applicable, including  the data scenarios we consider below to motivate our methods.  Therefore, we consider an asymptotic framework where $T$ remains fixed and the number of replicates of the point process increases.

While in general the intensity function of a doubly stochastic Poisson process cannot be consistently estimated as it is random, we demonstrate here that 
the situation is different for point processes conditional on a covariate, as we establish asymptotic consistency with rates of convergence for conditional intensity functions. We illustrate the implementation of the proposed point process regression with simulations and show that it leads to well interpretable results for the Chicago Divvy bike trips and the New York yellow taxi trips data. An  application to the earthquake aftershock process in Chile is presented in  section 6.3.

The main innovations presented in this paper are: (1) We develop the first fully nonparametric regression method that features point processes as responses with Euclidean predictors; (2) We obtain asymptotic rates of convergence for conditional intensity functions, while such a result is not achievable for intensity functions unconditionally; (3) Our approach does not require functional principal components and does not require distributional assumptions, as it is not utilizing likelihoods; (4) The proposed approach is shown to work well in relevant applications.

\section*{\sf 2. The space of intensity functions}\label{sFR}

Let $\{N(t),\, t\geq 0\}$ be a temporal point process where $N(t)$ represents the number of events that occur in the time interval $[0,t]$ and $N(0)=0$. We suppose that $N(t)$ is observed on the time window $[0,T]$ for some endpoint $T>0$, and is such that $m(t):=E(N(t))<\infty$ for $0 \le t \le T$.
In the context of replicated point processes that we consider here, it is natural to work within  the framework of a doubly stochastic Poisson process $(N,\Lambda)$ where one assumes that there is an underlying stochastic intensity process $\Lambda(t)$ that generates non-negative integrable functions on $[0,T]$ such that conditional on a realization $\Lambda=\lambda$, $N\vert \Lambda=\lambda$ is a non-homogeneous Poisson process with intensity function $\lambda$ \citep{cox:80}. A feature that greatly facilitates analysis  of such processes \citep{mull:13:1,gerv:17,gerv:19,pana:16} is the fact that a Poisson process has the order statistics property, i.e., conditional on $m$ events being observed in $[0,T]$, the successive event times are distributed as the order statistics of $m$ independent and identically distributed random variables with a  density that is proportional to the intensity function \citep{dale03}.

Denoting the space of intensity functions as
\begin{equation*}
\Omega=\left\{ \Lambda \colon [0,T]\to \mathbb{R}^+ \text{ such that} \ \int_{0}^T \Lambda(t) dt <\infty  \right\}, 
\end{equation*}
two key quantities for each $\Lambda \in \Omega$ are the {\it intensity factor}, the scalar $$\tau:= \int_{0}^T \Lambda(t) dt,$$ which is the expected number of events in $[0,T]$, conditional on $\Lambda(\cdot)$; and  the {\it shape function}
 $$f(t):=\frac{\Lambda(t)}{\tau}, \quad \text{where} \quad \int_0^T f(t)dt=1, \,\, f(t) \ge 0,  \quad t\in [0,T],$$ which is a density function.
Hence, $\Lambda(\cdot)=\tau f(\cdot)$ and since there is a one-to-one correspondence between $\Lambda$ and $(\tau,f)$ we may regard $\Omega$ as a product space $\Omega=\Omega_{\mathcal{T}} \times \Omega_{\mathcal{F}}$ where $ \Omega_{\mathcal{T}}=(0,\infty)$ and
\begin{gather*}
\Omega_{\mathcal{F}}=\{f\colon [0,T]\to \mathbb{R} \text{ such that} \ f\ge 0 \ \text{and} \ \int_0^T f(t)dt=1 \}.
\end{gather*}
 For our theoretical results, we focus on a subspace of $\Omega_{\mathcal{F}}$ consisting of densities that are well behaved and bounded away from zero, see assumptions (S1) and (S2) in section 4.

Furthermore, if we endow $\Omega_{\mathcal{T}}$ and $\Omega_{\mathcal{F}}$ with metrics $d_{\mathcal{T}}$ and $d_{\mathcal{F}}$, respectively, we may regard $(\Omega,d)$ as a product metric space $(\Omega_{\mathcal{T}},d_{\mathcal{T}})\times (\Omega_{\mathcal{F}},d_{\mathcal{F}})$,  where
\begin{equation}
d((\tau_1,f_1),(\tau_2,f_2)):=\sqrt{d_\mathcal{T}^2(\tau_1,\tau_2)+ d_{\mathcal{F}}^2(f_1,f_2)}.\label{metricdef}
\end{equation}
In the context of metric geometry such product metric spaces for which the distance arises as an $l^2$-type norm between the underlying metrics have been extensively studied. In particular, it is well known that $\Omega$ is a geodesic space if and only if $\Omega_{\mathcal{T}}$ and $\Omega_{\mathcal{F}}$ are geodesic spaces \citep{bura:01}. This decomposition enables us to measure differences in shape and magnitude separately. We choose the Euclidean metric $d_\mathcal{T}(\tau_1,\tau_2):=d_E(\tau_1,\tau_2)$
and the $2$-Wasserstein metric $d_{\mathcal{F}}(f_1,f_2):=d_W(\mu_1,\mu_2)$, which   for two probability measures $\mu_1, \mu_2$ on $[0,T]$ with associated density functions $f_1,f_2$ and quantile functions $Q_1, Q_2$ is defined as \cp{vill:03} 
\begin{equation*}
d_W^2(\mu_1,\mu_2)=d_{L^2}^2(Q_1,Q_2)=\int_{0}^1 (Q_1(t)-Q_2(t))^2 dt,
\end{equation*}
where we assume throughout that these quantities exist and are well defined. The Wasserstein metric  has been shown to be a most useful metric in practical applications that involve samples of distributions \cp{bols:03}.  The $2$-Wasserstein metric on the space of density functions $\Omega_\mathcal{F}$ has very rich geometrical interpretations due to  its connections  with optimal transport \citep{pana:19}. Although one could consider a metric based on the vertical alignment such as the $L^2$ metric, these metrics are not well suited for densities due to their inherent constraints $f\ge 0$ and $\int_0^T f(t)dt=1$,  which imply that  the space $\Omega_\mathcal{F}$, while convex, is not a  linear space  \citep{mull:16:1}.

A basic notion for statistical modeling is the mean of a random variable, where for random objects in a metric space  $(\tilde{\Omega}, \tilde{d}, P)$  it has proved advantageous to adopt the barycenter or Fr\'echet mean   \citep{frec:48}, defined as $w_\oplus:=\underset{w\in \tilde{\Omega}} {\arg\min} \ E({\tilde{d}}^2(Y,w))$ where $Y\in \tilde{\Omega}$ is a random object. 
The Fr\'echet mean may be regarded as an extension of the standard concept of mean in Euclidean space to abstract metric spaces in the sense that when $\tilde{\Omega}$ is a convex subset of the Euclidean space and $\tilde{d}$ is the Euclidean metric, then the ordinary mean and the Fr\'echet mean coincide. 
The barycenter and its estimation 
has attracted much interest for distribution spaces with the Wasserstein metric \citep{ague:11,caze:18, bigo:18,
pana:16,pana:19}; we adopt these spaces here for the shape part of the intensity function.

\section*{\sf 3. Framework for Intensity Function Regression}\label{s:LFRsetup}
\subsection*{\sf 3.1 Preliminaries}\label{s:LFRsetup3.1}

Our goal is to model the regression relation between random intensity functions $Y$  in the above  space $(\Omega,d)$ as responses and an Euclidean predictor $X\in \mathbb{R}$, for which we adopt the recently developed framework of Fr\'echet regression \citep{mull:18:3}, which can be viewed as a generalization of Fr\'echet means to the more general notion of conditional Fr\'echet means. Formally, define the regression or conditional intensity function $m_\oplus(x)$ as
\begin{gather*}
m_\oplus(x):=\underset{w\in \Omega} {\arg\min} \ M_{\oplus}(w,x),\quad M_{\oplus}(w,x):=E(d^2(Y,w)\vert X=x),
\end{gather*}
where $w=(\tau_0,f_0) \in \Omega=\Omega_{\mathcal{T}}\times \Omega_{\mathcal{F}}$ and $Y=(\tau,f)\in \Omega_{\mathcal{T}}\times \Omega_{\mathcal{F}}$, so that by (\ref{metricdef}), 
\begin{gather*}
M_{\oplus}(w,x)=E(d_\mathcal{T}^2(\tau,\tau_0)\vert X=x)+E(d_\mathcal{F}^2(f,f_0)\vert X=x).
\end{gather*}
Hence, the optimization problem is separable with optimal solution $m_\oplus(x)=(\tau_\oplus(x),f_\oplus(x))$,  where
\begin{equation}
\tau_\oplus(x) = \underset{\tau_0 \in \Omega_{\mathcal{T}}} {\arg\min} \ E(d_\mathcal{T}^2(\tau,\tau_0)\vert X=x)=\max\{E(\tau \vert X=x) ,0\};  \label{tau_plus} 
\end{equation}
\begin{equation}
f_\oplus(x) =\underset{f_0 \in \Omega_{\mathcal{F}}} {\arg\min} \ E(d_\mathcal{F}^2(f,f_0)\vert X=x).\label{f_plus}
\end{equation}

As we focus on a subspace of $\Omega_{\mathcal{F}}$ consisting of densities that are well behaved and bounded away from zero, the space of corresponding quantile functions $Q(\Omega_{\mathcal{F}})$ is a closed and convex subset of the Hilbert space $L^2([0,1])$ (see assumptions $(S1)$, $(S2)$ and Lemma \ref{Lemma_QOmegaf_closedconvex} in section 4.1). By equivalently casting the optimization problem (\ref{f_plus}) in terms of quantile functions, as per the definition of the $2$-Wasserstein metric, and employing properties of the $L^2$-inner product, Lemma S.12 in the Appendix implies the  existence and uniqueness of the solution to this program. The solution admits a closed form in terms of the quantile functions $Q_\oplus(x)$ corresponding to $f_\oplus(x) $, and is given by $Q_\oplus(x)=E(Q\vert X=x)$. This shows that under regularity conditions $m_\oplus(x)$ exists and is unique.
Note that since we measure the differences in the intensity factor from the changes in shape separately,  the order of magnitude between the two metrics is not relevant. More precisely, $m_\oplus(x)$ remains the same for weighted metrics  $d_\lambda^2=\alpha d_\mathcal{T}^2+\beta d_{\mathcal{F}}^2$ with $\alpha,\beta>0$.

The local Fr\'echet regression function $m_\oplus(x)$ operates directly on the space of intensity functions so that the regression is performed in the geometric space corresponding to the optimal transport geometry that is induced by the $2$-Wasserstein metric on the shape components. Thus $m_\oplus(x)$ provides a notion of conditional center that is shaped by  the underlying geometry of the data generating mechanism that produces random intensities. We show through a simulation example in section A.8 in the Appendix that the standard Euclidean conditional intensity $E(\Lambda\vert X=x)$ can be severely distorted,  while $m_\oplus(x)$  captures the underlying geometry of the problem and provides a better notion of center. As $m_\oplus(x)$  is a valid intensity function, it still enjoys the usual interpretation of rate of events per unit time,  while representing 
the (conditional) point process barycenter at predictor level $x$.

A basic difficulty is that the function $\Lambda(\cdot)$ is not observed. If sufficiently many arrival times are observed for each replication of the point process, then it is well known that consistent estimation of the density function $f$ may be achieved by classical density estimation techniques \cp{digg:85, pana:16}, while some work-arounds exist for sparsely observed point processes \cp{mull:13:1}.
However, the situation is much less benign regarding estimation of the intensity parameter $\tau$. It would be natural to employ the total count $N(T)$,  which is conditionally unbiased for $\tau$ in the sense that $E(N(T)\vert \Lambda)=\tau$ but is not (conditionally) consistent as $\text{Var}(N(T)\vert \Lambda)=\tau$. When just one replicate of a point process is observed, consistent estimation of the intensity function is therefore not possible. 
Further motivating  conditional intensity function modeling,  we show in the following that the situation is different when considering  conditional intensity functions. We demonstrate that the counts $N_i(T)$ can be used as initial estimates for the random intensity factors 
$\tau_i$, from which consistent estimators can then be derived. 
This phenomenon is analogous to classical linear regression modeling, where one has errors in the responses and yet consistent estimation of the conditional expectation that corresponds to the true regression function
is achieved. This  provides strong motivation for the proposed methods and the study  of conditional point processes.

\subsection*{\sf 3.2 Local Regression for Intensity Functions}\label{s:LFRestimates}

Suppose that a sample of replicates $(X_i,N_i,f_i,\tau_i)$ is drawn from the joint distribution of $(X,N,f,\tau)$, $i=1,\dots,n$, where $N\vert \Lambda=f\times \tau$ is a Poisson process with intensity function $\lambda=f\times \tau$. We employ empirical weights  from local linear regression \citep{fan:96} that are inherent to the local Fr\'echet regression approach \citep{mull:18:3}, and are given by
\begin{equation}
s_{in}(x, h) = \frac{1}{\hat{\sigma}_0^2}K_h(X_i - x)\left[\hat{u}_2 - \hat{u}_1(X_i-x)\right], \label{LFR_empirical_weight}
\end{equation}
where $\hat{u}_j=n \inv \sumin K_h(X_i-x)(X_i-x)^j$ with $j \in \{0,1,2\}$, $\hat{\sigma}_0^2=\hat{u}_0 \hat{u}_2-\hat{u}_1^2$  and
$K_h(\cdot)=h\inv K(\cdot/h)$, the kernel $K$ is a continuous and symmetric density function with support $[-1,1]$, and  $h=h_n$ is a sequence of bandwidths.

As we can factorize the response into a density function and an intensity factor, one regression component is the conditional mean $E(\tau\vert X=x)$. For estimating this quantity we employ local linear regression. If the $\tau_i$ were observed, then a naive estimator for (\ref{tau_plus}) would be given by
\begin{equation}\label{naive_tauplus}
\hat{\tau}_\oplus(x)=\max \{0, n^{-1} \sum_{i=1}^n s_{in}(x,h)\tau_i \},
\end{equation}
as local linear regression is a linear estimator, assigning  weights  $s_{in}(x,h)$ to the responses. However, this estimator is based on the intensity factors $\tau_i$, which are not observed; we only observe the counts of arrivals $N_i(T)$ for each replicate of the point process. This difficulty can be resolved by noting that the observed counts $N_i(T)$ satisfy the relationship $E(N_i(T) \vert \Lambda_i)=\tau_i$, which enables us to replace $\tau_i$ by $N_i(T)$ in (\ref{naive_tauplus})  since $E(N(T)\vert X)=E(E(N(T)\vert {X,\Lambda})\vert X)=E(\tau\vert X)$ is on target. Hence, under suitable regularity conditions one readily obtains well known non-parametric convergence rates for the corresponding locally weighted least squares estimator of $\tau_\oplus(x)$ (\ref{tau_plus}) \citep{fan:96}. As we require an increasing  asymptotic intensity framework, our final empirical estimate for the intensity factor part will be presented in section 4.2.

If the densities $f_i$ associated with point processes $N_i$ were completely observed, we could implement local Fr\'echet regression on the space of densities for (\ref{f_plus}) \citep{mull:18:3}, 
\begin{equation}\label{naive_fplus}
\hat{f}_\oplus(x)=\underset{f_0\in \Omega_\mathcal{F}} {\arg\min} \ n^{-1} \sum_{i=1}^n s_{in}(x,h)d_\mathcal{F}^2(f_i,f_0).
\end{equation}
 We however only observe the arrival times, from which the densities $f_i$ must be estimated, and this will induce an additional error that needs to be accounted for when analyzing the final estimator. Existence and uniqueness of the solution to the optimization program in  (\ref{naive_fplus}) can be obtained by considering corresponding quantile functions, which we discuss next.

As noted by \cite{cuca:08}, one of the main differences compared to classical density estimation techniques is that in the context of point processes we cannot let the number of observations go to infinity as it is a random feature of the point process itself. Instead it is useful to consider an asymptotic framework where the intensity factors $\tau_i$ diverge to infinity, while the observation window $[0,T]$ remains fixed; such frameworks have been considered before in the literature and allow to add information everywhere on $[0,T]$ as opposed to the common domain asymptotics $T\to \infty$, which are often not applicable,  see also \cite{pana:16} or \cite{digg:88}. This framework will be introduced in section 4. From now on, $\tau$ will denote a generic random  intensity factor such that $\tau_i=\int_0^T \Lambda_i(s)ds \overset{iid}{\sim} \tau$, \,\, $i=1,\dots,n$.

The minimization problem (\ref{naive_fplus}) is easily  solved by considering quantile functions. If $Q_i$ is the quantile function corresponding to $f_i$, $i=1,\dots,n$ and $\hat{Q}_\oplus(\cdot,x):[0,1]\to [0,T]$ is the quantile function corresponding to the density $\hat{f}_\oplus(x)$ in (\ref{naive_fplus}), 
\begin{equation*}
\hat{Q}_\oplus(\cdot,x)=\underset{q\in Q(\Omega_\mathcal{F})} {\arg\min} \ n^{-1} \sum_{i=1}^n s_{in}(x,h) \lvert \lvert Q_i-q  \rvert \rvert_{L^2([0,1])}^2,
\end{equation*}
where $Q(\Omega_\mathcal{F})$ is the space of quantile functions corresponding to densities in $\Omega_\mathcal{F}$. Standard properties of the $L^2([0,1])$ inner product imply (Proposition 1 in \cite{mull:18:3})
\begin{equation}\label{qfhat}
\hat{Q}_\oplus(\cdot,x)=\underset{q \in Q(\Omega_\mathcal{F})} {\arg\min}  \ \lvert \lvert q - n^{-1} \sum_{i=1}^n s_{in}(x,h) {Q}_i \rvert \rvert_{L^2([0,1])}^2.
\end{equation}
Existence and uniqueness of the solution of (\ref{qfhat}) and therefore of (\ref{naive_fplus}) is guaranteed  as $\hat{Q}_\oplus(\cdot,x)$ corresponds to the orthogonal projection of $n^{-1} \sum_{i=1}^n s_{in}(x,h) {Q}_i$ as an element of the Hilbert space $L^2([0,1])$ on  the closed and convex set $Q(\Omega_\mathcal{F})$ as shown in Lemma \ref{Lemma_QOmegaf_closedconvex} under regularity conditions on the space $\Omega_{\mathcal{F}}$.

To discuss estimation of the $Q_i$, which are needed in (\ref{qfhat}) but are not directly available, it is helpful to consider auxiliary probability measures $\hat{\mu}_i$ on $[0,T]$ that correspond to the empirical measure of the arrival times when the total count $N_i(T) \geq 1$ and to the uniform measure on $[0,T]$ otherwise (see \cite{pana:16}). That is,
\[\hat{\mu}_i= \begin{cases} 
      \displaystyle \frac{1}{N_i(T)} \sum_{j=1}^{N_i(T)} \delta_{Z_{ij}},  &\text{  if } N_i(T) \geq 1\\
      \frac{1}{T} \ \mathcal{L},  &\text{  if } N_i(T) =0, 
   \end{cases}
\]
where $Z_{ij}$ are the arrival times of the point process $N_i$ and $\mathcal{L}$ is the Lebesgue measure on $[0,T]$. For a probability measure $\mu$ on $[0,T]$ with cdf $F_\mu$ we consider its quantile function $Q_\mu(t):=\inf\{x\in [0,T]\colon F_{\mu}(x)\geq t\}$. Let $\hat{Q}_i:=Q_{\hat{\mu}_i}$, then replacing $Q_i$ by $\hat{Q}_i$ in (\ref{qfhat}) leads to the empirical estimate
\begin{equation} \label{qf_empirical}
\tilde{Q}_\oplus(\cdot,x):=\underset{q\in Q(\Omega_\mathcal{F})} {\arg\min}  \ \lvert \lvert q - n^{-1} \sum_{i=1}^n s_{in}(x,h) \hat{Q}_i \rvert \rvert_{L^2([0,1])}^2.
\end{equation}

\section*{\sf 4. Asymptotic Results}\label{s:As}
\subsection*{\sf 4.1 Convergence of the Shape Function Estimates}\label{s:As4.1}

Let $\tilde{f}_\oplus(x)$ be the density function corresponding to the quantile function $\tilde{Q}_\oplus(\cdot,x)$. Thus, $\tilde{f}_\oplus(x)$ corresponds to the empirical estimate for (\ref{naive_fplus}). We require the following assumptions, which guarantee that  $Q(\Omega_\mathcal{F})$ is a closed and convex subset of the Hilbert space $L^2([0,1])$, yielding  existence and uniqueness of the naive and empirical estimators in (\ref{qfhat}) and (\ref{qf_empirical}), respectively.

\begin{itemize}
\item[(S1)] Suppose that there exists $0<M<L<\infty$ such that $Q \in Q(\Omega_\mathcal{F})$  if
\begin{equation*}
	M \lvert x-y  \rvert \leq \lvert Q(x)-Q(y) \rvert \leq L \lvert x-y  \rvert ,
\end{equation*}
for all $x,y\in [0,1]$.
\item[(S2)] Suppose that for any $Q \in Q(\Omega_\mathcal{F})$ it holds that $Q(0)=0$ and $Q(1)=T$.
\end{itemize}
These assumptions are needed to ensure that the quantile functions do not increase too rapidly or too slowly, which is equivalent to constraining the corresponding density functions to be well behaved and bounded away from zero.

\begin{lemma}\label{Lemma_QOmegaf_closedconvex}
Under $(S1)-(S2)$, $Q(\Omega_\mathcal{F})$ is a closed and convex set on the Hilbert space $L^2([0,1])$.
\end{lemma}

Lemma \ref{Lemma_QOmegaf_closedconvex} guarantees  existence and uniqueness of  the local Fr\'echet regression function on shape space $f_\oplus(x)$ defined in (\ref{f_plus}). By employing the Hilbert space structure of $L^2([0,1])$ and properties of the $2$-Wasserstein metric, Lemma S.10 in the Appendix shows that $f_\oplus(x)$ corresponds to the density function with associated quantile  function $E(Q\vert X=x)$; the latter can be shown to reside in $Q(\Omega_\mathcal{F})$. Regarding the convergence of $\tilde{f}_\oplus(x)$ towards $f_\oplus(x)$ in the $2$-Wasserstein metric, by the triangle inequality
\begin{equation}\label{triang_ineq}
d_\mathcal{F}(f_\oplus(x),\tilde{f}_\oplus(x))\leq d_\mathcal{F}(f_\oplus(x),\hat{f}_\oplus(x))+d_\mathcal{F}(\hat{f}_\oplus(x),\tilde{f}_\oplus(x)),
\end{equation}
and for  the term $d_\mathcal{F}(\hat{f}_\oplus(x),\tilde{f}_\oplus(x))$ we observe that properties of the orthogonal projection on a closed and convex set in the Hilbert space $L^2([0,1])$ imply that
\begin{equation}\label{key_eqn_density}
d_\mathcal{F}(\tilde{f}_\oplus(x),\hat{f}_\oplus(x))= \lvert \lvert \tilde{Q}_\oplus(\cdot,x)- \hat{Q}_\oplus(\cdot,x)   \rvert \rvert_{L^2([0,1])} \leq  n^{-1} \sum_{i=1}^n \lvert s_{in}(x,h) \rvert \  \lvert \lvert  \hat{Q}_i - {Q}_i \rvert \rvert_{L^2([0,1])}.
\end{equation}
Therefore, convergence hinges on consistent estimation of the quantile functions $Q_i$. We consider the following asymptotic framework:
\begin{enumerate}
\item A first random mechanism generates pairs of predictors $X_i$ and intensity functions $\Lambda_i$,\newline $(X_1, \Lambda_1),\dots,(X_n,\Lambda_n)\overset{iid}{\sim} (X, \Lambda)$ which encapsulates the dependency between these random quantities. While the $X_i$ are observed, the $\Lambda_i$ are not observed.
\item Given the random intensity functions $\Lambda_i$, a second independent random mechanism then generates the observable number of arrivals $N_i^{(n)}(T)$ and the arrival times $Z_{i1}, \ldots, Z_{iN_i^{(n)}(T)}$ for the $i$-th point process $N_i^{(n)}$, $i=1,\dots,n$. 
\item Conditional on $N_i^{(n)}(T)$, the (unordered) arrival times $Z_{i1}, \ldots, Z_{iN_i^{(n)}(T)} \overset{iid}{\sim}f_i$.
\item Given the random intensity function $\Lambda_i$, $N_i^{(n)}(T)\sim \mathcal{P}(\alpha_n \tau_i)$ for a positive sequence $\alpha_n\to\infty$, where $\mathcal{P}(\eta)$ denotes a Poisson random variable with rate $\eta$.
\end{enumerate}

We note that conditions 1-3 are standard in the context of Cox processes while condition 4 allows for the observable number of arrivals $\Nin$ to diverge as $n$ increases and to avoid empty point processes \citep{pana:16}, which is the key for consistent estimation of $Q_i$ by using the empirical measure of the arrival times $\hmu_i$ with $\Nin$ in place of  $N_i(T)$. A similar fixed domain asymptotic point process framework was considered in \cite{pana:16} but without covariates.

The following result shows that the second term on the right hand side of (\ref{triang_ineq}) is $O_p(\alpha_n^{-1/4})$ provided that the support of $\tau$ is bounded away from zero and $\alpha_n$ grows fast enough.
\begin{itemize}
\item[(S3)] There exists a scalar $\kappa>0$ such that $\tau\geq \kappa$ almost surely.
\end{itemize}

\begin{proposition}\label{Proposition_consistentWassEst}
Suppose that $(S1)$, $(S2)$ and $(S3)$ hold, the marginal density $f_X$ of $X$ satisfies $f_X(\cdot)>0$ and is twice continuously differentiable, and $\displaystyle \frac{\alpha_n}{\log n}\to \infty$ as $n\to \infty$. Then
\begin{equation*}
d_\mathcal{F}(\hat{f}_\oplus(x),\tilde{f}_\oplus(x))=O_p(\alpha_n^{-1/4}). 
\end{equation*}
\end{proposition}

The term $d_\mathcal{F}(f_\oplus(x),\hat{f}_\oplus(x))$ on  the right hand side of   (\ref{triang_ineq}) was shown to be $O_p(n^{-2/5})$ under the following regularity condition  \citep{mull:18:3}:
\begin{itemize}
 \item[(L1)] The marginal density $f_X$ of $X$, as well as the conditional densities $g_y$ of \mbox{$X|\tilde{Y} = y$}, exist  for $\tilde{Y} \in \Omega_{\mathcal{F}}$ and are twice continuously differentiable, the latter for all $y \in \Omega_{\mathcal{F}}$, and $\sup_{x,y} |g_y''(x)| < \infty$.  Additionally, for any open $U\subset \Omega_{\mathcal{F}}$, $\int_U {\rm d}F_{\tilde{Y}|X}(x, y)$ is continuous as a function of $x$.
\end{itemize}

Summarizing these results, we obtain
\begin{theorem}\label{Theorem_densityEst}
Suppose that $(S1)$, $(S2)$, $(S3)$ and $(L1)$ hold, the density function satisfies $f_X(\cdot)>0$, $\displaystyle \frac{\alpha_n}{\log n}\to \infty$ as $n\to \infty$ and  $h=h_n\sim c_0 n^{-1/5}$ for some constant $c_0>0$.  Then
\begin{equation*}
d_\mathcal{F}(f_\oplus(x),\tilde{f}_\oplus(x))=O_p(n^{-2/5}+\alpha_n^{-1/4}).
\end{equation*}
\end{theorem}
Accordingly, consistent estimation in the $2$-Wasserstein metric for the shape part of the conditional intensity function can be achieved at the rate $O_p(n^{-2/5})$ as long as $n^{8/5}\alpha_n^{-1}$ is bounded above. If this assumption holds, the well known rate of convergence for local linear regression with real valued responses is thus obtainable.

\subsection*{\sf 4.2 Convergence of the Intensity Factor Estimates}\label{s:As4.2}

In the increasing asymptotics framework that was introduced in the previous section, we assumed that there is a common intensity factor multiplier $\alpha_n$ such that $\alpha_n \rightarrow \infty$ as $n \rightarrow \infty$.  This led to consistent estimation of the conditional intensity functions for the shape function part of the intensity function, where one works in the density space $\Omega_\mathcal{F}$. Since intensity functions can be factorized into a shape part, which corresponds to a density function, and an intensity factor, it remains to construct an estimator for the intensity factor (\ref{tau_plus}) conditional on predictors $X$. 

It turns out that this is a challenge, as an estimator for (\ref{naive_tauplus}) is not easily available. This is because in order to estimate the shape functions consistently, it is necessary to assume that the expected number of events increases without bound. 
We show in the following how this challenge can be overcome  and consistent estimation of $E(\tau\vert X)$ is 
nevertheless still possible up to the constant $E(\tau)$, so that relative intensities can be estimated consistently. The key to achieve this is to utilize the average observed number of arrivals $\bar{N}(T):=n^{-1} \sumin \Nin$. We require the following regularity conditions.

\begin{enumerate}
\item [(LL1)] The regression function $m(x)=E(\tau \vert X=x)$, the density function $f_X(x)>0$ of $X$ and $\sigma^2(x)=E(e^2 \vert X=x)$, where $e=\tau-m(X)$, are twice continuously differentiable in $x$. 
\item [(LL2)] For the bandwidth sequence $h$, $nh^5=O(1)$, as $n\to \infty$.
\item [(LL3)] There exists $\delta>0$ and $\bar{\sigma}>0$ such that $E(\lvert e \rvert^{2+\delta}\vert X)\leq \bar{\sigma}$, for all predictors $X$.
\end{enumerate}

We note that assumption $(LL1)$  is a basic smoothness assumption that is needed to expand the 
bias for local linear smoothing,  while  $(LL2)$ is also a common assumption and implies that as $n\to \infty$ and for $q\in \mathbb{N}$, we have $nh^q\to \infty$ if $0\leq q \leq 4$ and $nh^q\to 0$ for $q>5$. Assumption $(LL3)$ will be used for an application of the central limit theorem. 

Our main result on conditional intensity estimation is as follows. For two sequences $\beta_n$ and $\gamma_n$, denote by $\beta_n\asymp \gamma_n$ if $c_1\beta_n\le \gamma_n\le c_2 \beta_n$ for some constants $c_1,c_2>0$. Denote by $\tilde{\tau}_\oplus(x)$ the empirical and standardized estimate of the intensity factor part, which is given by
	\begin{equation}\label{empirical_tauplus}
		\tilde{\tau}_\oplus(x)=\max\left(0,\frac{n^{-1} \sum_{i=1}^n s_{in}(x,h) \Nin}{\bar{N}(T)} \right).
	\end{equation}

\begin{theorem} \label{Theorem_tauPlusEst}
Under  $(LL1)-(LL3)$ and $(S3)$, suppose that $\tau \leq M_1$ almost surely for a constant $M_1 \in [\kappa,\infty)$ with $\kappa$ as in assumption (S3). If $\psi(\alpha_n) \asymp n^{2/5}$  for some function $\psi:\bbR^+\to \bbR$, $h=h_n= c_0 n^{-1/5}$ for some constant $c_0>0$ and $\displaystyle \frac{\alpha_n}{\log n}\to \infty$  as $n\to \infty$, then
\begin{equation*}
\tilde{\tau}_\oplus(x)= \frac{1}{E(\tau)} \tau_\oplus(x) + O_p(n^{-2/5}).
\end{equation*}
\end{theorem}

This means that  $\tau_\oplus(x)$ can still be consistently estimated up to the constant $E(\tau)$ by using the observable numbers of arrivals $\Nin$ of each replication of the point process instead of the true intensity factors $\tau_i$, which are not observed. Furthermore, as the observed counts $\Nin$ grow with $\alpha_n$ as $n\to\infty$, we can stabilize the local linear estimator by employing comparisons against the average number of arrivals $\bar{N}(T)$. We remark that even though the quantity $E(\tau)$ is unknown, relative intensity factors at different covariate levels can still be recovered consistently, which is a key result of interest in our framework.

The assumptions require that $\psi(\alpha_n)$ does not increase faster than $n^{2/5}$ for some function $\psi:\bbR\to \bbR$. This is due to the fact that local linear regression estimators with real valued responses are employed. These have a well known optimal rate of convergence $O_p(n^{-2/5})$ under mild assumptions, which is obtained under our assumptions for general growth rates of $\alpha_n$. For example, if $\alpha_n=c_1 n^\rho$ has a polynomial growth rate, where $c_1,\rho>0$, then our assumptions are satisfied by taking $\psi \colon t\to t^{2/(5\rho)}$, $t>0$, and leads to the optimal rate $O_p(n^{-2/5})$. Similarly, if $\alpha_n=c_1 \exp(n \gamma)$ has an exponential growth rate, where $c_1,\gamma>0$, then the conditions are satisfied by taking $\psi\colon t\to(\log(t)/\gamma)^{2/5}$, $t>0$.

\subsection*{\sf 4.3 Convergence of the Conditional Intensity Function Estimates}\label{s:As4.3}

We are now in position to construct an estimate for the conditional intensity function by combining our previous results. Recall that the regression or conditional intensity function satisfies $m_\oplus(x)=\tau_\oplus(x) f_\oplus(x)$ where $\tau_\oplus(x)$ and $f_\oplus(x)$ are defined in (\ref{tau_plus}) and (\ref{f_plus}), respectively, and 
\begin{equation*}
\tilde{\tau}_\oplus(x)=\max\left(0,\frac{n^{-1} \sum_{i=1}^n s_{in}(x,h) \Nin}{\bar{N}(T)} \right),
\end{equation*}
which corresponds to the estimate of $\tau_\oplus(x)$, up to the constant $E(\tau)$, as per Theorem~\ref{Theorem_tauPlusEst}. 

Since $\Lambda_\oplus(x)=\tau_\oplus(x) {f}_\oplus(x)$, we obtain an estimate of $\Lambda_\oplus(x)$ by plugging in the previously obtained estimates of the intensity factor $\tau_\oplus(x)$ and of the 
shape function $ {f}_\oplus(x)$, leading to 
\begin{equation}\label{estimate_cond_intensity}
\hat{\Lambda}_\oplus(x)=\tilde{\tau}_\oplus(x) \tilde{f}_\oplus(x).
\end{equation}
Here  $\tilde{f}_\oplus(x)$ is the density corresponding to the quantile function defined in (\ref{qf_empirical}). This estimator is consistent for the conditional intensity function up to $E(\tau)$, as per the following result. 

\begin{corollary}\label{cor1}
Under the regularity conditions of Theorems \ref{Theorem_densityEst} and \ref{Theorem_tauPlusEst}, the estimate $\hat{\Lambda}_\oplus(x)$ of the conditional intensity function $m_\oplus(x)$ is consistent up to the constant $E(\tau)$ in the sense that
\begin{equation*}
d(m_\oplus(x),E(\tau)\hat{\Lambda}_\oplus(x) )=O_p(n^{-2/5}+\alpha_n^{-1/4}),
\end{equation*}
where $ \alpha_n/\log n\to \infty$ as $n\to \infty$ and $\psi(\alpha_n) \asymp n^{2/5}$  for some function $\psi:\bbR^+\to \bbR$.
\end{corollary}
This result follows directly from Theorems \ref{Theorem_densityEst} and \ref{Theorem_tauPlusEst}. If the sequence $\alpha_n$ has a polynomial growth rate $\alpha_n/n^{\rho} \to C$ as $n\to\infty$ for some $\rho>0$ and $C>0$, then the convergence rate in the Corollary is $O_p(n^{-\rho/4})$ if $\rho\in (0,8/5)$ while the well known non-parametric rate  for local linear regression with real valued responses $O_p(n^{-2/5})$ is achieved whenever $\rho\ge 8/5$. The fastest convergence rate achievable is obtained when $\rho$ is at least $8/5$ which leads to $d(m_\oplus(x),E(\tau)\hat{\Lambda}_\oplus(x) )=O_p(n^{-2/5})$. In this case, both the estimation of the intensity factor part, up to the constant $E(\tau)$, and the shape part of the conditional intensity function can be recovered at the rate $O_p(n^{-2/5})$.

We remark that in the special case where  $\Var(\tau)=0$  and the distribution of  the random density $f$ corresponds to a point mass in the Wasserstein space of probability distributions $\Omega_{\mathcal{F}}$ endowed with the $2$-Wasserstein metric, one has that $f=g$ almost surely for some density $g$ with corresponding quantile function $Q_g\in Q(\Omega_{\mathcal{F}})$ and $\tau=\eta_0$ almost surely for some constant $\eta_0\in [\kappa,M_1]$ with $\kappa,M_1$ as in Theorem \ref{Theorem_tauPlusEst}. This setting corresponds to the situation when there is no regression of the point process on $X$ and is still covered by Corollary \ref{cor1}. Moreover, in this special case our framework is equivalent to that of replicated Poisson processes as the underlying intensity functions are non-random and identical. Lemma S.12 in the Appendix shows that $\tilde{\tau}_\oplus(x)=\eta_0$ and $\tilde{f}_\oplus(x)=g$ so that $m_\oplus(x)$ is equivalent to the underlying common intensity. Thus the problem translates into one of consistent estimation of the common underlying intensity function across independent replications of a single Poisson process, which we obtain up to a constant. In this direction, several works exist such as \cite{hend:03} where a non-parametric estimate of the underlying intensity function is considered and pointwise as well as MSE convergence results are derived. Parametric approaches have also been extensively studied;  see \cite{Lewi:76,Lee:91,kuhl:97,kuhl:00,Kao:88} for further details. Non-parametric approaches using wavelets have also been explored in \cite{kuhl:00:02} and semiparametric approaches in \cite{kuhl:01}. When replications of a non-homogeneous Poisson process with a common and non-random underlying intensity function are available \citep{hend:03}, one can readily exploit this fact so that an asymptotic infill framework is not required in this situation; rather, one can pull observations together to estimate the common shape or intensity factor components, however the situation is different  in the regression framework that we study here.

It is often of interest to study the association between categorical predictors and point processes, for which the previously studied local regression approaches are not applicable as continuity of the predictor $X$ is required. 
The next section is devoted to the construction of a second regression model that hinges on a generalization of the classical parametric multivariate linear regression model in the Euclidean case, and allows to address this problem.

\subsection*{\sf 4.4 Global Regression Framework for Intensity Functions}\label{s:As4.4}

We briefly demonstrate here a generalization of multiple linear regression to the case where responses are point processes that allows the inclusion of categorical predictors while responses are objects residing in intensity space $(\Omega,d)$. The key is a characterization of multiple linear regression as a weighted sum of the responses, which can then be generalized to the case of weighted Fr\'echet means \citep{mull:18:3}.

Consider an Euclidean predictor $X\in \mathbb{R}^p$ and assume that $\mu:=E(X)$ and $\Sigma:=\text{Var}(X)$ exist, with $\Sigma$ positive definite. In particular, this allows to consider either continuous or categorial predictors. The standard linear regression setting for $(X,\tilde{Y})\in \mathbb{R}^p \times \mathbb{R}$ is that the regression function $E(\tilde{Y}\vert X=x)=\beta_0+\beta_1^T (x-\mu)$ is linear in $x$, where $\beta_0$ and $\beta_1$ are the scalar intercept and slope vector, respectively. \cite{mull:18:3} recharacterized the linear regression function as $E(\tilde{Y}\vert X=x)=\underset{y\in \mathbb{R}} {\arg\min}\ E(s(X,x) d_E^2(\tilde{Y},y))$, where $s(X,x):=1+(X-\mu)^T \Sigma^{-1}(x-\mu)$ are weights that vary with $x$ and $d_E$ is the Euclidean metric. This allows a direct generalization to linear regression in intensity space $(\Omega,d)$ by simply replacing $\tilde{Y}$ by the object $Y\in\Omega$ and the standard Euclidean distance $d_E$ by the metric $d$ in intensity space, which inherits properties of the standard linear regression setup as we show below.

The global regression function of  $Y\in \Omega$ on $X$ is given by $$m_{\text{G} \oplus}(x):= \underset{w\in \Omega} {\arg\min}\ E(s(X,x) d^2(Y,w)).$$ Although $\Omega$ is not a linear space due to the non-negative nature of the intensity functions, the global regression curve $m_{\text{G} \oplus}(x)$ passes through the Fr\'echet mean of $Y$ at $x=\mu$ since $s(X,\mu)=1$, a feature inherent to linear regression models. Moreover, the weights $s(X,x)$ can be negative, do not necessarily decay to zero away from $x$,  and do not depend on a tuning parameter like local methods do. Arguments similar to those outlined in section 3.1 show that $m_{\text{G} \oplus}(x)=(\tau_{\text{G} \oplus}(x), f_{\text{G} \oplus}(x))$, where
\begin{gather}
\tau_{\text{G} \oplus}(x) = \underset{\tau_0 \in \Omega_{\mathcal{T}}} {\arg\min} \ E(s(X,x)d_\mathcal{T}^2(\tau,\tau_0))=\max\{E(s(X,x) \tau) ,0\};  \label{tau_gfr_plus} \\
f_{\text{G} \oplus}(x) =\underset{f_0 \in \Omega_{\mathcal{F}}} {\arg\min} \ E(s(X,x) d_\mathcal{F}^2(f,f_0)).\label{f_gfr_plus}
\end{gather}

Lemma \ref{Lemma_QOmegaf_closedconvex} guarantees  existence and uniqueness of  the global Fr\'echet regression function on shape space $f_{\text{G} \oplus}(x)$. Similarly as in the local framework, Lemma S.11 in the Appendix shows that the Hilbert space structure allows to characterize the quantile function corresponding to $f_{\text{G} \oplus}(x)$ as the orthogonal projection of $Q_x=E(s(X,x)Q)$ as an element of $L^2([0,1])$ on the closed and convex set $Q(\Omega_\mathcal{F})$. Suppose that a sample of replicates $(X_i,N_i,f_i,\tau_i)\overset{iid}{\sim}(X,N,f,\tau)$, where $N\vert \Lambda=f\times \tau$ is a Poisson process with intensity function $\lambda=f\times \tau$, is available and consider the same asymptotic framework as outlined in section 4.1. To obtain empirical estimates, define $s_{in}(x):=1+(X_i-\bar{X})^T \hat{\Sigma}\inv (x-\bar{X})$, where $\bar{X}:=n\inv \sumin X_i$ and $\hat{\Sigma}:=n\inv \sumin (X_i-\bar{X})(X_i-\bar{X})^T$. The next result shows that the global regression function in $\Omega_{\mathcal{T}}$ space, $\tau_{\text{G} \oplus}(x)$, can be consistently estimated up to the constant $E(\tau)$.
\begin{theorem} \label{Theorem_gfr_tauPlusEst}
Suppose that (S3) holds and $\tau \leq M_1$ almost surely for some constant $M_1\in [\kappa,\infty)$ with $\kappa$ as in assumption (S3). If $\psi(\alpha_n)\asymp n^{1/2}$ for some function $\psi:\bbR^+\to\bbR$ and $\alpha_n/\log n\to \infty$ as $n\to\infty$, then
\begin{equation*}
\max\left(0,\frac{n\inv \sumin s_{in}(x) \Nin}{\bar{N}(T)} \right)= \frac{1}{E(\tau)} \tau_{\text{G} \oplus}(x) + O_p(n^{-1/2}).
\end{equation*}
\end{theorem}
Theorem \ref{Theorem_gfr_tauPlusEst} shows that the parametric $\sqrt{n}$-convergence rate can be obtained under mild conditions on the growth of $\alpha_n$. Thus faster rates are obtained compared to the local setting. For example, if $\alpha_n=c_1 n^\rho$ has a polynomial growth rate, where $c_1,\rho>0$, then the  assumptions are satisfied by taking $\psi \colon t\to t^{1/(2\rho)}$, $t>0$, which leads to the optimal $\sqrt{n}$-rate. Similarly, if $\alpha_n=c_1 \exp(n \gamma)$ has an exponential growth rate, where $c_1,\gamma>0$, then the conditions are satisfied by taking $\psi\colon t\to(\log(t)/\gamma)^{1/2}$, $t>0$.

Similarly as in the local regression setup, the shape components $f_i$ remain unobserved and must be estimated from the arrival times across each replication. We consider the same estimation scheme for the shape functions as outlined in section 3.2 but replacing the local weights $s_{in}(x,h)$ by the global weights $s_{in}(x)$. This leads to the empirical estimate $\tilde{f}_{G\oplus}(x)$ of $f_{\text{G} \oplus}(x)$. The following result shows consistency of the estimated global regression function in $\Omega_{\mathcal{F}}$ space.
\begin{theorem} \label{Theorem_gfr_fPlusEst}
Suppose that $(S1)$, $(S2)$ and $(S3)$ hold, and $\displaystyle \frac{\alpha_n}{\log n}\to \infty$  as $n\to \infty$. Then
\begin{equation*}
d_\mathcal{F}(f_{G \oplus}(x),\tilde{f}_{G\oplus}(x))=O_p(n^{-1/2}+\alpha_n^{-1/4}).
\end{equation*}
\end{theorem}
Thus, if $\alpha_n$ has a polynomial growth rate $\alpha_n/n^\rho\to C$ as $n\to\infty$ for some $\rho>0$ and $C>0$, we obtain the $\sqrt{n}$-rate as long as $\rho\geq 2$. If $\rho\in(0,2)$, then the rate achieved is $O_p(n^{-\rho/4})$. The following corollary summarizes the consistency, up to the constant $E(\tau)$, of the empirical estimate $\hat{\Lambda}_{G\oplus}(x):=\max\left(0,\frac{n\inv \sumin s_{in}(x) \Nin}{\bar{N}(T)} \right) \tilde{f}_{G\oplus}(x)$ of the global regression function $m_{G\oplus}(x)$.
\begin{corollary}\label{cor2}
Under the regularity conditions of Theorems \ref{Theorem_gfr_tauPlusEst} and \ref{Theorem_gfr_fPlusEst}, the estimate $\hat{\Lambda}_{G\oplus}(x)$ of the conditional intensity function $m_{G\oplus}(x)$ is consistent up to the constant $E(\tau)$ in the sense that
\begin{equation*}
d(m_{G\oplus}(x),E(\tau)\hat{\Lambda}_{G\oplus}(x) )=O_p(n^{-1/2}+\alpha_n^{-1/4}),
\end{equation*}
where $\alpha_n/\log n\to \infty$  as $n\to\infty$ and $\psi(\alpha_n)\asymp n^{1/2}$ for some function $\psi:\bbR^+\to\bbR$.
\end{corollary}
Thus, when  $\alpha_n$ has a polynomial growth rate $\alpha_n/n^\rho\to C$ as $n\to\infty$ for some $\rho>0$ and $C>0$, the parametric $\sqrt{n}$-rate is achieved whenever $\rho\ge 2$ and otherwise the rate is $O_p(n^{-\rho/4})$ if $\rho\in(0,2)$, which is attained by the estimation of the corresponding shape component part.

Similarly as in section 4.3, the special case of no regression on $X$ when $\tau=\eta_0$ almost surely for some positive constant $\eta_0\in [\kappa,M_1]$ and $f=g$ almost surely for some density $g$ with corresponding quantile function $Q_g\in Q(\Omega_{\mathcal{F}})$ is still covered by Corollary \ref{cor2}. Lemma S.13 in the Appendix shows that in this case $\tau_{\text{G} \oplus}(x)=\eta_0$ and $f_{\text{G} \oplus}(x) =g$. Thus the global Fr\'echet regression function coincides with the local version and is equivalent to the common underlying intensity function across independent replications of a single Poisson process. Here convergence towards $m_{G\oplus}(x)=\eta_0 g$, up to a constant, can be achieved at the parametric rate if $\alpha_n$ grows faster than $n^2$.

\section*{\sf 5. Simulations}\label{s:simulations}

\subsection*{\sf 5.1 Numerical approximation to the shape component estimates}

The minimization problem for the local Fr\'echet regression on the shape component as in (\ref{qf_empirical}) is solved numerically and similar to the quadratic optimization problem considered in \cite{mull:18:3}. Recall that
\begin{equation*}
	\tilde{Q}_\oplus(\cdot,x)=\underset{q\in Q(\Omega_\mathcal{F})} {\arg\min}  \ \lvert \lvert q - n^{-1} \sum_{i=1}^n s_{in}(x,h) \hat{Q}_i \rvert \rvert_{L^2([0,1])}^2.
\end{equation*}
Let $r_j$, $j=1,\dots,\nu$, be an equispaced grid in $(0,1)$, where $\Delta r_\nu=1/(\nu+1)$ is the grid spacing and $\nu$ is a positive integer. Denote the objective function by $\mathcal{M}(q)= \lvert \lvert q - W \rvert \rvert_{L^2([0,1])}^2$, where $W(\cdot)=n^{-1} \sum_{i=1}^n s_{in}(x,h) \hat{Q}_i(\cdot)$ and $q\in Q(\Omega_\mathcal{F})$. We employ Riemann sum approximations to numerically solve (\ref{qf_empirical}) as follows. Letting  $w_j:=W(r_j)=n^{-1} \sum_{i=1}^n s_{in}(x,h) \hat{Q}_i(r_j)$, $j=1,\dots,\nu$, the Riemann sum approximation of  $\mathcal{M}(q)$ is given by $\mathcal{M}_\nu(q):=\sum_{j=1}^\nu (q_j-w_j)^2 \Delta r_\nu$, where $q_j=q(r_j)$. Replacing $\mathcal{M}(q)$ by $\mathcal{M}_\nu(q)$ in (\ref{qf_empirical}) leads to an intermediate discretized optimization problem
\begin{equation*}
	\mathcal{S}_\nu:=\underset{q\in Q(\Omega_\mathcal{F})} {\arg\min}  \ \mathcal{M}_\nu(q).
\end{equation*}
Here $\mathcal{S}_\nu$ is the set of all such optimal solutions in $ Q(\Omega_\mathcal{F})$. This set is non-empty, which can be seen by considering an auxiliary quadratic convex optimization program
\begin{equation}
	q_\nu^{*}:=\underset{\tilde{q}\in \mathbb{R}^\nu} {\arg\min}  \ \lVert \tilde{q}- w \rVert_{E}^2  \label{numericalQP},
\end{equation}
subject to the constraints $0<\tilde{q}_1\leq \cdots \leq \tilde{q}_\nu<T$, $M \Delta r_\nu\leq \tilde{q}_{j+1}-\tilde{q}_{j}\leq L\Delta r_\nu$, $j=1,\dots,{\nu-1}$, $M \Delta r_\nu \leq \tilde{q}_{1}\leq L\Delta r_\nu$, and $M \Delta r_\nu \leq 1-\tilde{q}_{\nu}\leq L\Delta r_\nu$, where $w=(w_1,\dots,w_\nu)^T$. Therefore any function $q\in Q(\Omega_\mathcal{F})$ that interpolates the values $q_\nu^*$ at the grid points $r_1,\dots,r_\nu$ belongs to $\mathcal{S}_\nu$. The next proposition shows that $\tilde{Q}_\oplus(\cdot,x)$ can be well recovered in the $L^2$-norm by choosing a sufficiently fine grid. Let $Q_{\oplus \nu}$ be any (fixed) element in $\mathcal{S}_\nu$, which can be selected by the axiom of choice.
\begin{proposition}\label{Proposition_RiemannSum}
	Suppose that $(S1)$ and $(S2)$ hold. Then
	\begin{equation*}
		\lvert \lvert \tilde{Q}_\oplus(\cdot,x)- Q_{\oplus \nu} \rvert \rvert_{L^2([0,1])}=o(1),
	\end{equation*}
	as $\nu\to\infty$.
\end{proposition}
A natural element in $\mathcal{S}_\nu$ corresponds to the standard linear interpolation function constructed from $q_\nu^{*}$, which is given by $Q_\nu^*(t)=q_{\nu,j}^*+(t-r_j)(q_{\nu,j+1}^*-q_{\nu,j}^*)/\Delta r_\nu$ for $t\in [r_j,r_{j+1})$, $j=0,\dots,\nu$, where $q_{\nu,j}^{*}$ is the $j$th coordinate of $q_\nu^{*}$, $j=1,\dots,\nu$, and $q_{\nu,0}^{*}=0$, $r_0=0$, $q_{\nu, \nu+1}^{*}=T$ and $r_{\nu+1}=1$. By continuity, we define $Q_\nu^*(1):=\lim_{t\to1^{-}} Q_\nu^*(t)=T$ as the left-limit. Lemma S.6 in the Appendix shows that $Q_\nu^*\in Q(\Omega_\mathcal{F})$ and thus $Q_\nu^*$ lies in $\mathcal{S}_\nu$. In practice, $M,L$ are taken as very small/large constants, $M=10^{-10}$ and $L=10^{10}$. This choice works very well in practice. The optimization problem (\ref{numericalQP}) is a quadratic convex program (QP) with linear constraints similar to the one considered in \cite{mull:18:3} but slightly modifying the constraint matrix associated with the QP, and can be solved using state of the art optimization routines. The linear interpolation $Q_\nu^*$ of the optimal discrete solution $q_\nu^{*}$ corresponds to a discretized version of $\tilde{Q}_\oplus(\cdot,x)$ which is then mapped back to density space to obtain a discrete approximation of the corresponding density function $\tilde{f}_\oplus(\cdot,x)$. The latter step is performed by first constructing the cdf associated with $q_\nu^{*}$ and then utilizing local linear smoothing methods \citep{fan:96:3}. The implementation of the global regression is similar.

\subsection*{\sf 5.2 Simulations for local Fr\'echet regression}

To assess the finite sample performance of the proposed conditional intensity function estimates, we constructed a generative model that produces simulated random intensity functions $\Lambda(\cdot)=f(\cdot) \tau$ along with an Euclidean predictor $X\in\mathbb{R}$. First, to generate a random density function $f$ we consider the transformation to a Hilbert space approach using the log quantile density transformation (LQD) \citep{mull:16:1}, where a Karhunen-Loève (KL) decomposition is employed for the transformed density, which is an element of $L^2$, and the latter curve is mapped back to density space. Specifically, denoting by  $\psi:f\rightarrow-\log[f(Q(\cdot))]$ the LQD transform of $f$, where $Q$ is the quantile function corresponding to the density $f$, we consider a truncated KL decomposition \citep{hsin:15} conditionally on $X=x$
\begin{align}
	\psi(f)(\cdot)&=\mu(\cdot,x)+\sum_{k=1}^K \xi_k(x) \phi_k(\cdot)\label{LQDtransf},
\end{align}
where $\mu(\cdot,x)$ is the (conditional) mean function of the $L^2$ process $\psi$, which we assume Gaussian, the $\xi_k(x)$ are independent across $k$ such that $\xi_k(x)\sim N(0,\upsilon_k(x))$, where the eigenvalues $\upsilon_1(x)$  and $\upsilon_K(x)$ are strictly positive for all $x$  in the support of $F_X$, and the eigenfunctions $\phi_k$ are orthonormal in $L^2$. Thus the mean function and the scores are allowed to vary with $x$ while the eigenfunctions are independent of $x$, which provides better interpretability of the dependency of $\psi(f)$ on $X=x$. Performing the KL decomposition in the transformed Hilbert space rather than in density space is well suited due to the former being a linear vector space whereas the latter lacks linearity structure, and therefore a truncated KL expansion applied directly to the density process may not reside in density space \citep{mull:16:1}.

 The data generation mechanism for the shape part of the intensity function is as follows: First generate the covariate $X\sim \mathcal{U}(0,1)$. Then a random element $\mathcal{X}(s)$, $s\in (0,1)$, in $L^2$ is generated from (\ref{LQDtransf}) by sampling the Functional Principal Component (FPC) scores $\xi_k$. The random density function $f$ with support $[0,T]$, where $T=1$, is obtained by mapping back $\mathcal{X}(\cdot)$ to density space using the inverse LQD transform, i.e. $Q(t)=\theta_\mathcal{X}\inv \int_0^t \exp(\mathcal{X}(s))ds$, where $\theta_\mathcal{X}\inv=\int_0^1 \exp(\mathcal{X}(s))ds$ and $f$ is the density corresponding to the quantile function $Q$.

For the intensity factor $\tau$, we consider a linear regression setting $E(\tau \vert X=x)= a_1+b_1 x$ such that the values on the right hand side are all positive. The conditional intensity factors $\tau$ for covariate level $X$ were obtained through a linear regression model $\tau=a_1+b_1 X+\varepsilon$, where $\varepsilon$ is independent of $X$ and has a truncated normal distribution with mean zero, variance $\sigma_1^2$ and support $[c_1,d_1]$. The choice of the constants above are such that $a_1+b_1 x+\varepsilon>0$ for all $x$ in the support of $F_X$.

Next, random samples of data $(X_i,\tau_i,Q_i)$, $i=1,\dots,n$, were generated following the above procedure, where $a_1=1$, $b_1=0.2$, $c_1=-0.2$, $d_1=0.2$, $\sigma_1=1.5$, $K=2$, $\mu(s,x)=\exp(1.5 x)+\exp(1.5 s)$, $\phi_1(s)=-\sqrt{2} \cos{(\pi s)}$, $\phi_2(s)=\sqrt{2} \sin{(\pi s)}$, $s\in[0,1]$, $\upsilon_1^2(x)=3+2 x$ and $\upsilon_2^2(x)=(2-x)^2$. A triangular array of point processes was then obtained as follows: Conditional on $(X_i,\tau_i,Q_i)$, the observable number of arrivals for the $i$-{th} point process $\Nin$ was sampled from a Poisson distribution with rate $\alpha_n \tau_i$, where $\alpha_n=40 n^{4/5}$. 
Then, conditional on $\Nin$ and $Q_i$, the arrival times were generated as an i.i.d. sample of size $\Nin$ from $Q_i$. For this step, we utilize inverse sampling method by generating $\Nin$ i.i.d. uniform in $(0,1)$ random variables $u_1,\dots,u_{\Nin}$, independent of all other random quantities, and then the arrival times are obtained from the $Q(u_k)$, $k=1,\dots,\Nin$. We generate $Q_i$ over a dense grid on $(0,1)$ and used the bandwidth sequence $h=h_n=n^{-1/5}$ for the local Fr\'echet regression step. 

\begin{figure}[H]
\includegraphics[width=0.45\textwidth]{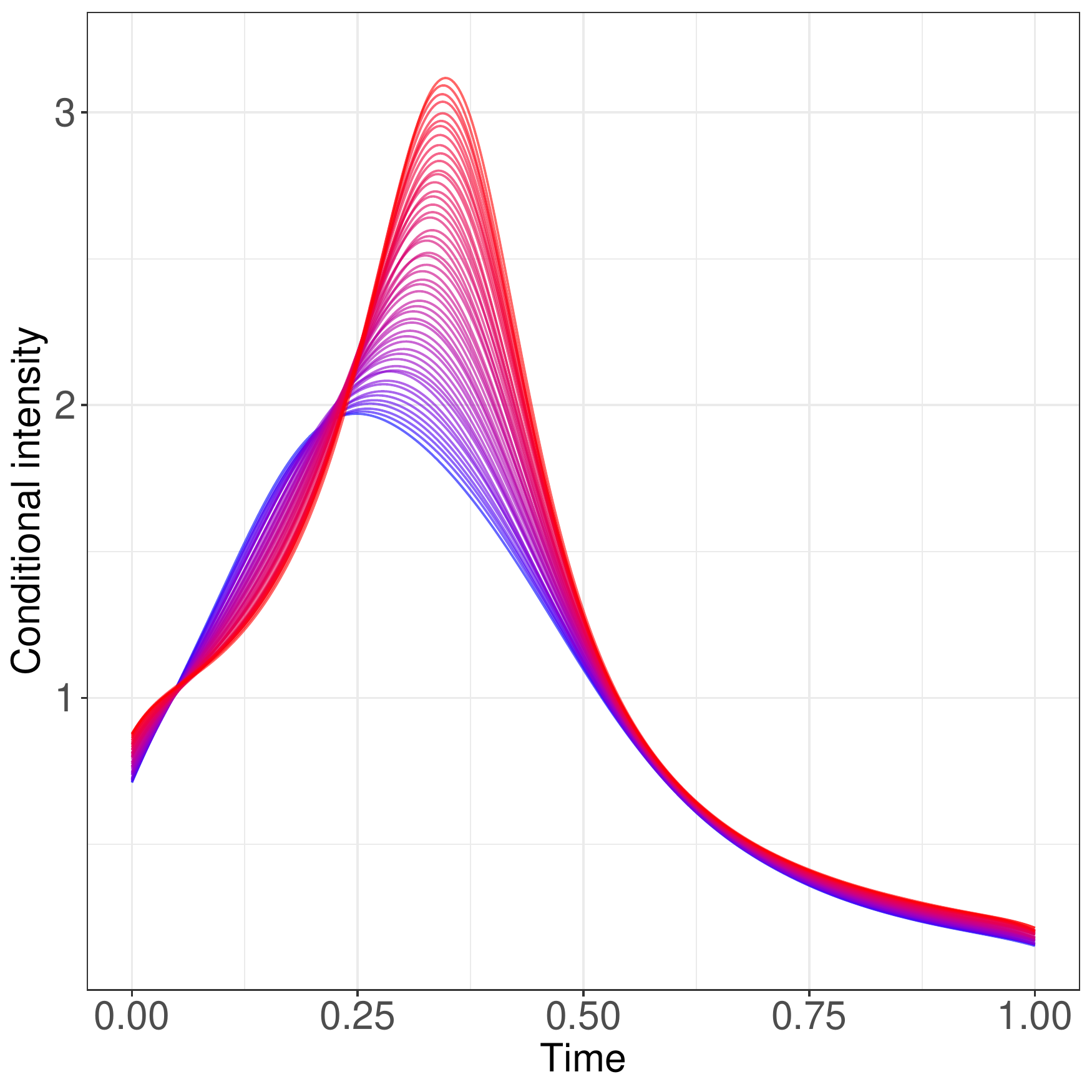}%
\includegraphics[width=0.45\textwidth]{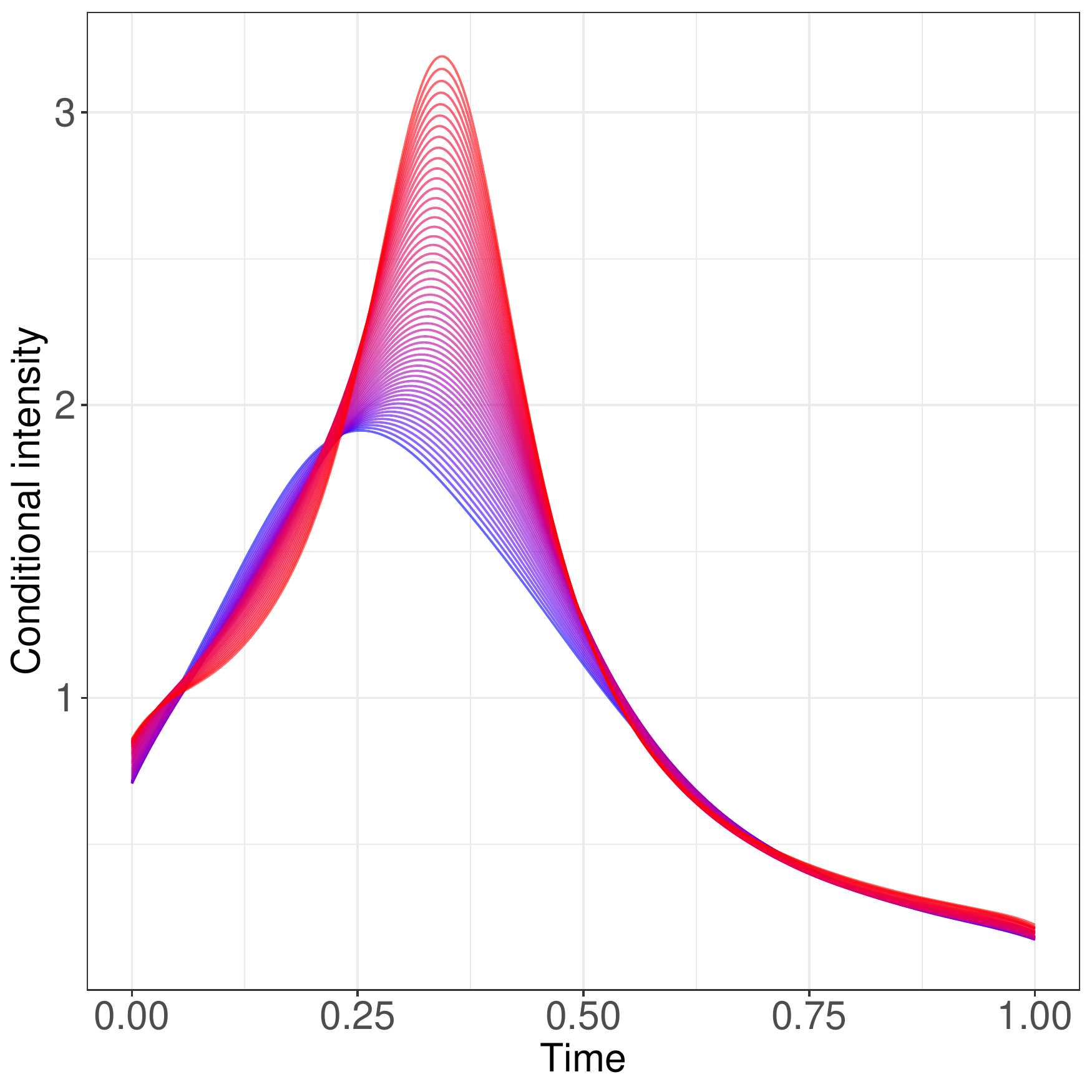}
\centering
\caption{Conditional intensity functions in the simulation setting over a dense grid of predictor levels $x$, displayed in blue when $x=0$ to red when $x=1$. The left panel shows the approximation of $\Lambda_\oplus(x)$ through a Monte Carlo approach while the right panel shows the estimate $\hLambda_\oplus(x)$ adjusted by the constant $E(\tau)$ using $n=2000$.}\label{fig:oracleRegresss}
\end{figure}

Figure~\ref{fig:oracleRegresss} shows the ``oracle'' regression function with intensity factors $E(\tau\vert X=x)$ and shape (density) function defined through the  corresponding quantile function $E(Q(\cdot) \vert X=x)$, where we consider a grid of $50$ equispaced predictor values in $(0,1)$. Here $E(Q(\cdot) \vert X=x)$ is approximated through a Monte Carlo approach where we average across random quantiles $Q_i$ generated at predictor level $x$.

\begin{figure}[H]
	\centering
	\subfloat[]{\includegraphics[width=0.4\textwidth]{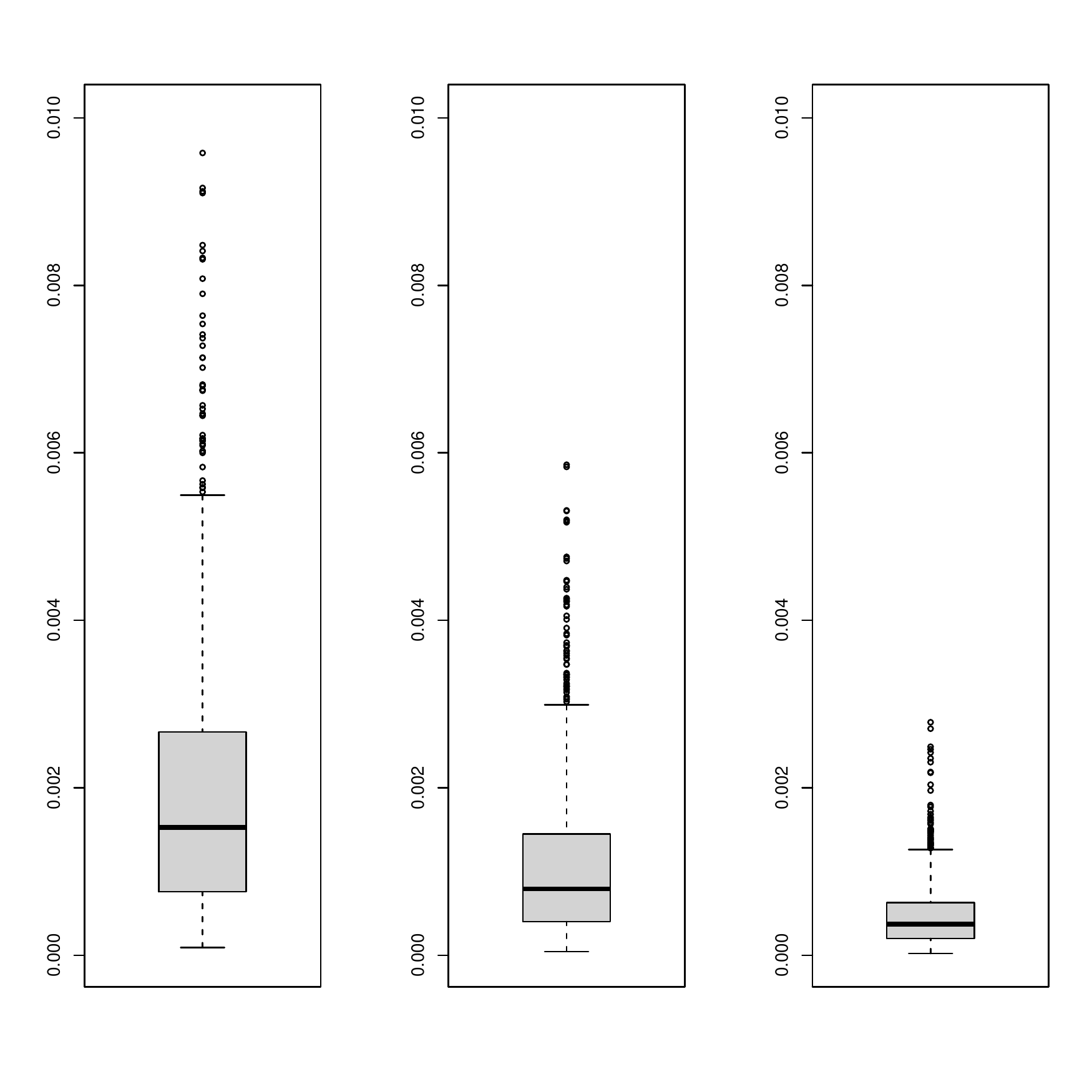}}
	\hfil
	\subfloat[]{\includegraphics[width=0.4\textwidth]{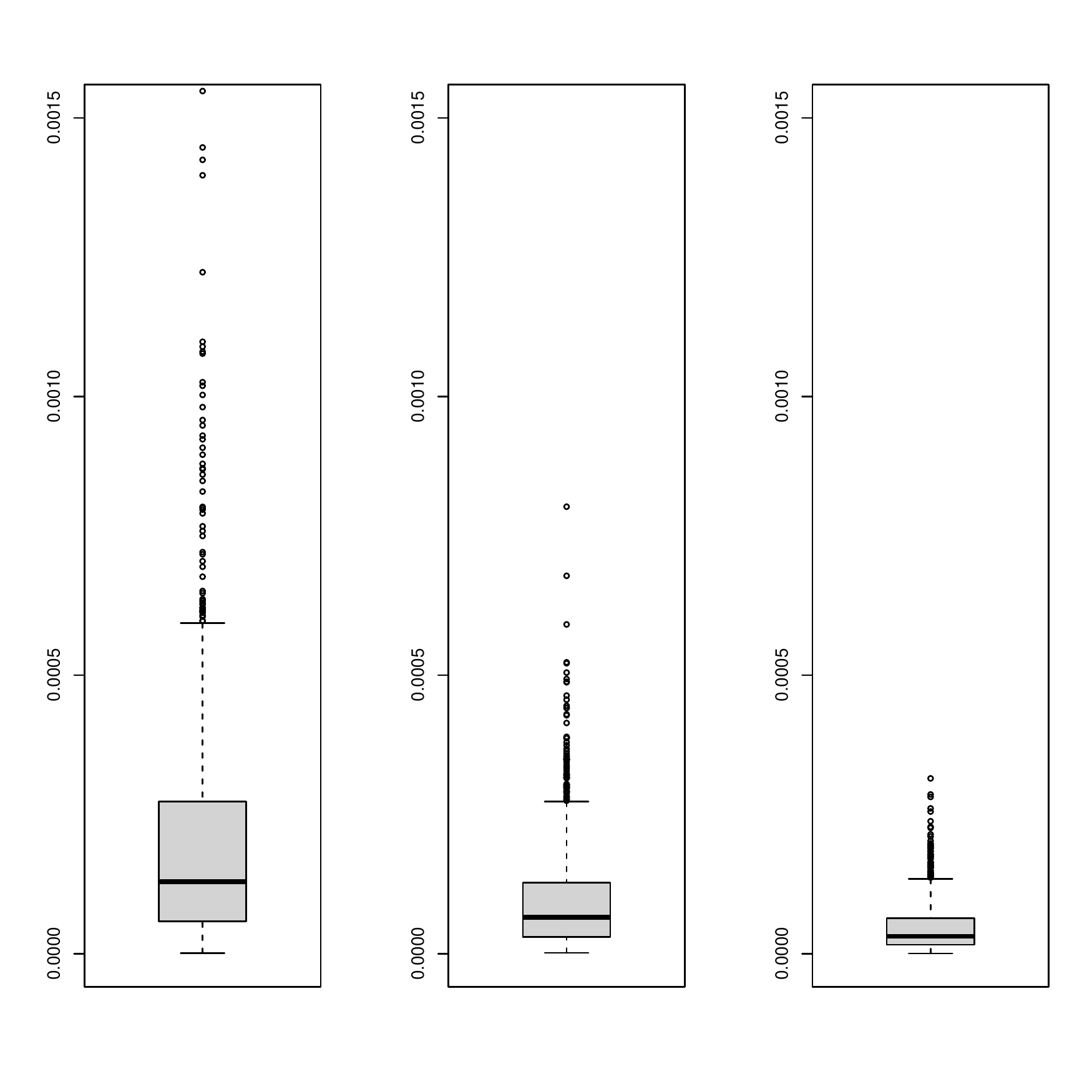}}
	\caption{Boxplots of the errors for the conditional shape function estimates $ISE_r^{\mathcal{F}}$ (left panels) and the conditional intensity factors $ISE_r^{\mathcal{T}}$ (right panels), in the simulation setting for $n=100$ (left), $ n=200$ (middle) and $n=500$ (right). Here four and three outliers were removed for the shape and intensity factor boxplots when $n=100$, 
	respectively.}\label{fig:boxplotISE}
\end{figure}

We ran $1000$ simulations for sample sizes $n=100, 200$ and $500$. For the $r^{th}$ simulation, we measure the performance of the method by comparing against the ``oracle'' conditional intensity function as defined before. Denoting by $\tilde{f}_\oplus^r(x)$ and $\tilde{\tau}_\oplus^r(x)$ the empirical estimates for the shape function and intensity factor parts of the conditional intensity function in the $r$-{th} simulation, respectively, 
we measured the quality of the estimation by integrated squared errors similar to \cite{mull:18:3}, using the metric as in (\ref{metricdef}). Since $\tau_{\oplus}(x)$ can be consistently estimated up to the constant $E(\tau)$, we expect the estimates and $\tau_\oplus(x)$ to differ by a positive constant and the estimates $\tilde{\tau}_\oplus^r(x)$ of $\tau_\oplus(x)$ as in Theorem \ref{Theorem_tauPlusEst} by $E(\tau)$. This leads to
\begin{align*}
ISE_r &=\int_0^1 \left\{ d_{L^2}^2(\tilde{Q}_\oplus(x),E(Q(\cdot) \vert X=x))+d_E^2(E(\tau) \tilde{\tau}_\oplus^r(x),E(\tau \vert X=x))\right\} dx\, \\
& =\,\,ISE_r^{\mathcal{F}}\,+\,ISE_r^{\mathcal{T}}.
\end{align*}
The previous integrals are obtained numerically over a dense grid of predictor values consisting of $200$ equidistant points in $(0,1)$, where $E(Q(\cdot) \vert X=x)$ is obtained through a Monte Carlo approach for each $x$ in the dense grid as explained before. The boxplots of $ISE_r^{\mathcal{F}}$ and $ISE_r^{\mathcal{T}}$
are presented in Figure~\ref{fig:boxplotISE}. As sample size increases, these error estimates are seen to decrease towards $0$. This indicates that the estimated conditional intensity functions converge to their true counterparts, up to the constant $E(\tau)$.

%\begin{figure}[H]
%	\includegraphics[width=0.4\textwidth]{./simulation/LFR_ShapeBoxplot}
%	\includegraphics[width=0.4\textwidth]{./simulation/LFR_tauBoxplot}
%	\centering
%	\caption{Boxplots of the errors for the conditional shape function estimates $ISE_r^{\mathcal{F}}$ (upper panels) and the conditional intensity factors $ISE_r^{\mathcal{T}}$ (lower panels), in the simulation setting for $n=100$ (left), $ n=200$ (middle) and $n=500$ (right). Here four and three outliers were removed for the shape and intensity factor boxplots when $n=100$, 
%	respectively.}\label{fig:boxplotISE}
%\end{figure}

\subsection*{\sf 5.3 Simulations for global Fr\'echet regression}

\begin{figure}[H]
	\includegraphics[width=0.4\textwidth]{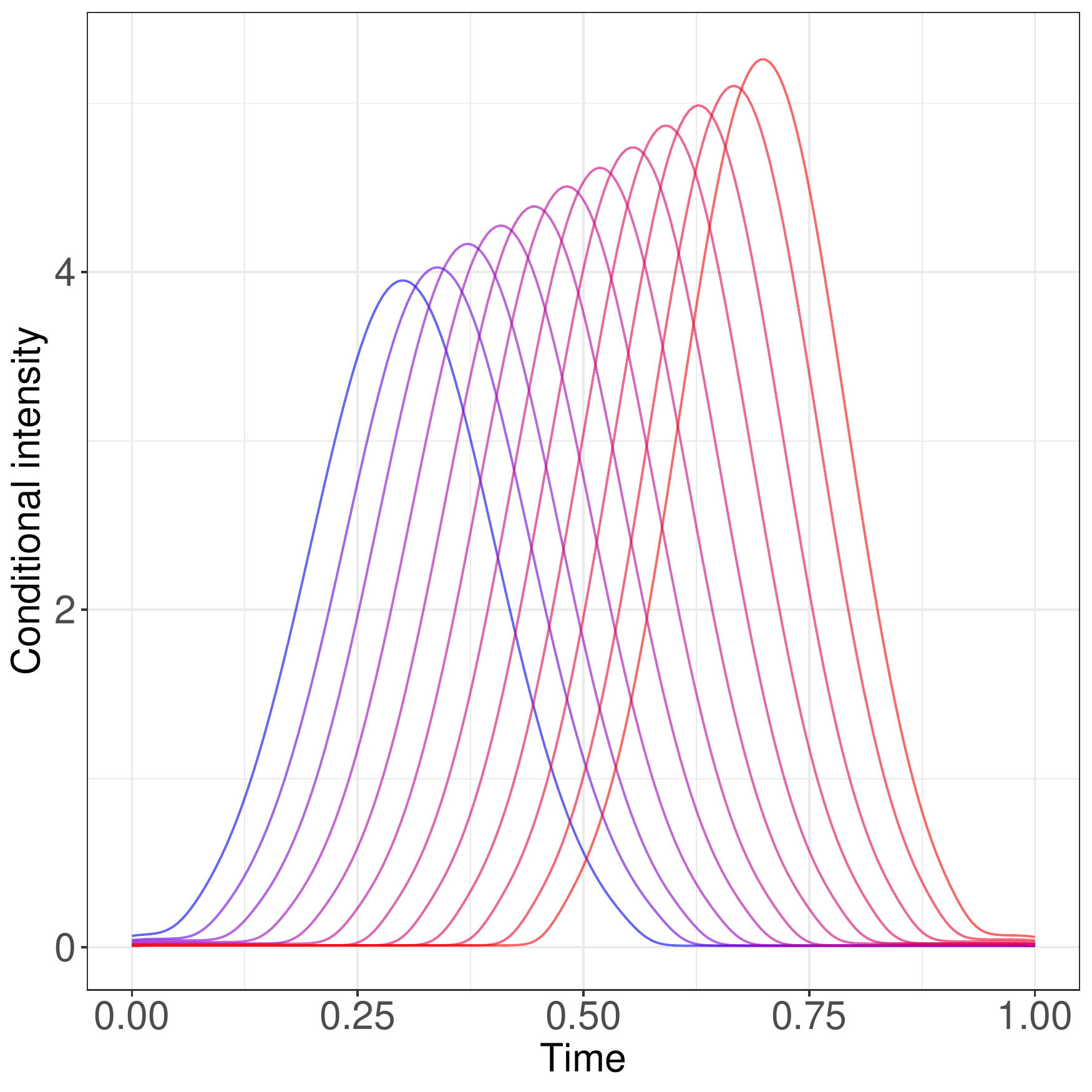}%
	\includegraphics[width=0.4\textwidth]{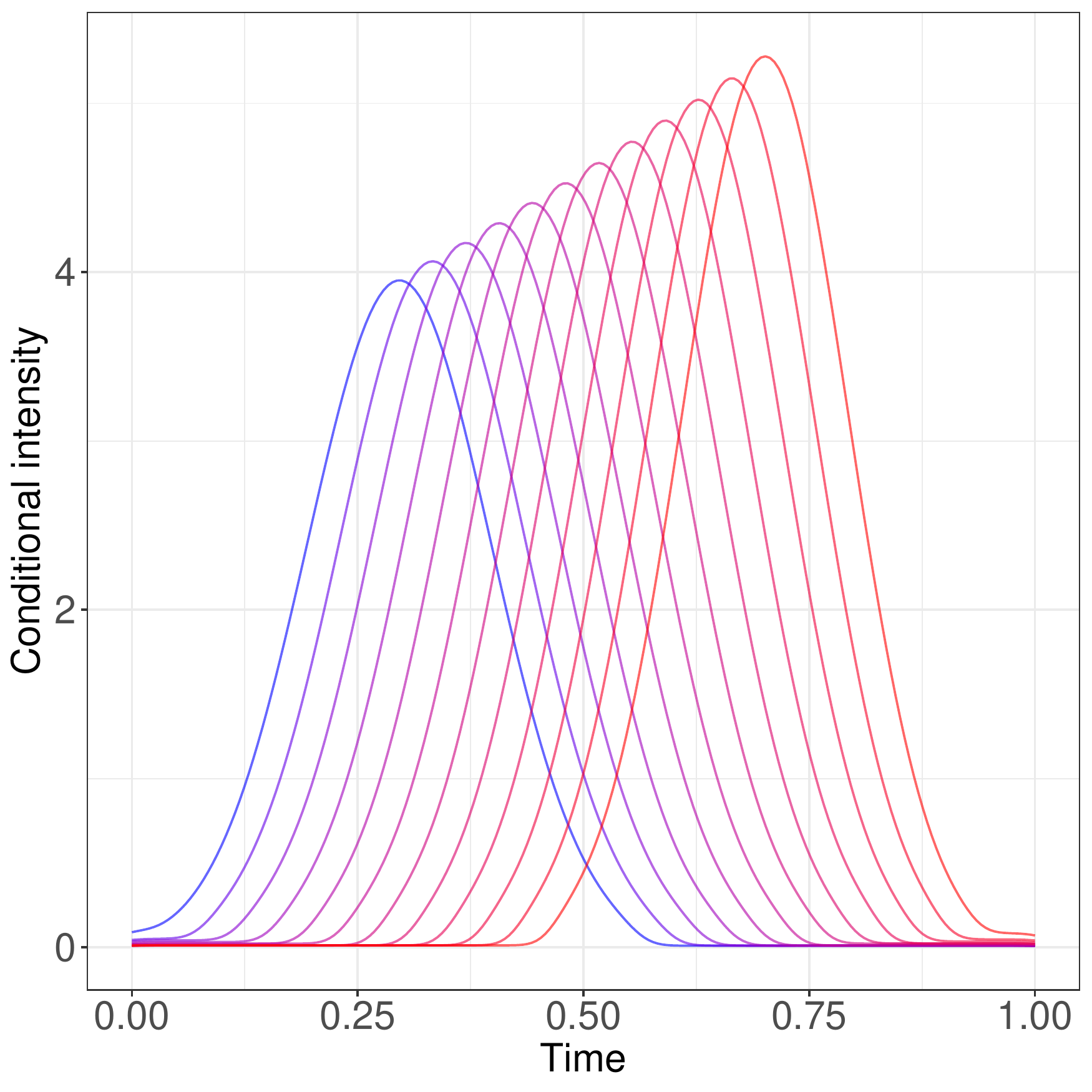}
	\centering
	\caption{Conditional global intensity functions $m_{G\oplus}(x)$ in the simulation setting over a dense grid of predictor levels $x$, displayed in blue when $x=0$ to red when $x=1$. The left panel shows the approximation of $m_{G\oplus}(x)$ through a Monte Carlo approach while the right panel shows the estimate $\hat{\Lambda}_{G\oplus}(x) $ adjusted by the constant $E(\tau)$ using $n=2000$.}\label{fig:oracleGlobalRegresss}
\end{figure}

In this section we assess the finite sample performance of the global Fr\'echet regression estimates. The data generation mechanism is as follows. First generate the covariate $X\sim \mathcal{U}(0,1)$. Then the random density $f$ corresponds to a truncated normal random variable with support $[0,T]$, $T=1$, mean $\mu(x)=a_2+b_2 x +\varepsilon_1$ and standard deviation $\sigma(x)=a_3+b_3 x +\varepsilon_2$, where $\varepsilon_k$, $k=1,2$, is independent of all other random quantities and has a truncated normal distribution with mean zero, standard deviation $\sigma_k$ and support $[e_k,f_k]$ such that $\sigma(x)$ is positive for all $x$ in the support of $F_X$. Thus both the mean and standard deviation change linearly with $x$. We choose $a_2=0.3$, $b_2=0.4$, $a_3=0.1$, $b_3=-0.01$, $e_1=-0.1$, $f_1=0.1$, $e_2=-0.01$, $f_2=0.01$ and $\sigma_1=\sigma_2=0.5$. These settings reflect a  situation where the shape components are Gaussian and pushed to the right as $x$ increases while the intensity factor becomes larger. Figure~\ref{fig:oracleGlobalRegresss} shows the  ``oracle'' global Fr\'echet regression function over a dense grid of predictor values and the estimated counterpart adjusted by the constant $E(\tau)$. Lemma S.11 in the appendix shows that $Q(f_{G \oplus}(x))$ is the orthogonal projection of $Q_{G}(\cdot,x)=E(s(X,x)Q(\cdot))$ onto $Q(\Omega_\mathcal{F})$, where $Q$ is the quantile function corresponding to the generic random density $f$. We obtain $Q_{G}(\cdot,x)$ at each value of $x$ in the grid by employing a Monte Carlo approach similarly as in section 5.2 by averaging across random trajectories $s(X,x)Q_i$.

\begin{figure}[H]
	\centering
	\subfloat[]{\includegraphics[width=0.4\textwidth]{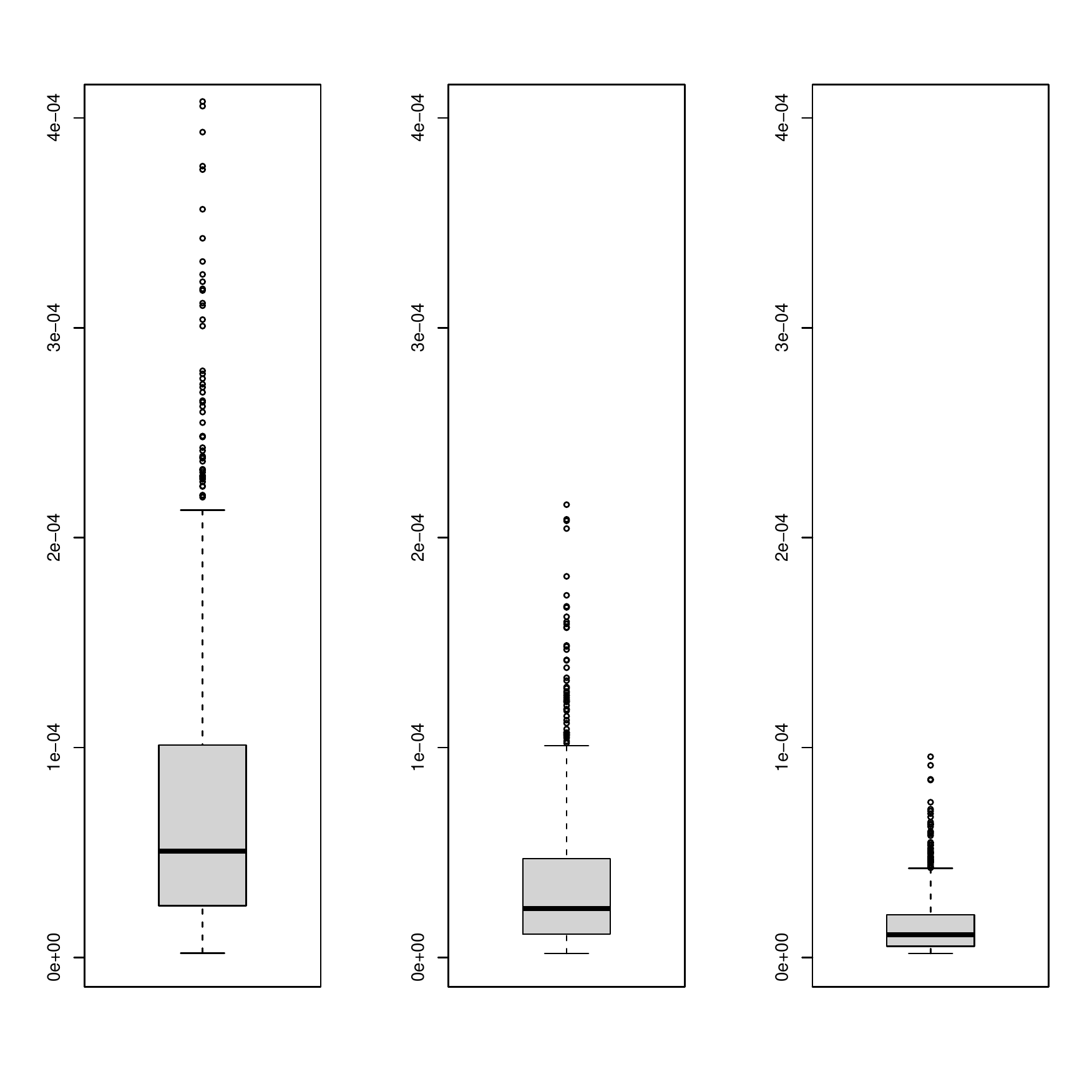}}
	\hfil
	\subfloat[]{\includegraphics[width=0.4\textwidth]{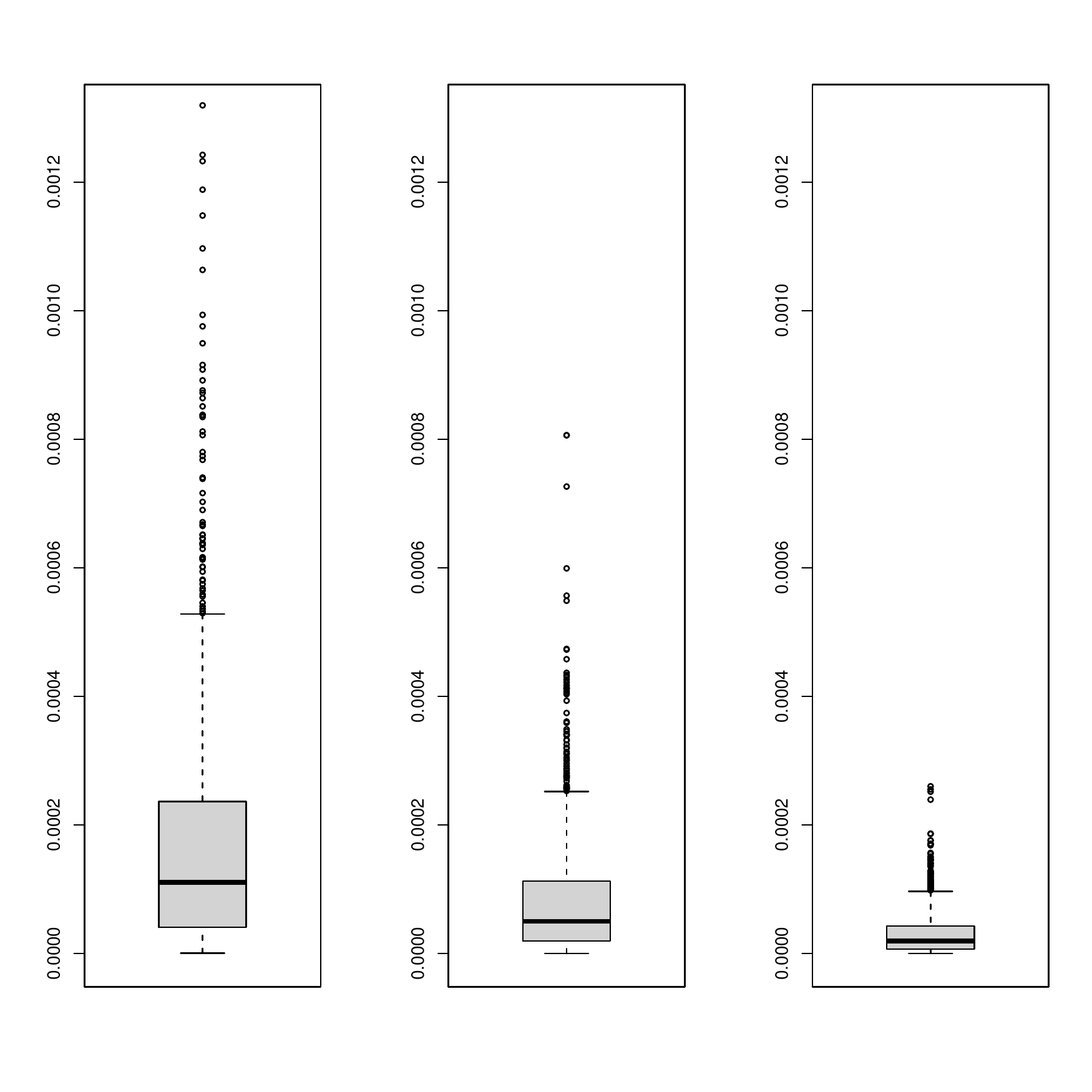}}
	\caption{Boxplots of the errors for the conditional shape function estimates $ISE_r^{\mathcal{F}}$ (left panels) and the conditional intensity factors $ISE_r^{\mathcal{T}}$ (right panels), in the simulation setting for $n=100$ (left), $ n=200$ (middle) and $n=500$ (right). Five outliers were removed for both the shape and intensity factor boxplots when $n=100$, 
		respectively.}\label{fig:boxplotGFR_ISE}
\end{figure}

We ran $1000$ simulations for sample sizes $n=100, 200$ and $500$. For the $r^{th}$ simulation, we measure the performance of the method by comparing against the ``oracle'' global intensity function as defined before and using integrated squared errors analogous to  the ones outlined in section 5.2. The boxplots of the error metrics against the ``oracle'' global intensity function 
are presented in Figure~\ref{fig:boxplotGFR_ISE} and are clearly seen to converge to zero as sample size increases.

%\begin{figure}[H]
%	\centering
%	\includegraphics[width=0.4\textwidth]{./simulation/GFR_ShapeBoxplot}%
%	
%	\includegraphics[width=0.4\textwidth]{./simulation/GFR_tauBoxplot}
%	\centering
%	\caption{Boxplots of the errors for the conditional shape function estimates $ISE_r^{\mathcal{F}}$ (upper panels) and the conditional intensity factors $ISE_r^{\mathcal{T}}$ (lower panels), in the simulation setting for $n=100$ (left), $ n=200$ (middle) and $n=500$ (right). Five outliers were removed for both the shape and intensity factor boxplots when $n=100$, 
%	respectively.}\label{fig:boxplotGFR_ISE}
%\end{figure}

\section*{\sf 6. Data Applications and Extensions}\label{s:data}
\subsection*{\sf 6.1 Chicago's Divvy Bike System}

We illustrate our approach for the bike trips records of the Chicago Divvy bike system, which is publicly available at \texttt{https://www.divvybikes.com/system-data}.   The bike trip records contain information such as the bike pickup and drop-off location, date and time, between more than $600$ bike rental stations in Chicago.
In the context of replicated temporal Poisson processes, \cite{gerv:19} analyzed this dataset by adapting an additive principal component model to the log-intensity functions of daily pickups and estimating model parameters through a likelihood based approach, and such 
bike sharing systems  have been extensively studied  \citep{borg:11}.
We considered the point process of daily pickups of bikes in a cluster consisting of $6$  stations not far from each other in the Chicago Divvy system during weekdays of $2017$, consisting of a station on East South Water street and the five nearest bike rental stations south of the Chicago river.

\begin{figure}[!ht]
\subfloat[]{%
\includegraphics[width=0.49\textwidth]{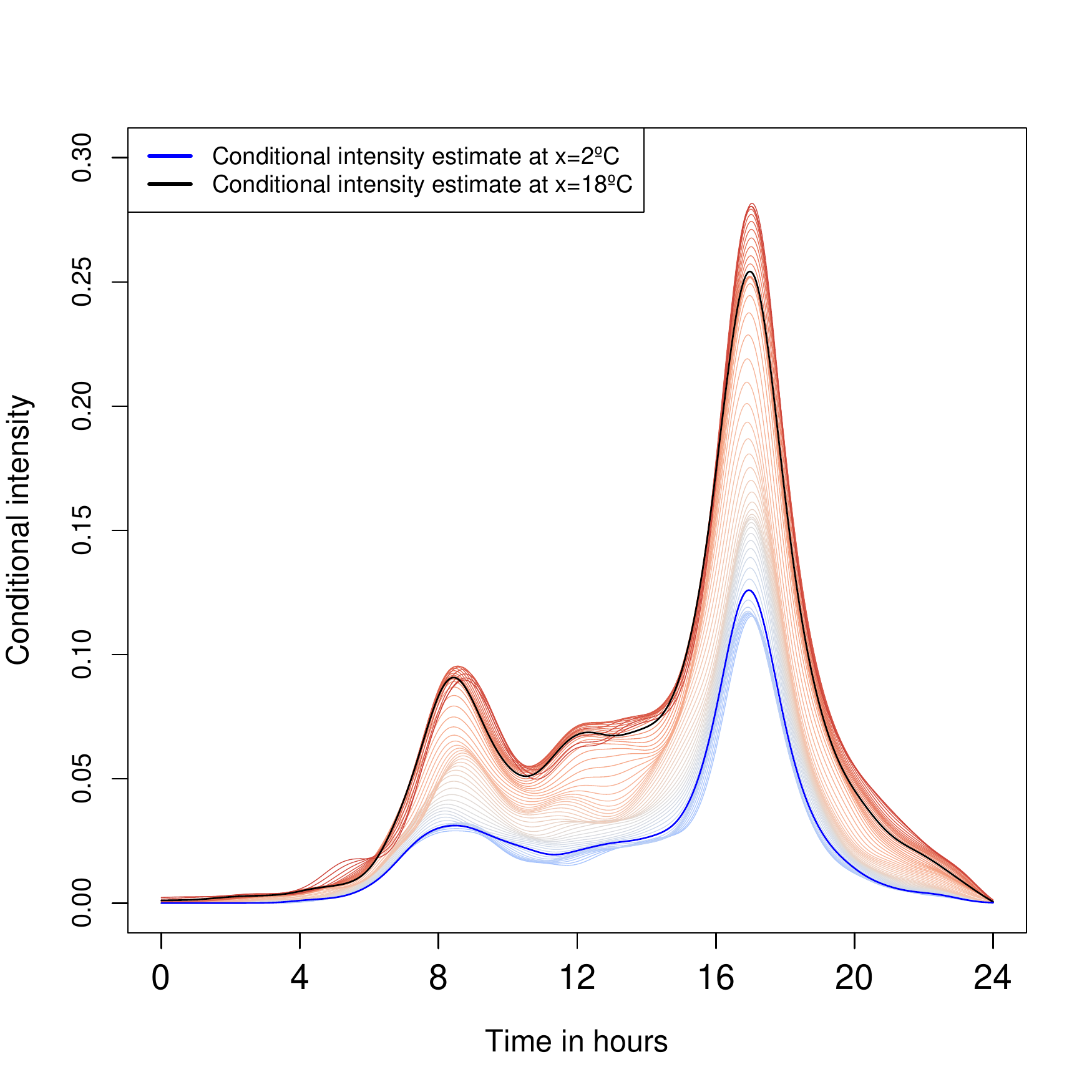}}
\centering
\hfill
\subfloat[]{%
\includegraphics[width=0.49\textwidth]{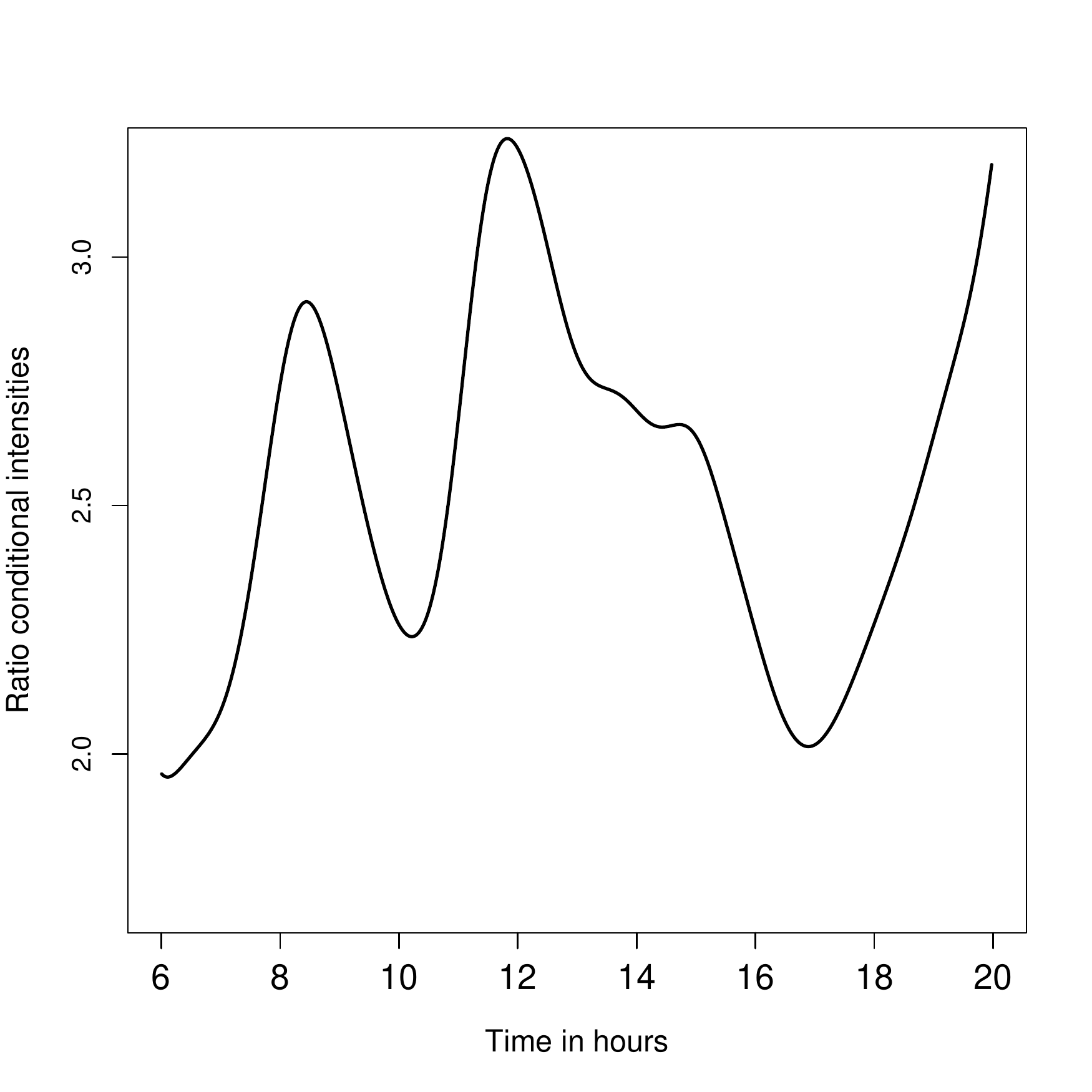}}
\centering
\caption{(a) Estimated conditional intensity functions in dependence on temperature values between $-2 ^\circ$C (blue) and $24^\circ$C (red), using  bandwidth $h=1.5^\circ$C, with highlighted  conditional intensity functions at temperatures $2^\circ$C and $18^\circ$C. (b) Ratio between the estimated conditional intensity functions at temperature $18^\circ$C to that at $2^\circ$C.}\label{fig: bike_results}
\end{figure}

To study the effect of the temperature on the demand of bikes, we obtained the daily observed temperature in Chicago as recorded at the weather station `Northerly island' from \texttt{https://www.ncdc.noaa.gov} and fitted a local Fr\'echet regression model to obtain the conditional intensity functions of the bike rentals, using estimates $\hat{\Lambda}_\oplus(x)$ as in (\ref{estimate_cond_intensity}), where we used a bandwidth of $1.5^\circ$C. The results are presented in Figure \ref{fig: bike_results}.

A clear difference emerges  between days with temperature above $10^\circ$C which have a uniformly higher intensity function compared to days with lower temperature. In both cases, the shape of the intensity function appears to be bimodal with peaks at $9$am and $5$pm, which are likely  due to bike rentals for the purpose of commuting to the workplace. Moreover, the conditional intensity function estimate is higher around $5$pm compared to the early peak at $9$am, which may be explained by the fact that it is warmer and easier to bike in the afternoon than early in the morning, so perhaps commuters use public or shared transport in the morning and a bike in the afternoon.

There appears to be a ``shoulder" or minor peak of bike rental demand at around $12$pm on warm days only, which is likely related to more or less optional lunch break related bicycle travel, for leisure or to catch some food. Overall, we find that increasing temperature boosts the bike rental demand in this region of Chicago. Figure \ref{fig: bike_results} (b) shows the quotient between the estimated conditional intensity function at $x=2^\circ$C and $x=18^\circ$C. We observe that the ratio is much higher around noon compared to the ratios at $9$am and $5$pm;  moreover, this ratio is higher for the morning commute compared to the evening commute, indicating that  the afternoon demand is not just a reflection of the morning demand.  This ratio can be characterized as the degree to which bike travel is optional, where the obvious alternatives to making  a trip by  bike are  not making a trip or making the trip by other means of transportation.

\subsection*{\sf 6.2 New York Yellow Taxi System}
The New York yellow taxi trip records is a rich and large scale database that contains information such as the taxi pickup and drop-off latitude and longitude locations as well as the date and time, among several other variables. The data is available from the NYC Taxi and Limousine Commission (TLC) at \texttt{https://www1.nyc.gov/site/tlc/about/data.page}.
Using point processes, this data has been studied by several authors in an applied setting; see for example \cite{Saya:16} and the references therein for a review and comparison of different intensity models. The Poisson process as a working model for the taxi pickups at a fixed location is well justified theoretically as a superposition of many independent and sufficiently sparse point processes, similar to the case of call arrivals at a telephone exchange \citep{cox:80}. It is of interest to study how the demand of taxis is associated with the day of the week and for this we employ a regression approach.

We consider the point process of daily pickups of yellow taxis that occurred at Penn station in Manhattan during $2017$. Penn station is a major train station located in Midtown Manhattan that serves commuters from New York City, New Jersey and Long Island, and connects NYC with several other cities. We view these data as a sample of  replicated point processes, as  each day produces a replication of the underlying data generation mechanism. To study the effect of weekdays and weekends on the demand of taxis at Penn station, we consider a categorical predictor $X$ that indicates whether the day corresponds to a Monday-Thursday, Friday, Saturday or Sunday. Since local smoothing does not apply to indicator type predictors, we instead consider the global regression framework that was introduced in section 4.4 and is  well suited for categorical predictors.

\begin{figure}[!ht]
\includegraphics[width=0.5\textwidth]{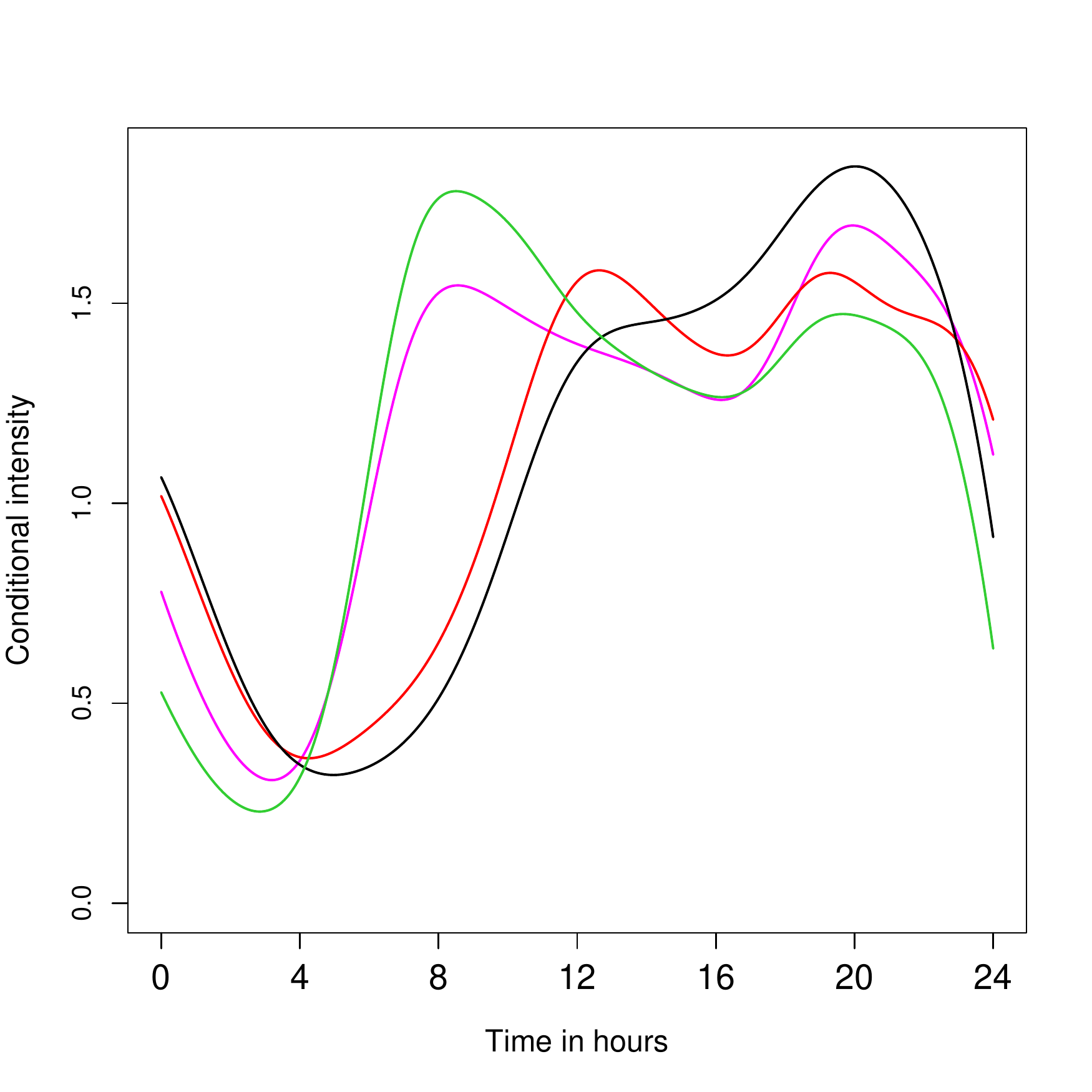}
\centering
\caption{Estimated intensity functions using global regression with type of day as predictors.  The fitted response functions are shown for Monday-Thursday (green), Friday (magenta), Saturday (orange) and Sunday (black). The Y-axis is scaled by a factor of $10^{-5}$.}\label{fig: taxi_results}
\end{figure}

Fitting a global regression model for the intensity function on the day of week $X$ by using the estimates $\hat{\Lambda}_{G\oplus}(x)$ defined after Theorem \ref{Theorem_gfr_fPlusEst} leads to the results as  presented in Figure \ref{fig: taxi_results}. For Sundays, the intensity function is highest late in the day and after 4pm is higher than on all other days, likely due to people returning to New York City from an out of town  trip. The weekday (Monday through Thursday) intensity function is bimodal with a higher first mode.  These modes likely correspond to the commuter traffic, where the 8am mode would be due to commuters who live outside New York City and arrive for work in the City in the morning,  while the evening mode likely corresponds to people who return from out of town at Penn Station and hail a taxi there. On Fridays, the same modes are present but with a reversal of their height, as the second mode is now higher than the first mode,  likely corresponding to reverse commuters who live in New York City and return from outside, perhaps having a work place away from the City.

The patterns for Saturday are also bimodal but with different locations and levels of the modes, indicating that relative large numbers of people arrive at Penn Station around noon and at 8pm, perhaps indicating leisure and shopping trips.

\subsection*{\sf 6.3 Aftershock Earthquake Process in Chile}

The proposed Cox process regression also has applications in   seismology, where the times earthquakes strike naturally forms a point processes. So it is not surprising that 
earthquake activity  has met with major  interest in the point process literature \citep{dale03}. Chile is widely known for its strong seismic activity in both frequency and intensity. Our goal is to study the aftershock process that follows a major  earthquake occurring at time $t=0$, where the major earthquake that may trigger aftershocks is  referred to as the  mainshock. We focus on  the arrival of aftershocks that occur in a time window of $2$ months after the mainshock so that the aftershock process  $N(t)$ is observed in $[0,T]$ for $T=2$ months.

To demonstrate the proposed regression methods, we considered the mainshocks that occurred between $1980$ and $2017$ in Chile. These  include some strong earthquakes such as the magnitude $8.8$ earthquake on the moment magnitude scale in February $2010$, $8.2$ in April $2014$, $8.0$ in March $1985$, and other strong earthquakes. The data was obtained from the U.S. Geological Survey web page \texttt{https://earthquake.usgs.gov/earthquakes/search/} which provides information concerning the location, magnitude and date of the earthquakes.
We classified each earthquake in terms of its magnitude category and selected $15$ mainshocks at random from each category in the order strongest to weakest. This enables us to consider the strongest earthquakes above magnitude $7$; see Table~\ref{tab:freqtable}. 
Table~\ref{tab:eq2010} shows some of the strongest aftershocks along with their arrival time after the $2010$ earthquake. 

In order to avoid including an aftershock that could correspond to more than one mainshock, we adopted the following selection scheme: If we select an earthquake that occurs at calendar time $t_0$ as mainshock, then we cannot choose any earthquake that occurs in the interval $(t_0-T,t_0+T)$ as mainshock.  Furthermore, we consider an earthquake to be a mainshock if the sequence of earthquakes that occur during the following $2$ months after its arrival time have strictly smaller magnitudes.

\begin{table}[H]
	\begin{center}
		\begin{tabular}{| c | c | c |}
			\hline
			Magnitude & Number of Earthquakes \\ \hline
			$[5.0,5.5)$ & $1,679$ \\ \hline
			$[5.5,6.0)$ & $476$ \\ \hline
			$[6,6.5)$ & $160$ \\ \hline
			$[6.5,7.0)$ & $53$ \\ \hline
			$[7.0,9.0)$ & $25$ \\ \hline
		\end{tabular}
		\caption{\label{tab:freqtable} Distribution of earthquakes in Chile between $1980$ and $2017$.}
	\end{center}
\end{table}

\begin{table}[H]
	\begin{center}
		\begin{tabular}{| c | c | c |c |c |c |c |c |}
			\hline
			Magnitude &$6.2$& $6.0$ &  $6.0$& $7.4$& $6.1$& $6.0$& $6.0$                \\ \hline
			Aftershock Time [hours] & $0.3$& $0.6$& $1.1$&$1.5$& $1.9$& $3.9$&$3.9$    \\ \hline
		\end{tabular}
		\single\caption{\label{tab:eq2010} Some of the strongest aftershocks within $4$ hours from the magnitude $8.8$ earthquake on 27 February, $2010$, Chile.}
	\end{center}
\end{table}

We implemented conditional intensity function estimation using the magnitude of the mainshock as a one-dimensional predictor. Figure~\ref{fig: earthquake_intensity_all} shows the estimated conditional intensity functions (\ref{estimate_cond_intensity}) for different levels of magnitudes between $5$ and $8.5$, along with the conditional intensity for the mean magnitude level $x=6.3$ and for a strong earthquake at magnitude $x=8.0$.
We observe that the conditional intensity function for the strong earthquake has an exponential decay, suggesting  that most aftershocks occur closer to the mainshock. The shape of the regression curve  agrees with the traditional model for aftershock sequences that correspond to large earthquakes, with intensity function declining as a power law, which is known as  the modified Omori law \citep{dale03,utsu:95}.

The regression curve for a magnitude $6.3$ earthquake is mostly flat but presents a slow decay towards the end of the time window of observation. This could be due to the fact that medium or low magnitude earthquakes do not tend to produce substantially more aftershocks compared to the natural seismic background activity in Chile. In fact, the conditional intensity function for the strong earthquake is uniformly higher than the one for the magnitude $6.3$ mainshock. Finally, the conditional intensity functions for earthquakes below magnitude $6$ are mostly flat, which indicates that those earthquakes tend to produce aftershocks that are more uniformly scattered and at least partly correspond to the background seismic activity in Chile.\\

\section*{\sf 7. Conclusions}\label{s:conclusion}

We develop here a novel fully non-parametric regression method that features point processes as responses coupled with Euclidean predictors $X\in \mathbb{R}^p$ by establishing  a connection to  conditional barycenters. Crucially, our model is based on the availability of repeated realizations of the same point process. The random objects for which we construct conditional barycenters are the  intensity functions of  Cox processes,  which we can represent as taking values in a product metric space.  A novelty in point processes is that  the regression setting makes it possible to achieve 
consistent estimation of intensity functions (up to a constant scale factor for the intensity that is common to all observed realizations of the point process).

Obtaining such consistency has been an elusive goal and in fact is  not possible when one has one realization of the point process over a fixed domain.
What we show here is that this lack of consistency can be overcome a regression setting where one can harness concepts of conditional barycenters that have been developed for Fr\'echet regression. For each point process, one may have a continuous one-dimensional or general vector predictor that is a random variable associated with the point process. In the former case we can use a nonparametric smoothing method under minimal assumptions, while in the latter case we target a global model that is akin to multiple linear regression and makes it possible to include indicators as predictors.

Our approach relies on straightforward computations and does not require the use of functional principal components, a tool that is not well suited for intensity functions as they do not reside in a linear space due to their non-negative nature. We show that the proposed regression model is applicable to many data situations where one is interested to study  the behavior of point processes in dependence on covariates, including applications in transportation and seismology.

\clearpage

\newgeometry{letterpaper,left=0.57in,right=0.57in,top=0.57in,bottom=0.57in}
\appendix

\section*{\sf  \centering Appendix A }

\subsection*{\sf A.1 Proofs of Results in Section 4.1}
\vspace{.3cm} 

\noindent
\textbf{Proof of Lemma \ref{Lemma_QOmegaf_closedconvex}}\\
Let $Q_1,Q_2\in Q(\Omega_\mathcal{F})$, $\lambda \in [0,1]$ and $Q_\lambda=\lambda Q_1+(1-\lambda)Q_2$ and $x,y\in [0,1]$. It is clear that $Q_\lambda(0)=0$, $Q_\lambda(1)=T$, and by the triangle inequality $\lvert Q_\lambda(x)-Q_\lambda(y)\rvert \leq L \lvert x-y \rvert$.
Suppose now that $x\leq y$, then since $Q_\lambda$ is non-decreasing we have $\lvert Q_\lambda(x)-Q_\lambda(y)\rvert =Q_\lambda(y)-Q_\lambda(x)$. Furthermore, for $M$ as in (S1), 
\begin{gather*}
Q_\lambda(y)-Q_\lambda(x)=\lambda (Q_1(y)-Q_1(x))+ (1-\lambda) (Q_2(y)-Q_2(x))\\
\geq \lambda M (y-x) +  (1-\lambda) M (y-x) =M (y-x),
\end{gather*}
which implies $\lvert Q_\lambda(x)-Q_\lambda(y)\rvert \geq M \lvert x-y \rvert$ for all $x,y\in [0,1]$. Hence, $Q(\Omega_\mathcal{F})$ is convex and it is clearly a subset of $L^2([0,1])$ since the quantile functions are bounded.
Next, let $Q_1,Q_2, \dots$ be a sequence in $Q(\Omega_\mathcal{F})$ such that $Q_n \overset{L^2}{\to} Q \in L^2([0,1])$ as $n\to \infty$. We show that $Q \in Q(\Omega_\mathcal{F})$. In fact, since the family $\{ Q_n \}_{n=1}^\infty$ has a common Lipschitz constant $L$, it follows that it is uniformly equicontinuous. Moreover, since $[0,1]$ is compact and $Q_n \overset{L^2}{\to} Q \in L^2([0,1])$ then $Q_n\to Q$ uniformly as $n \to \infty$. This implies $Q(0)=0$ and $Q(1)=T$. Next, for any $x,y\in [0,1]$ and $\epsilon>0$ we have that $\lvert \lvert Q_n-Q \rvert \rvert_\infty\leq \epsilon/2$ for $n$ large enough and 
\begin{gather*}
\lvert Q(x)-Q(y) \rvert \leq \lvert Q(x) - Q_n(x) \rvert + \lvert Q_n(x) - Q_n(y) \rvert + \lvert Q_n(y) - Q(y) \rvert \\
\leq 2 \lvert \lvert Q_n-Q \rvert \rvert_\infty + L \lvert x-y \rvert\leq \epsilon + L \lvert x-y \rvert,
\end{gather*}
for large enough $n(\epsilon)$, using that $Q_n\in Q(\Omega_\mathcal{F})$ by assumption and the functions in this space satisfy the Lipschitz condition with constant $L$.   Taking $\epsilon \downarrow 0$ we obtain that $Q$ is also Lipschitz with constant $L$. Similarly, for $n$ large enough
\begin{align*}
M \lvert x-y \rvert &\leq \lvert Q_n(x) - Q_n(y) \rvert 
\leq 2 \lvert \lvert Q_n-Q \rvert \rvert_\infty + \lvert  Q(x)-Q(y) \rvert \leq \epsilon + \lvert  Q(x)-Q(y) \rvert, 
\end{align*}
so that $Q$ satisfies condition $(S1)$. Therefore, $Q(\Omega_\mathcal{F})$ is closed.\newline

\single \begin{figure}[H]
	\includegraphics[width=0.60\textwidth]{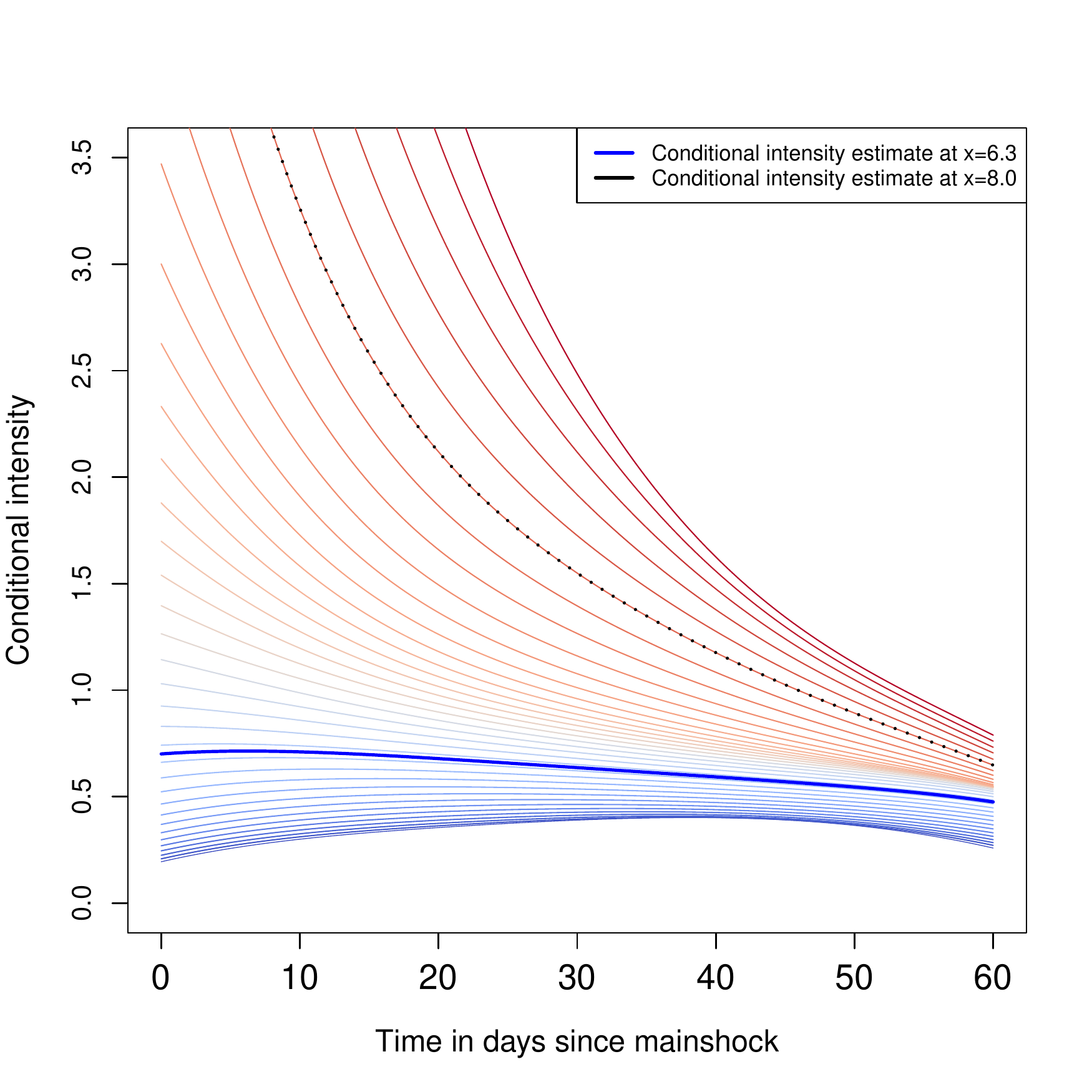}
	\centering
	\caption{\label{fig: earthquake_intensity_all} Estimated conditional intensity functions for the aftershock process for different levels of magnitude between $5.0$ (blue) and $8.5$ (red), with specific conditional intensity functions at magnitudes $6.3$ and $8.0$, and bandwidth $0.8$ in magnitude. The Y-axis is scaled by a factor of $10^{-3}$.}
\end{figure}

The following lemma shows that there are no empty point processes and follows by adopting analogous arguments as the ones outlined in the proof of Lemma 3  in \cite{pana:16}. We present it here only for completeness and without proof. In what follows, we introduce auxiliary quantities $s_i(x,h)=\sigma_0^{-2} K_h(X_i-x)[\mu_2 -\mu_1 (X_i-x)]$, $i=1,\dots,n$, where $\mu_j=E(K_h(X_1-x)(X_1-x)^j)$, $j\in\{0,1,2\}$ and $\sigma_0^2=\mu_0\mu_2-\mu_1^2$.
\vspace{.3cm} 

\noindent
\textbf{Lemma S.1 } \, {\it Suppose that $\tau \geq \kappa>0$ almost surely with $\kappa$ as in assumption $(S3)$ and $\alpha_n/\log n \to \infty$ as $n\to \infty$. Then}
\begin{align*}
	\liminf_{n \to \infty} \frac{\min_{1\leq i \leq n} N_i^{(n)}(T)}{\alpha_n} &\geq \frac{ (1-e^{-1})\kappa}{2}  \quad \text{a.s.}.
\end{align*}
\noindent

\noindent
\textbf{Proof of Proposition \ref{Proposition_consistentWassEst}}.\\
From (\ref{key_eqn_density}) we have
\begin{align}
	d_\mathcal{F}(\hat{f}_\oplus(x),\tilde{f}_\oplus(x))\leq n^{-1} \sum_{i=1}^n \lvert s_{in}(x,h) \rvert \  \lvert \lvert  \hat{Q}_i - {Q}_i \rvert \rvert_{L^2([0,1])}  \label{eq:local_bound_0}.
\end{align}
From the proof of Lemma $2$ in  \cite{mull:18:3}, we have $s_{in}(x,h)-s_i(x,h)=W_{0n} K_h(X_i-x)+W_{1n}K_h(X_i-x)(X_i-x)$, where $W_{0n}=\hmu_2/\hsigma_0^2 -\mu_2/\sigma_0^2=O_p((nh)^{-1/2})$ and $W_{1n}=\hmu_1/\hsigma_0^2 -\mu_1/\sigma_0^2=O_p((nh^3)^{-1/2})$. Thus
\begin{align}
	n^{-1} \sum_{i=1}^n \lvert s_{in}(x,h) \rvert \  \lvert \lvert  \hat{Q}_i - {Q}_i \rvert \rvert_{L^2([0,1])}
	&\le 
	n^{-1} \sum_{i=1}^n \lvert s_{in}(x,h)-s_i(x,h) \rvert \  \lvert \lvert  \hat{Q}_i - {Q}_i \rvert \rvert_{L^2([0,1])}
	+
	n^{-1} \sum_{i=1}^n \lvert s_{i}(x,h) \rvert \  \lvert \lvert  \hat{Q}_i - {Q}_i \rvert \rvert_{L^2([0,1])}\nonumber \\
	&\le  W_{0n} \ n^{-1} \sum_{i=1}^n  K_h(X_i-x)  \lvert \lvert  \hat{Q}_i - {Q}_i \rvert \rvert_{L^2([0,1])}\nonumber\\
	& \quad +
	W_{1n} \ n^{-1} \sum_{i=1}^n  K_h(X_i-x) \lvert X_i-x\rvert \ \lvert \lvert  \hat{Q}_i - {Q}_i \rvert \rvert_{L^2([0,1])}\nonumber\\
	&\quad \quad + 
	\lvert \mu_2/\sigma_0^2\rvert \ n^{-1} \sum_{i=1}^n  K_h(X_i-x)  \lvert \lvert  \hat{Q}_i - {Q}_i \rvert \rvert_{L^2([0,1])}\nonumber\\
	&\quad \quad \quad + 
	\lvert \mu_1/\sigma_0^2\rvert \ n^{-1} \sum_{i=1}^n  K_h(X_i-x) \lvert X_i-x\rvert \  \lvert \lvert  \hat{Q}_i - {Q}_i \rvert \rvert_{L^2([0,1])}. \label{eq:local_bound}
\end{align}
Let $\mu_i$ be the probability measure on $[0,T]$ with corresponding quantile function $Q_i$. Note that $\lvert \lvert  \hat{Q}_i - {Q}_i \rvert \rvert_{L^2([0,1])}= d_W(\hmu_i,\mu_i) 1_{(\Nin>0)} + d_W(\hmu_i,\mu_i) 1_{(\Nin=0)} \le d_W(\hmu_i,\mu_i) 1_{(\Nin>0)} + T 1_{(\Nin=0)} \le d_W(\hmu_i,\mu_i) 1_{(\Nin>0)}$ almost surely and for large enough $n$, where the last inequality is due to Lemma S.1 above using the condition $\alpha_n/\log n \to \infty$ as $n\to \infty$. This along with the fact that $E(d_W(\hmu_i,\mu_i) 1_{(\Nin>0)}\vert \Lambda_i,X_i)=E(d_W(\hmu_i,\mu_i) 1_{(\Nin>0)}\vert \Lambda_i)\le (T/(2\kappa)^{1/4}) \alpha_n^{-1/4}$ for large enough $n$, which follows by similar arguments as the ones outlined in the proof of Theorem 2 in \cite{pana:16}, shows that
\begin{align*}
	E(\lvert \lvert  \hat{Q}_i - {Q}_i \rvert \rvert_{L^2([0,1])}\vert \Lambda_i,X_i) &=O(\alpha_n^{-1/4}), 
\end{align*}
for large enough $n$, where the $O(\alpha_n^{-1/4})$ term is uniform in $i$. Thus, for $j\in\{0,1\}$ and by a conditioning argument, we have
\begin{align*}
	E\left( n^{-1} \sum_{i=1}^n  K_h(X_i-x) \lvert X_i-x\rvert^j \  \lvert \lvert  \hat{Q}_i - {Q}_i \rvert \rvert_{L^2([0,1])}\right) &=
	n^{-1} \sum_{i=1}^n E\left( K_h(X_i-x) \lvert X_i-x\rvert^j \  	E(\lvert \lvert  \hat{Q}_i - {Q}_i \rvert \rvert_{L^2([0,1])}\vert \Lambda_i,X_i)   \right)\\
	&=  O(\alpha_n^{-1/4}) \ n^{-1} \sum_{i=1}^n E\left( K_h(X_i-x) \lvert X_i-x\rvert^j  \right)=O(h^j \alpha_n^{-1/4}),
\end{align*}
which implies
\begin{align}
	n^{-1} \sum_{i=1}^n  K_h(X_i-x) \lvert X_i-x\rvert^j \  \lvert \lvert  \hat{Q}_i - {Q}_i \rvert \rvert_{L^2([0,1])}=O_p(h^j \alpha_n^{-1/4}) \label{eq:local_bound_2}.
\end{align}
Define auxiliary quantities $\gamma_{k}=\int_{-1}^1 u^k K(u)du$, $k=0,1,2$. By Taylor expansion it is easy to see that $u_0=f(x)+h \int_{-1}^1 u K(u) f'(\xi_{0u}) du$, $u_1=h^2 (\int_{-1}^1 u^2 K(u) f'(\xi_{1u}) du )$, $u_2=h^2  \gamma_2 f(x) + h^3 \int_{-1}^1 u^3 K(u) f'(\xi_{2u}) du$, where $\xi_{1u}$, $\xi_{2u}$  and $\xi_{3u}$ are between $x$ and $x+uh$. Since $a_k:= \int_{-1}^1 u^{k+1} K(u) f'(\xi_{ku}) du $, $k=0,1,2$, satisfies $\lvert a_k\rvert \le \sup_{s\in [x-1,x+1]} \lvert f'(s)\rvert <\infty$ for large enough $n$, it follows that
\begin{align*}
	\lvert \mu_2/\sigma_0^2\rvert &= (\gamma_2 f(x)+ h a_2)/( \gamma_2 f^2(x) +h(a_2f(x)+a_0\gamma_2 f(x))+h^2(a_0a_2-a_1^2))=O(1),\\
	\lvert \mu_1/\sigma_0^2\rvert &= a_1/( \gamma_2 f^2(x) +h(a_2f(x)+a_0\gamma_2 f(x))+h^2(a_0a_2-a_1^2))=O(1),
\end{align*}
as $n\to\infty$. Combining with (\ref{eq:local_bound}), (\ref{eq:local_bound_2}) and the fact that $W_{0n}=O_p((nh)^{-1/2})$ and $W_{1n}=O_p((nh^3)^{-1/2})$ leads to
\begin{gather*}
	n^{-1} \sum_{i=1}^n \lvert s_{in}(x,h) \rvert \  \lvert \lvert  \hat{Q}_i - {Q}_i \rvert \rvert_{L^2([0,1])} =O_p(\alpha_n^{-1/4}),
\end{gather*}
and the result then follows from (\ref{eq:local_bound_0}).

\vspace{.3cm} 

\noindent
\textbf{Proof of Theorem \ref{Theorem_densityEst}.}\\
Under the assumptions $(K0)$, $(L1)$ and $f_X>0$ with unbounded support and the choice of $h_n$, we have from Corollary $1$ in \cite{mull:18:3} that
\begin{equation*}
	d_\mathcal{F}(f_\oplus(x),\hat{f}_\oplus(x))=O_p(n^{-2/5}).
\end{equation*}
Next, from Proposition \ref{Proposition_consistentWassEst} we obtain $	d_\mathcal{F}(\hat{f}_\oplus(x),\tilde{f}_\oplus(x))=O_p(\alpha_n^{-1/4})$. Combining with  the triangle inequality  (\ref{triang_ineq}) leads to
\begin{equation*}
	d_\mathcal{F}(f_\oplus(x),\tilde{f}_\oplus(x))\leq d_\mathcal{F}(f_\oplus(x),\hat{f}_\oplus(x))+d_\mathcal{F}(\hat{f}_\oplus(x),\tilde{f}_\oplus(x))=O_p(n^{-2/5}+\alpha_n^{-1/4}),
\end{equation*}
which shows the result.\\

\subsection*{\sf A.2 Proof of Results in Section 4.2}

We begin by establishing four auxiliary lemmas.\vspace{.3cm} 

\noindent
\textbf{Lemma S.2} \, {\it Suppose that there exists $M_1<\infty$ such that $\tau \leq M_1$ almost surely and let $i\in \{1,\dots,n\}$. Then, conditionally on a realization of $\Lambda_i$, $\Nin/\alpha_n-\tau_i$ is mean zero with (conditional) variance bounded above by $M_1/\alpha_n$.}

\vspace{.3cm}

\noindent
\textbf{Proof of Lemma S.2.} 

Since $N_{i}^{(n)}$ is a Cox process, it is generated by two independent random mechanisms: First the generation of the intensity function $\Lambda_i$, and then, conditional on $\Lambda_i$, of the realizations of $N_{i}^{(n)}$ corresponding to those of a Poisson process $N_{i}^{(n)}(\cdot \vert \Lambda_i)$ with intensity function $\alpha_n \Lambda_i$ \citep{dale03}. Thus, we may regard the probability space associated with the generation of the intensity function and the Poisson process as a product probability space $\mathcal{W}_1 \times \mathcal{W}_2$ such that $\Lambda_i=\Lambda_i(w_1)$, where  $w_1 \in \mathcal{W}_1$ leads to  a realization of the intensity function and then $N_{i}^{(n)}(\cdot)=N_{i}^{(n)}(\cdot,w_2)$, where $w_2 \in \mathcal{W}_2$ leads to a  realization of a Poisson process with intensity function $\alpha_n \Lambda_i$. For $\tau_i=\int_{0}^T \Lambda_i(s)ds$, we have that given $\Lambda_i(w_1)$,   $\Nin\sim \mathcal{P}(\alpha_n \tau_i(w_1))$, with (conditional) variance $\alpha_n \tau_i(w_1)$. Thus $E_{\mathcal{W}_2}((\Nin/\alpha_n-\tau_i(w_1))\vert \Lambda_i(w_1))=0$ and $E_{\mathcal{W}_2}((\Nin/\alpha_n-\tau_i(w_1))^2\vert \Lambda_i(w_1))=\alpha_n^{-2} \text{Var}_{\mathcal{W}_2}(\Nin\ \vert \Lambda_i(w_1))=\tau_i(w_1)/\alpha_n\le M_1/\alpha_n$, which shows that $\text{Var}_{\mathcal{W}_2}(\Nin/\alpha_n-\tau_i(w_1) \vert \Lambda_i(w_1) )\le M_1/\alpha_n$. The result follows.

\vspace{.3cm} 

\noindent
\textbf{Lemma S.3} \, {\it Suppose that $(S3)$ holds. If there exists $M_1<\infty$ such that $\kappa \leq \tau \leq M_1$ almost surely with $\kappa$ as in assumption $(S3)$, then}
\begin{equation*} \label{lemmaAsymp_3}
	\sqrt{\alpha_n} \  n^{-1} \sum_{i=1}^n s_{in}(x,h) \left(\frac{\Nin}{\alpha_n}-\tau_i \right)=O_p((nh)^{-1/2}).
\end{equation*}

\vspace{1cm} 
\noindent \textbf{Proof of Lemma S.3.}

From the proof of Lemma $2$ in  \cite{mull:18:3}, we have $s_{in}(x,h)-s_i(x,h)=W_{0n} K_h(X_i-x)+W_{1n}K_h(X_i-x)(X_i-x)$, where $W_{0n}=\hmu_2/\hsigma_0^2 -\mu_2/\sigma_0^2=O_p((nh)^{-1/2})$ and $W_{1n}=\hmu_1/\hsigma_0^2 -\mu_1/\sigma_0^2=O_p((nh^3)^{-1/2})$. Thus
\begin{align}
	\Big \lvert \sqrt{\alpha_n} \  n^{-1} \sum_{i=1}^n s_{in}(x,h) \left(\frac{\Nin}{\alpha_n}-\tau_i \right) \Big \rvert
	&\le 
	\Big \lvert \sqrt{\alpha_n} \  n^{-1} \sum_{i=1}^n  (s_{in}(x,h)-s_i(x,h))\  \left(\frac{\Nin}{\alpha_n}-\tau_i\right) \Big \rvert\nonumber \\
	&\quad +\Big \lvert \sqrt{\alpha_n} \  n^{-1} \sum_{i=1}^n  s_i(x,h)\  \left(\frac{\Nin}{\alpha_n}-\tau_i\right) \Big \rvert\nonumber\\
	&\le 
	\Big \lvert \sqrt{\alpha_n} W_{0n} \  n^{-1} \sum_{i=1}^n  K_h(X_i-x) \  \left(\frac{\Nin}{\alpha_n}-\tau_i\right) \Big \rvert\nonumber \\ 
	&\quad +
	\Big \lvert \sqrt{\alpha_n} W_{1n} \  n^{-1} \sum_{i=1}^n  K_h(X_i-x) (X_i-x) \  \left(\frac{\Nin}{\alpha_n}-\tau_i\right) \Big \rvert\nonumber \\ 
	&\quad \quad +
	\Big \lvert \sqrt{\alpha_n} \  n^{-1} \sum_{i=1}^n  s_i(x,h)\  \left(\frac{\Nin}{\alpha_n}-\tau_i\right) \Big \rvert  \label{eq:key_lemmaS2_0}.
\end{align}
Defining $Z_{jn}:=n^{-1} \sum_{i=1}^n  K_h(X_i-x) (X_i-x)^j  [\Nin/\alpha_n-\tau_i]$, $j=0,1$, then by independence and a conditioning argument we obtain
\begin{align*}
	E(Z_{jn}^2)&=
	E\left( n^{-2} \sum_{i=1}^n  K_h^2(X_i-x)  (X_i-x)^{2j}  \left(\frac{\Nin}{\alpha_n}-\tau_i\right)^2 \right)\\
	&+
	n^{-2} \sum_{i=1}^n \sum_{k\neq i} 
	E\Big[ K_h(X_i-x) (X_i-x)^j   \left(\frac{\Nin}{\alpha_n}-\tau_i\right) \Big] 
	E\Big[ K_h(X_k-x) (X_k-x)^j  \left(\frac{N_k^{(n)}(T)}{\alpha_n}-\tau_k\right)\Big]\\
	&=
	E\left( n^{-2} \sum_{i=1}^n  K_h^2(X_i-x)  (X_i-x)^{2j}  E\left( (\Nin/\alpha_n-\tau_i)^2\vert \Lambda_i,X_i \right)\right)\\
	&\le (M_1/\alpha_n)
	E\left( n^{-2} \sum_{i=1}^n  K_h^2(X_i-x)  (X_i-x)^{2j}  \right)\\
	&=
	(M_1 h/(n \alpha_n)) \int_{-1}^1  u^{2j} K^2(u) f(x+uh) du \\
	&=O(h^{2j-1}/(n\alpha_n)),
\end{align*}
where the second equality and third inequality are due to Lemma S.2 above and using that $E((\Nin/\alpha_n-\tau_i)\vert \Lambda_i,X_i)=E((\Nin/\alpha_n-\tau_i)\vert \Lambda_i)=0$. This implies $Z_{0n}=O_p((nh\alpha_n)^{-1/2})$ and $Z_{1n}=O_p((n\alpha_n/h)^{-1/2})$. Thus
\begin{align}
	\sqrt{\alpha_n} W_{0n} \  n^{-1} \sum_{i=1}^n  K_h(X_i-x) \  \left(\frac{\Nin}{\alpha_n}-\tau_i\right)
	&= O_p((nh)\inv),  \label{eq:key_lemmaS2_1}\\ 
	\sqrt{\alpha_n} W_{1n} \  n^{-1} \sum_{i=1}^n  K_h(X_i-x) (X_i-x) \  \left(\frac{\Nin}{\alpha_n}-\tau_i\right) 
	&=O_p((nh)\inv) \label{eq:key_lemmaS2_2}.
\end{align}
Similarly, due to  $s_i(x,h)=\sigma_0^{-2} K_h(X_i-x)[\mu_2 -\mu_1 (X_i-x)]$, 
\begin{align*}
	\sqrt{\alpha_n} \  n^{-1} \sum_{i=1}^n  s_i(x,h)\  \left(\frac{\Nin}{\alpha_n}-\tau_i\right)
	&= 
	(\mu_2/\sigma_0^2) \sqrt{\alpha_n} Z_{0n}-(\mu_1/\sigma_0^2) \sqrt{\alpha_n} Z_{1n}=O_p((nh)^{-1/2}),
\end{align*}
where the last equality is due to $\mu_2/\sigma_0^2=O(1)$ and $\mu_1/\sigma_0^2=O(1)$, which were shown in the proof of Proposition \ref{Proposition_consistentWassEst}. This along with (\ref{eq:key_lemmaS2_0}), (\ref{eq:key_lemmaS2_1}) and (\ref{eq:key_lemmaS2_2}) leads to the result.\\

For the following, recall that $\bar{N}(T):=n^{-1} \sumin \Nin$.\\

\noindent
\textbf{Lemma S.4} \,  {\it Suppose that the same assumptions as in Lemma S.3 hold. If $\alpha_n\to \infty$ as $n\to \infty$, then}
\begin{equation}
	\sqrt{n} \left(\frac{\bar{N}(T)}{\alpha_n} - E(\tau) \right)=O_p(1). \label{LemmaS.4}
\end{equation}\\
\noindent
\textbf{Proof of Lemma S.4.}\newline
The result follows from an application of a central limit theorem for triangular arrays. First consider the case when $\Var{(\tau)}>0$. Let $a_i=\Nin/\alpha_n$, then by conditioning on $\Lambda_i$ it follows that $E(a_i)=E(\tau)$ and $\sigma_i^2:=\text{Var}(a_i)=\text{Var}(\tau)+E(\tau)/\alpha_n$. Setting $s_n^2=\sumin \sigma_i^2=n (\text{Var}(\tau)+E(\tau)/\alpha_n)$, we show that 
\begin{equation} \label{eq:asn1}
	\frac{1}{s_n} \sumin (a_i-E(\tau)) \overset{\mathcal{L}}{\to} N(0,1), 
\end{equation}
whence  we may infer $\sqrt{n} \left(\frac{\bar{N}(T)}{\alpha_n} - E(\tau) \right)/ \sqrt{\text{Var}(\tau)+E(\tau)/\alpha_n } =O_p(1)$ and furthermore (\ref{LemmaS.4}),   since $\text{Var}(\tau)+E(\tau)/\alpha_n$ is bounded above as $\tau$ is uniformly bounded and the positive sequence $\alpha_n$ satisfies $1/\alpha_n\to 0$ as $n \to \infty$. The Lyapunov condition 
\begin{equation}
	\lim_{n\to\infty} \frac{1}{s_n^4} \sumin E(\Nin/\alpha_n-E(\tau))^4=0
\end{equation}
implies \eqref{eq:asn1} and will hold  if we show  $\lim_{n\to\infty} \frac{1}{ n \alpha_n^4 }  E(N_1^{(n)}(T)-\alpha_n E(\tau))^4 = 0$. 
Noting that $N_1^{(n)}(T) \vert \Lambda_1 \sim \mathcal{P}(\alpha_n \tau_1)$ and
\begin{gather}
	(N_1^{(n)}(T)-\alpha_n E(\tau))^4 = [N_1^{(n)}(T)]^4-4 [N_1^{(n)}(T)]^3\alpha_n E(\tau)+6 [N_1^{(n)}(T)]^2 \alpha_n^2 E(\tau)^2 \nonumber \\
	\quad - 4 [N_1^{(n)}(T)] \alpha_n^3 E(\tau)^3+\alpha_n^4 E(\tau)^4, \label{eq:cond_moments}
\end{gather}
conditional on $\Lambda_1$, the higher order moments of $N_1^{(n)}(T)\vert \Lambda_1$ are given by $E([N_1^{(n)}(T)]^4 \vert \Lambda_1)= (\alpha_n \tau_1)^4+6(\alpha_n \tau_1)^3+7(\alpha_n \tau_1)^2+(\alpha_n \tau_1)$, $E([N_1^{(n)}(T)]^3 \vert \Lambda_1 )=(\alpha_n \tau_1)^3+3(\alpha_n \tau_1)^2+(\alpha_n \tau_1)$ and $E([N_1^{(n)}(T)]^2\vert \Lambda_1)=(\alpha_n \tau_1)^2+(\alpha_n \tau_1)$. Thus, by taking expectation of the conditional moments and using the fact that $\tau_1$ is uniformly bounded along with equation (\ref{eq:cond_moments}) leads to $\lim_{n\to\infty} \frac{1}{ n \alpha_n^4 }  E( N_1^{(n)}(T)-\alpha_n E(\tau))^4=0$, completing the proof for the case when $\Var{(\tau)}>0$. Next, if $\Var{(\tau)}=0$, then $\tau=\tau_0$ almost surely, for some $\tau_0\in [\kappa,M_1] $. By a conditioning argument we obtain
\begin{align*}
	\Var{(\bar{N}(T))}&= n\inv \Var(N_1^{(n)}(T))=n\inv \{E[\Var(N_1^{(n)}(T)\vert \Lambda_1)]+\Var(E[N_1^{(n)}(T)\vert \Lambda_1])\}
	=n\inv \alpha_n \tau_0.
\end{align*}
Letting $\upsilon_n=\sqrt{n} [\bar{N}(T)/\alpha_n - E(\tau) ]$, due to $E(\bar{N}(T)/\alpha_n)=E(\tau)=\tau_0$, it follows that $E(\upsilon_n^2)=\tau_0/\alpha_n$. This implies $\upsilon_n=O_p(1)$, which shows the result.\\

\noindent
\textbf{Lemma S.5} \, {\it Suppose that the same assumptions as in Lemma S.3 hold. If $\psi(\alpha_n) =O(n^{1/2})$ for some function $\psi:\bbR\to\bbR$ such that $\psi(\alpha_n) \to \infty$ as $n\to\infty$ and $\displaystyle \frac{\alpha_n}{\log n}\to \infty$ as $n\to \infty$, then}
\begin{equation}\label{lemmaAsympS.5}
	\psi(\alpha_n)  (\bar{N}(T)/\alpha_n)^{-1} E(\tau \vert X=x) =\psi(\alpha_n)  E(\tau \vert X=x)/E(\tau) + O_p(1).
\end{equation}\\
\noindent
\textbf{Proof of Lemma S.5.}\newline
From Lemma S.4 we have $\bar{N}(T)/\alpha_n=E(\tau)+O_p(n^{-1/2})$, and a Taylor expansion leads to 
\begin{align*}
	(\bar{N}(T)/\alpha_n)^{-1} &= \frac{1}{E(\tau)}-\frac{1}{E(\tau)^2} \left(\frac{\bar{N}(T)}{\alpha_n}-E(\tau)\right)+o_p(n^{-1/2}).
\end{align*}
With  $m(x)=E(\tau \vert X=x)$ one obtains 
\begin{align*}
	\psi(\alpha_n)  (\bar{N}(T)/\alpha_n)^{-1} m(x) = \frac{\psi(\alpha_n) }{E(\tau)} m(x)- \frac{m(x)}{E(\tau)^2} \frac{\psi(\alpha_n) }{\sqrt{n}} \sqrt{n} \left(\frac{\bar{N}(T)}{\alpha_n}-E(\tau)\right)+o_p \left( \psi(\alpha_n) /\sqrt{n}\right).
\end{align*}
The results follows since $\psi(\alpha_n) =O(n^{1/2})$  implies $\psi(\alpha_n)  /n^{1/2}=O(1)$ as $n\to \infty$ and by using Lemma S.4.\\

\noindent
\textbf{Proof of Theorem \ref{Theorem_tauPlusEst}.}\\
Observe 
\begin{align*}
	& \psi(\alpha_n) \  \frac{n^{-1} \sum_{i=1}^n s_{in}(x,h) \Nin}{\bar{N}(T)}
	=
	\psi(\alpha_n)  (\bar{N}(T)/\alpha_n)^{-1} n^{-1} \sum_{i=1}^n s_{in}(x,h) (\Nin/\alpha_n-\tau_i) \\
	& \quad\quad\quad\quad\quad\quad\quad\quad\quad\quad\quad\quad\quad\quad  +\psi(\alpha_n)  (\bar{N}(T)/\alpha_n)^{-1} n^{-1} \sum_{i=1}^n s_{in}(x,h) \tau_i \\
	& \ = (\psi(\alpha_n) /\sqrt{\alpha_n})  (\bar{N}(T)/\alpha_n)^{-1} O_p((nh)^{-1/2})
	+
	(\psi(\alpha_n) /\sqrt{n h_n})(\bar{N}(T)/\alpha_n)^{-1} ( \sqrt{nh_n} E(\tau \vert X=x) +O_p(1))\\
	& \ =
	O_p(\alpha_n^{-1/2})
	+
	\psi(\alpha_n)  (\bar{N}(T)/\alpha_n)^{-1} E(\tau \vert X=x)
	+
	(\psi(\alpha_n) /\sqrt{n h_n}) (\bar{N}(T)/\alpha_n)^{-1} O_p(1)\\
	& \ = \psi(\alpha_n) \ \frac{E( \tau \vert X=x)}{E(\tau)}+O_p(1), 
\end{align*}
where the second equality follows from Lemma S.3 and by applying Lemma S.6  below with $h=h_n= c_0 n^{-1/5}$ for some constant $c_0>0$; and the third and last equalities follows from Lemma S.5 and since $\psi(\alpha_n) =O(\sqrt{nh_n})$ with $nh_n=n^{4/5}\to\infty$ as $n\to\infty$. The result then follows using that $\psi(\alpha_n)\asymp n^{2/5}$.\\

\subsection*{\sf A.3 Consistency of Local Linear Estimator}

In this section we include  for completeness a well known result regarding the asymptotic normality of the local linear regression estimate for the conditional mean function. Suppose that $(X_1,Y_1),\dots, (X_n,Y_n) \overset{iid}{\sim} (X,Y)$ where $X$ and $Y$ are real valued. Let $m(x)=E(Y\vert X=x)$ be the regression function and $f(x)>0$ be the design density function. Then, the local linear regression estimate \citep{fan:96} $\displaystyle \hat{m}(x)$ of $m(x)$ is given by
\begin{equation*}
\hat{m}(x)=b_1^T \left(\frac{1}{n} \mathbf{X}^T\mathbf{W}\mathbf{X}\right)^{-1} \left(\frac{1}{n} \mathbf{X}^T\mathbf{W}\mathbf{Y}\right) ,\quad b_1=(1 \ 0)^T,
\end{equation*}
where $\mathbf{W}=\text{diag}(K_h(X_i-x))$, $K_h(\cdot)=K(\cdot)/h$ with $K(\cdot)$ a kernel function, $\mathbf{Y}=(Y_1,\dots,Y_n)^T$, and the $i^{th}$ row of $\mathbf{X}$ is given by $(1,X_i-x)$, $i=1,\dots,n$. Now, if we let $\hat{u}_j:=n\inv \sumin K_h(X_i-x) (X_i-x)^j$ and $\hat{\sigma}_0^2:=\hat{u}_0\hat{u}_2-\hat{u}_1^2$, 

\[\displaystyle \left(\frac{1}{n} \mathbf{X}^T\mathbf{W}\mathbf{X} \right)^{-1}=
\begin{bmatrix}
    \frac{\hat{u}_2}{\hat{\sigma}_0^2} & -\frac{\hat{u}_1}{\hat{\sigma}_0^2} \\
	-\frac{\hat{u}_1}{\hat{\sigma}_0^2} & \frac{\hat{u}_0}{\hat{\sigma}_0^2} 
\end{bmatrix}
\]
and
\[ \frac{1}{n} \mathbf{X}^T\mathbf{W}\mathbf{Y}=
\begin{bmatrix}
    n\inv \sumin K_h(X_i-x)Y_i  \\
	n\inv \sumin K_h(X_i-x)(X_i-x)Y_i
\end{bmatrix},
\]
whence
\begin{equation}\label{LLR_beta0}
\hat{m}(x)= 
\begin{bmatrix}
\frac{\hat{u}_2}{\hat{\sigma}_0^2} & -\frac{\hat{u}_1}{\hat{\sigma}_0^2}
\end{bmatrix}
\begin{bmatrix}
    n\inv \sumin K_h(X_i-x)Y_i  \\
	n\inv \sumin K_h(X_i-x)(X_i-x)Y_i
\end{bmatrix}.
\end{equation}
We make the following assumptions:
\begin{enumerate}
\item [(A1)] The regression function $m(x)=E(Y \vert X=x)$, the design density function $f(x)>0$ and $\sigma^2(x)=E(e^2 \vert X=x)$, where $e=Y-m(X)$, are twice continuously differentiable.
\item [(A2)] The kernel $K(\cdot)$ is bounded and corresponds to a density function which is symmetric around zero and has compact support $[-1,1]$.
\item [(A3)] As $n\to \infty$, $nh^5=O(1)$.
\item [(A4)] There exists $\delta>0$ and $\bar{\sigma}>0$ such that $E(\lvert e_i \rvert^{2+\delta}\vert X_i)\leq \bar{\sigma}$, $i=1,\dots,n$, where $e_i:=Y_i-m(X_i)$.
\end{enumerate}
Assumption $(A3)$ implies that as $n\to \infty$ we have $h\to 0$, $nh^q\to \infty$ for $q\in \{0,1,2,3,4  \}$ and $nh^p\to 0$ for $p\geq 6$.

By a second order Taylor expansion of $m(\cdot)$ around $x$
\begin{align*}
Y_i &=m(X_i)+e_i\\
&=m(x)+m'(x)(X_i-x)+\frac{m''(x)}{2}(X_i-x)^2+ \frac{(m''(\xi(X_i,x))-m''(x))}{2}(X_i-x)^2+e_i, 
\end{align*}
where $\xi(X_i,x)$ lies between $x$ and $X_i$. By replacing the previous expression in (\ref{LLR_beta0}) and after some algebra we obtain
\begin{equation}\label{LLR_beta0_taylor}
\hat{m}(x)= m(x)+
\begin{bmatrix}
\frac{\hat{u}_2}{\hat{\sigma}_0^2} & -\frac{\hat{u}_1}{\hat{\sigma}_0^2}
\end{bmatrix}
\begin{bmatrix}
    \hat{u}_2 \frac{m''(x)}{2}+n\inv \sumin K_h(X_i-x)e_i+R_1  \\
	\hat{u}_3 \frac{m''(x)}{2}+n\inv \sumin K_h(X_i-x)(X_i-x)e_i+R_2
\end{bmatrix},
\end{equation}
where the remainder terms are given by
\begin{gather*}
R_1= n\inv \sumin K_h(X_i-x) \frac{(m''(\xi(X_i,x))-m''(x))}{2}(X_i-x)^2,\\
R_2= n\inv \sumin K_h(X_i-x) \frac{(m''(\xi(X_i,x))-m''(x))}{2}(X_i-x)^3.
\end{gather*}

We now study the asymptotic distribution of $\hat{m}(x)$ with a suitably chosen scaling factor. For this, we introduce the following quantities. Let $i,j$ be non-negative integers and define $k_{ij}:=\int_{-1}^1 K(u)^i u^j du$. The auxiliary lemma on the  asymptotic normality of $\hat{m}(x)$ is listed for completeness only.  Its  proof follows by standard arguments. See for example Theorem 5.2 in \cite{fan:96} and the references therein.\vspace{.25cm}

\noindent
\textbf{Lemma S.6} \, {\it Under assumptions $(A1)-(A4)$,}
\begin{equation*}
\sqrt{nh}\left(\hat{m}(x)-m(x)-h^2 \frac{m''(x)}{2} k_{12}\right)\overset{\mathcal{L}}{\to} N\left(0,k_{20} \frac{\sigma^2(x)}{f(x)}\right),
\end{equation*}
as $n\to\infty$.

\subsection*{\sf A.4 Proofs of Results in Section 4.4}

We need the following auxiliary lemma.\\

\noindent
\textbf{Lemma S.7} \, {\it Suppose that the conditions of Theorem \ref{Theorem_gfr_tauPlusEst} hold. Then}
\begin{align*}
	n\inv \sumin s_i(x) \left( \frac{\Nin}{\alpha_n}-\tau_i\right) =O_p((n\alpha_n)^{-1/2}),
\end{align*}
where $s_i(x):=1+(X_i-\mu)^T \Sigma^{-1} (x-\mu)$, $i=1,\dots,n$.\\
\noindent
\textbf{Proof of Lemma S.7.}\newline
Letting $S_n=n\inv \sumin s_i(x) (\Nin/\alpha_n-\tau_i)$, by independence we have
\begin{align*}
	E(S_n^2)&= n^{-2} \sumin E(s_i^2(x) (\Nin/\alpha_n-\tau_i)^2  )
	+ n^{-2} \sumin \sum_{k\neq i} E(s_i(x) (\Nin/\alpha_n-\tau_i)) E(s_k(x) (N_k^{(n)}(T)/\alpha_n-\tau_k))\\
	&= n^{-2} \sumin E(s_i^2(x) (\Nin/\alpha_n-\tau_i)^2 )\\
	&= n\inv E(s_1^2(x) (N_1^{(n)}(T)/\alpha_n-\tau_1)^2 ),
\end{align*}
where the second equality is due to $E(s_i(x) (\Nin/\alpha_n-\tau_i))=E[s_i(x)E( \Nin/\alpha_n-\tau_i\vert \Lambda_i X_i)]=0$. Next, from Lemma S.2 and by a conditioning argument, it follows that 
\begin{align*}
	E(s_1^2(x) (N_1^{(n)}(T)/\alpha_n-\tau_1)^2 )&=
	E(s_1^2(x) E[(N_1^{(n)}(T)/\alpha_n-\tau_1)^2\vert \Lambda_1,X_1 ])\\
	&\le (M_1/\alpha_n) E(s_1^2(x))\\
	&= (M_1/\alpha_n) (1+(x-\mu)^T \Sigma\inv (x-\mu)).
\end{align*}
This implies $E(S_n^2)=O((n\alpha_n)\inv)$ and the result follows.\\

\noindent
\textbf{Proof of Theorem \ref{Theorem_gfr_tauPlusEst}.}\\
Consider the auxiliary quantities $s_i(x):=1+(X_i-\mu)^T \Sigma^{-1} (x-\mu)$, $W_{0n}(x):=\bar{X}^T \hSigma^{-1}(x-\bar{X})-\mu^T \Sigma^{-1}(x-\mu)$ and $W_{1n}(x):=\Sigma^{-1}(x-\mu)-\hat{\Sigma}^{-1}(x-\bar{X})$. The arguments in the proof of Theorem 1 in \cite{mull:18:3} show that $W_{0n}=O_p(n^{-1/2})$, $\lVert W_{1n} \rVert_2=O_p(n^{-1/2})$ and $s_{in}(x)-s_i(x)=-W_{0n}-W_{1n}^TX_i$. Thus
\be  \nonumber 
&&n\inv \sumin s_{in}(x) \Nin/\alpha_n\\
&&\quad\quad =
-W_{0n} \bar{N}(T)/\alpha_n- n\inv \sumin X_i^T (\Nin/\alpha_n) W_{1n}+n\inv \sumin s_i(x) \Nin/\alpha_n. \nonumber\ee
Next, note that $W_{0n} \bar{N}(T)/\alpha_n=W_{0n} (\bar{N}(T)/\alpha_n-E(\tau)) + W_{0n} E(\tau)=O_p(n^{-1/2})$ by Lemma S.4. Further, from the fact that $E\left(\lVert X_1 \rVert_2  N_1^{(n)}(T)/\alpha_n \right)\leq \left(E\lVert X_1 \rVert_2^2  \ E(N_1^{(n)}(T)/\alpha_n )^2 \right)^{1/2}=\sqrt{(\text{tr}(\Sigma)+\lVert \mu \rVert_2^2 ) (E(\tau^2)+o(1))}$ is uniformly bounded and observing  the inequality $\lVert n\inv \sumin X_i^T (\Nin/\alpha_n) \rVert_2\leq n\inv \sumin (\Nin/\alpha_n)  \lVert X_i \rVert_2$, it follows that \newline  $n\inv \sumin X_i^T (\Nin/\alpha_n)=O_p(1)$ and thus $n\inv \sumin X_i^T (\Nin/\alpha_n) W_{1n}=O_p(n^{-1/2})$. This shows that
\begin{equation}
n\inv \sumin s_{in}(x) \Nin/\alpha_n=n\inv \sumin s_i(x) \Nin/\alpha_n+O_p(n^{-1/2})\label{gfr_2}.
\end{equation}
Next, note that
\begin{align*}
n\inv \sumin s_i(x) \Nin/\alpha_n&=n\inv \sumin s_i(x) (\Nin/\alpha_n-\tau_i)+n\inv \sumin s_i(x) \tau_i\\
&=n\inv \sumin s_i(x) (\Nin/\alpha_n-\tau_i)+E(s(X,x)\tau)+O_p(n^{-1/2})\\
&=  E(s(X,x)\tau)+O_p(n^{-1/2}),
\end{align*}
where the second equality follows from the central limit theorem and the third is due to Lemma S.7 along with the fact that $\alpha_n\to\infty$ as $n\to\infty$. Combining this  with (\ref{gfr_2}) and the fact that $\psi(\alpha_n)\asymp n^{1/2}$ leads to
\begin{align*}
	\psi(\alpha_n) \ n\inv \sumin s_{in}(x) \Nin/\alpha_n&= \psi(\alpha_n) E(s(X,x)\tau)+O_p(1).
\end{align*}
Finally, arguments similar to those  in the proof of Lemma S.5 lead to
\begin{equation*}
\psi(\alpha_n) \  \frac{n\inv \sumin s_{in}(x) \Nin}{\bar{N}(T)}=\psi(\alpha_n)  \ \frac{E(s(X,x)\tau)}{E(\tau)} +O_p(1),
\end{equation*}
whence  the result follows since $\psi(\alpha_n)\asymp n^{1/2}$.\\
\noindent
\textbf{Proof of Theorem \ref{Theorem_gfr_fPlusEst}.}\\
The proof follows by arguments  similar to those in the proof of Proposition \ref{Proposition_consistentWassEst} and Theorem 2 in \cite{mull:18:3} and is therefore omitted.\\

\subsection*{\sf A.5 Proofs of Results in Section 5.1}

We require the following auxiliary lemma.\vspace{.3cm} 

\noindent
\textbf{Lemma S.8} \, {\it Suppose that $(S1)$ and $(S2)$ hold. Let $\nu$ be a positive integer and $r_j$, $j=1,\dots,\nu$, be an equispaced grid in $(0,1)$, where $\Delta r_\nu=1/(\nu+1)$ is the grid spacing. Then
\begin{equation*}
	\sup_{Q_1,Q_2\in Q(\Omega_\mathcal{F})} \Big \lvert \int_0^1 (Q_1(t)-Q_2(t))^2 dt- \sum_{j=1}^{\nu} (Q_1(r_j)-Q_2(r_j))^2 \Delta r_\nu  \Big\rvert =o(1),
\end{equation*}
as $\nu\to\infty$.}\\
\noindent
\textbf{Proof of Lemma S.8.}\newline
Since $r_j=j/(\nu+1)$ and denoting by $r_0=0$, we have
\begin{align*}
	\Big \lvert \int_0^1 (Q_1(t)-Q_2(t))^2 dt- \sum_{j=1}^{\nu} (Q_1(r_j)-Q_2(r_j))^2 \Delta r_\nu  \Big\rvert
	&=
	\Big \lvert \sum_{j=1}^{\nu} \Big[\int_{r_{j-1}}^{r_j} (Q_1(t)-Q_2(t))^2 dt- (Q_1(r_j)-Q_2(r_j))^2 \Delta r_\nu\Big] \\
	&\quad \quad +\int_{r_{\nu}}^{1} (Q_1(t)-Q_2(t))^2 dt  \Big\rvert\\
	&\le \frac{1}{\nu+1}+
	\sum_{j=1}^{\nu} \Big\lvert \int_{r_{j-1}}^{r_j}  (Q_1(t)-Q_2(t))^2 dt- (Q_1(r_j)-Q_2(r_j))^2 \Delta r_\nu \Big\rvert.
\end{align*}
Next, using that $Q_1,Q_2 \in Q(\Omega_\mathcal{F})$ along with simple calculations shows that
\begin{align*}
	\Big\lvert \int_{r_{j-1}}^{r_j}  (Q_1(t)-Q_2(t))^2 dt- (Q_1(r_j)-Q_2(r_j))^2 \Delta r_\nu \Big\rvert
	&= 
	\Big\lvert \int_{r_{j-1}}^{r_j}  (Q_1(t)-Q_2(t))^2 -(Q_1(r_j)-Q_2(r_j))^2 dt \Big\rvert\\
	&\le 4L
	\int_{r_{j-1}}^{r_j}  \lvert Q_1(t)-Q_1(r_j)+Q_2(r_j)-Q_2(t)\rvert dt\\
	&\le 8 L^2 \int_{r_{j-1}}^{r_j} (r_j-t) dt= 4  L^2 \Delta r_\nu^2.
\end{align*}
Thus
\begin{align*}
	\sum_{j=1}^{\nu} \Big\lvert \int_{r_{j-1}}^{r_j}  (Q_1(t)-Q_2(t))^2 dt- (Q_1(r_j)-Q_2(r_j))^2 \Delta r_\nu \Big\rvert
	&\le 4  L^2 \Delta r_\nu,
\end{align*}
whence the result  follows.\newline

\noindent
\textbf{Proof of Proposition \ref{Proposition_RiemannSum}}\newline
We will show convergence along subsequences, which is a similar idea as in Proposition 4.1 in \cite{peyr:19} or \cite{berthe:20}. Recall that $Q_{\oplus \nu}$ is any (fixed) element in $\mathcal{S}_\nu$, which can be selected by the axiom of choice. Note that any sequence $q_n\in Q(\Omega_\mathcal{F})$ is uniformly bounded and uniformly equicontinuous. Since the $q_n$ are continuous functions defined on $[0,1]$, an application of the Arzela-Ascoli theorem shows that $Q(\Omega_\mathcal{F})$ is a compact set in $L^2([0,1])$. Consider a sequence of positive integers $\nu=\nu_m$ such that $\nu_m\to\infty$ as $m\to\infty$. Note that $\lvert \mathcal{M}(q)-\mathcal{M}_{\nu_m}(q)\rvert=o(1)$ as $m\to\infty$. Since $Q(\Omega_\mathcal{F})$ is compact, there exists a subsequence $(Q_{\oplus \nu_{m_k}})_{k\in\mathbb{N}}$ of $(Q_{\oplus \nu_m})_{m\in\mathbb{N}}$ which converges to an element $Q^* \in Q(\Omega_\mathcal{F})$ as $k\to\infty$, i.e.,  $\lVert Q_{\oplus \nu_{m_k}}-Q^* \rVert_{L^2([0,1])}\to 0$ as $k\to\infty$. Next, as $k\to\infty$ we have
\begin{align*}
	\mathcal{M}(\tilde{Q}_\oplus(\cdot,x))&=\mathcal{M}_{\nu_{m_k}}(\tilde{Q}_\oplus(\cdot,x))+o(1)\ge \mathcal{M}_{\nu_{m_k}}(Q_{\oplus \nu_{m_k}})+o(1),
\end{align*}
where the inequality is due to the fact that $Q_{\oplus \nu_{m_k}}\in \mathcal{S}_{\nu_{m_k}}$ minimizes $\mathcal{M}_{\nu_{m_k}}(\cdot)$ over the class $Q(\Omega_\mathcal{F})$ and $\tilde{Q}_\oplus(\cdot,x)$ is an element of the latter space. Note that
\begin{align*}
	\mathcal{M}_{\nu_{m_k}}(Q_{\oplus \nu_{m_k}})&= \sum_{j=1}^{\nu_{m_k}} (Q_{\oplus \nu_{m_k}}(r_j)-w_j)^2 \Delta r_{\nu_{m_k}}\\
	&=\sum_{j=1}^{\nu_{m_k}} (Q_{\oplus \nu_{m_k}}(r_j)-Q^*(r_j))^2 \Delta r_{\nu_{m_k}}+
	\sum_{j=1}^{\nu_{m_k}} (Q^*(r_j)-w_j)^2 \Delta r_{\nu_{m_k}}\\
	&\quad +2\sum_{j=1}^{\nu_{m_k}} (Q_{\oplus \nu_{m_k}}(r_j)-Q^*(r_j)) (Q^*(r_j)-w_j) \Delta r_{\nu_{m_k}}\\
	&= \lVert Q_{\oplus \nu_{m_k}}-Q^* \rVert_{L^2([0,1])}+ \mathcal{M}(Q^*)+o(1) + O(\mathcal{M}(Q^*) \lVert Q_{\oplus \nu_{m_k}}-Q^* \rVert_{L^2([0,1])})\\
	&= \mathcal{M}(Q^*)+o(1),
\end{align*}
as $k\to\infty$, where the third equality follows by using the uniform Riemann sum integrability over the class $Q(\Omega_\mathcal{F})$ shown in Lemma S.8 above along with the Cauchy-Schwarz inequality, and the last equality is due to $\lVert Q_{\oplus \nu_{m_k}}-Q^* \rVert_{L^2([0,1])}\to 0$ as $k\to\infty$. This shows that
\begin{align*}
	\mathcal{M}(\tilde{Q}_\oplus(\cdot,x))&\ge \mathcal{M}(Q^*)+o(1),
\end{align*}
and taking $k\to\infty$ leads to $\mathcal{M}(\tilde{Q}_\oplus(\cdot,x))\ge \mathcal{M}(Q^*)$. Since $\tilde{Q}_\oplus(\cdot,x)$ is the unique solution to the optimization problem (\ref{qf_empirical}) involving $\mathcal{M}$, then $Q^*(\cdot)=\tilde{Q}_\oplus(\cdot,x)$. The previous arguments show that all convergent subsequences of $(Q_{\oplus \nu_m})_{m\in\mathbb{N}}$ converge to the same limit  $\tilde{Q}_\oplus(\cdot,x)$. Since $Q_{\oplus \nu_m}\in Q(\Omega_\mathcal{F})$ for all $m\ge 1$  and $Q(\Omega_\mathcal{F})$ is compact, this implies that $Q_{\oplus \nu_m}$ converges to $\tilde{Q}_\oplus(\cdot,x)$ in the $L^2$ norm. The result follows.\newline

Recall that $Q_\nu^*(t)=q_{\nu,j}^*+(t-r_j)(q_{\nu,j+1}^*-q_{\nu,j}^*)/\Delta r_\nu$ for $t\in [r_j,r_{j+1})$, $j=0,\dots,\nu$, where $q_{\nu,j}^{*}$ is the $j$th coordinate of $q_\nu^{*}$, $j=1,\dots,\nu$, $q_{\nu,0}^{*}=0$, $r_0=0$, $q_{\nu, \nu+1}^{*}=1$ and $r_{\nu+1}=1$. By continuity we define $Q_\nu^*(1):=\lim_{t\to1^{-}} Q_\nu^*(t)=1$. Lemma S.6 below shows that $Q_\nu^*$ is in the quantile space $Q(\Omega_\mathcal{F})$.\vspace{.3cm}

\noindent
\textbf{Lemma S.9} \, {\it Suppose that $(S1)$ and $(S2)$ hold. The linear interpolation function $Q_\nu^*$  satisfies $Q_\nu^*\in Q(\Omega_\mathcal{F})$.} \\
\noindent
\textbf{Proof of Lemma S.9.}\newline
Let  $I_j=[r_j,r_{j+1})$, $j=0,\dots,\nu$, and $t_1,t_2\in (0,1)$. If  $t_1,t_2\in I_j$, then  $\lvert Q_\nu^*(t_2)-Q_\nu^*(t_1)\rvert =(q_{\nu,j+1}^*-q_{\nu,j}^*) \lvert t_2-t_1\rvert/\Delta r_\nu$ and the constraints of the optimization problem (\ref{numericalQP}) imply $M \lvert t_2-t_1\rvert \le\lvert Q_\nu^*(t_2)-Q_\nu^*(t_1)\rvert \le L \lvert t_2-t_1\rvert$. Next, consider the case when $t_1\in I_j$ and $t_2\in I_k$ for $j<k$. Note that $Q_\nu^*(r_{j+1})-Q_\nu^*(t_1)= q_{\nu,j+1}^*-q_{\nu,j}^*-(t_1-r_j)(q_{\nu,j+1}^*-q_{\nu,j}^*)/\Delta r_\nu= (q_{\nu,j+1}^*-q_{\nu,j}^*)(1-(t_1-r_j)/\Delta r_\nu)$, which implies $ M (r_{j+1}-t_1)\le Q_\nu^*(r_{j+1})-Q_\nu^*(t_1) \le L (r_{j+1}-t_1)$. Also
\begin{align*}
	Q_\nu^*(t_2)-Q_\nu^*(t_1)&=Q_\nu^*(t_2)-Q_\nu^*(r_k) +  \sum_{l=j+2}^k (q_{\nu,l}^*-q_{\nu,l-1}^*)	+Q_\nu^*(r_{j+1})-Q_\nu^*(t_1),
\end{align*}
where $\sum_{l=j+2}^k (q_{\nu,l}^*-q_{\nu,l-1}^*)$ is defined as zero whenever $j=k-1$, and $M \Delta r_\nu \le q_{\nu,l}^*-q_{\nu,l-1}^* \le L \Delta r_\nu$, which is due to the constraints in (\ref{numericalQP}). Combining this with $M (t_2-r_k) \le Q_\nu^*(t_2)-Q_\nu^*(r_k) \le L(t_2-r_k)$, which is due to $t_2,r_k \in I_k$, leads to
\begin{align*}
	M (t_2-t_1) \le Q_\nu^*(t_2)-Q_\nu^*(t_1) \le L(t_2-t_1).
\end{align*}
Interchanging the role of $t_1$ and $t_2$ shows that $M \lvert t_2-t_1\rvert \le \lvert Q_\nu^*(t_2)-Q_\nu^*(t_1)\rvert \le L \lvert t_2-t_1\rvert$ for any $t_1,t_2\in(0,1)$. Finally, by construction it is clear that $Q_\nu^*(0)=0$ and $Q_\nu^*(1)=\lim_{t\to1^{-}} Q_\nu^*(t)=T$. Thus $Q_\nu^*\in Q(\Omega_\mathcal{F}) $ and the result follows. \newline

\subsection*{\sf A.6 Additional Theoretical results}

Lemma S.10  and S.11 below present the explicit solution for  the local Fr\'echet regression shape component $f_\oplus(x)$ and   the corresponding global regression, respectively.\vspace{.3cm} 

\noindent \textbf{Lemma S.10} \, {\it Suppose that $(S1)$ and $(S2)$ hold. The solution $f_\oplus(x)$ to the local Fr\'echet regression problem on the shape component
	\begin{equation*}
		f_\oplus(x) =\underset{f_0 \in \Omega_{\mathcal{F}}} {\arg\min} \ E(d_\mathcal{F}^2(f,f_0)\vert X=x),
	\end{equation*}
	is given by  the density function with corresponding quantile function $E(Q\vert X=x)$.}\\
\noindent
\textbf{Proof of Lemma S.10.}\newline
Denoting by $Q$ and $Q_0$ the quantile functions corresponding to $f$ and $f_0$, respectively, and $Q_x:=E(Q\vert X=x)$, then similarly as in the proof of Proposition $1$ in \cite{mull:18:3} it follows that
\begin{align*}
	E(d_\mathcal{F}^2(f,f_0)\vert X=x)&=E(\lVert Q-Q_0\rVert_{L^2([0,1])}^2\vert X=x)\\
	&= E(\lVert Q-Q_x\rVert_{L^2([0,1])}^2\vert X=x) + \lVert Q_x-Q_0\rVert_{L^2([0,1])}^2 +2 E(\langle Q-Q_x, Q_x-Q_0 \rangle_{L^2([0,1])}\vert X=x) \\
	&= E(\lVert Q-Q_x\rVert_{L^2([0,1])}^2\vert X=x) + \lVert Q_x-Q_0\rVert_{L^2([0,1])}^2,
\end{align*}
and thus the optimal solution $Q_0$ is achieved by setting $Q_0=Q_x$, provided that we can show that $Q_x$ lies in the space $Q(\Omega_\mathcal{F})$. Indeed, since $Q\in Q(\Omega_\mathcal{F})$ we have $M\lvert t-s \rvert \le \lvert Q(t)-Q(s) \rvert \le L \lvert t-s\rvert $, $t,s\in(0,1)$. It is then easy to show that $M\lvert t-s \rvert \le \lvert Q_x(t)-Q_x(s) \rvert \le L \lvert t-s\rvert $. It is also clear that $Q_x(0)=0$ and $Q_x(1)=T$, and the  result follows.\\

\noindent
\textbf{Lemma S.11} \, {\it Suppose that $(S1)$ and $(S2)$ hold. The solution $f_{\text{G} \oplus}(x)$ to the global Fr\'echet regression problem on the shape component
	\begin{align*}
		f_{\text{G} \oplus}(x) &=\underset{f_0 \in \Omega_{\mathcal{F}}} {\arg\min} \ E(s(X,x) d_\mathcal{F}^2(f,f_0))
	\end{align*}
	is given by the density function whose corresponding quantile function is equal to the $L^2$-orthogonal projection of $E(s(X,x)Q)$ on  $Q(\Omega_\mathcal{F})$.}\\
\noindent
\textbf{Proof of Lemma S.11.}\newline
Denoting by $Q$ and $Q_0$ the quantile functions corresponding to $f$ and $f_0$, respectively, and $Q_x:=E(s(X,x)Q)$, then similarly as in the proof of  Lemma S.10 or Proposition $1$ in \cite{mull:18:3} we have
\begin{align*}
	E(s(X,x) d_\mathcal{F}^2(f,f_0))
	&= 
	E(s(X,x) \lVert Q-Q_x\rVert_{L^2([0,1])}^2) + \lVert Q_x-Q_0\rVert_{L^2([0,1])}^2.
\end{align*}
Since $Q(\Omega_\mathcal{F})$ is closed and convex in $L^2([0,1])$ due to Lemma \ref{Lemma_QOmegaf_closedconvex}, it follows that the optimal solution $Q_0$ exists and is unique, and corresponds to the orthogonal projection of $Q_x$ on $Q(\Omega_\mathcal{F})$, and the result follows.\\

The following lemma presents explicit solutions of the local Fr\'echet regression function for the special case where the distributions associated to the random intensity factor and shape functions are point masses.\vspace{.3cm}

\noindent
\textbf{Lemma S.12} \, {\it Suppose that $(S1)$-$(S3)$ hold and there exists $M_1<\infty$ such that $\kappa \leq \tau \leq M_1$ almost surely with $\kappa$ as in assumption $(S3)$. Also, suppose that $\Var(\tau)=0$  and the distribution of the random density $f$ corresponds to a point mass in the space of probability distributions $\Omega_\mathcal{F}$ endowed with the $2$-Wasserstein metric. Then $f=g$ almost surely for some density $g$ with corresponding quantile function $Q_g\in Q(\Omega_{\mathcal{F}})$, $\tau=\eta_0$ almost surely for some positive constant $\eta_0\in [\kappa,M_1]$ and  the local Fr\'echet regression function satisfies}
\begin{align*}
	\tilde{f}_\oplus(x)=g,\quad 
	\tilde{\tau}_\oplus(x)=\eta_0.
\end{align*}
\noindent
\textbf{Proof of Lemma S.12}\newline
Since the probability distribution of the random density $f$ is a point mass in $\Omega_\mathcal{F}$, there exists a density function $g$ with corresponding quantile function $Q_g\in Q(\Omega_{\mathcal{F}})$ such that $f=g$ almost surely. Similarly, $\tau=\eta_0$ a.s. for some $\eta_0\in[\kappa,M_1]$. Let $f_0$ be a density with corresponding quantile function $Q_0\in Q(\Omega_{\mathcal{F}})$ and let $Q$ be the quantile function associated with $f$. Then
\begin{align*}
	E(d_\mathcal{F}^2(f,f_0)\vert X=x)&= E(\lvert \lvert Q-Q_0\rvert \rvert_{L^2([0,1])}^2)
	=\lvert \lvert Q_g-Q_0\rvert \rvert_{L^2([0,1])}^2.
\end{align*}
Thus, from (\ref{f_plus}) the minimizer has  $Q_0=Q_g$ which implies $\tilde{f}_\oplus(x)=g$, and the first result follows. Next, from (\ref{tau_gfr_plus}) we have $\tau_\oplus(x)=\max\{E(\tau \vert X=x) ,0\}=\eta_0$, implying  the second result.\\

The following lemma shows the corresponding explicit solutions when considering the global Fr\'echet regression framework and the point mass probability distribution on the components $\tau$ and $f$.\vspace{.3cm} 

\noindent
\textbf{Lemma S.13} \, {\it Suppose that the same regularity conditions as in Lemma S.12 hold. Then $f=g$ almost surely for some density $g$ with corresponding quantile function $Q_g\in Q(\Omega_{\mathcal{F}})$, $\tau=\eta_0$ almost surely for some positive constant $\eta_0\in [\kappa,M_1]$ and the following relations hold for the global Fr\'echet regression function:}
\begin{align*}
	f_{\text{G} \oplus}(x) =g,\quad 
	\tau_{\text{G} \oplus}(x) =\eta_0.
\end{align*}
\noindent
\textbf{Proof of Lemma S.13}\newline
Analogously as in the proof of Lemma S.12 we have $f=g$ almost surely for some density $g$ with corresponding quantile function $Q_g\in Q(\Omega_{\mathcal{F}})$ and $\tau=\eta_0$ almost surely for some positive constant $\eta_0\in [\kappa,M_1]$. Denote by $Q$ the quantile function associated with $f$ and let $f_0$ be a density function with corresponding quantile $Q_0\in Q(\Omega_{\mathcal{F}})$. Since $E(s(X,x) d_\mathcal{F}^2(f,f_0)=E(s(X,x) \lvert \lvert Q-Q_0\rvert \rvert_{L^2([0,1])}^2 )=E(s(X,x) ) \lvert \lvert Q_g-Q_0\rvert \rvert_{L^2([0,1])}^2= \lvert \lvert Q_g-Q_0\rvert \rvert_{L^2([0,1])}^2$, which is due to $E(s(X,x))=1$, then the minimizer  is attained when $Q_0=Q_g\in Q(\Omega_{\mathcal{F}})$. From (\ref{f_gfr_plus})  it then follows that $f_{\text{G} \oplus}(x) =g$. Next, from (\ref{tau_gfr_plus}) we have $\tau_{\text{G} \oplus}(x)=\max\{E(s(X,x) \tau) ,0\}=\eta_0$ by again using that $E(s(X,x))=1$. The result follows.\\

\subsection*{\sf A.7 Local Fr\'echet regression when $\alpha_n$ does not grow with sample size $n$}

\begin{figure}[H]
	\includegraphics[width=0.5\textwidth]{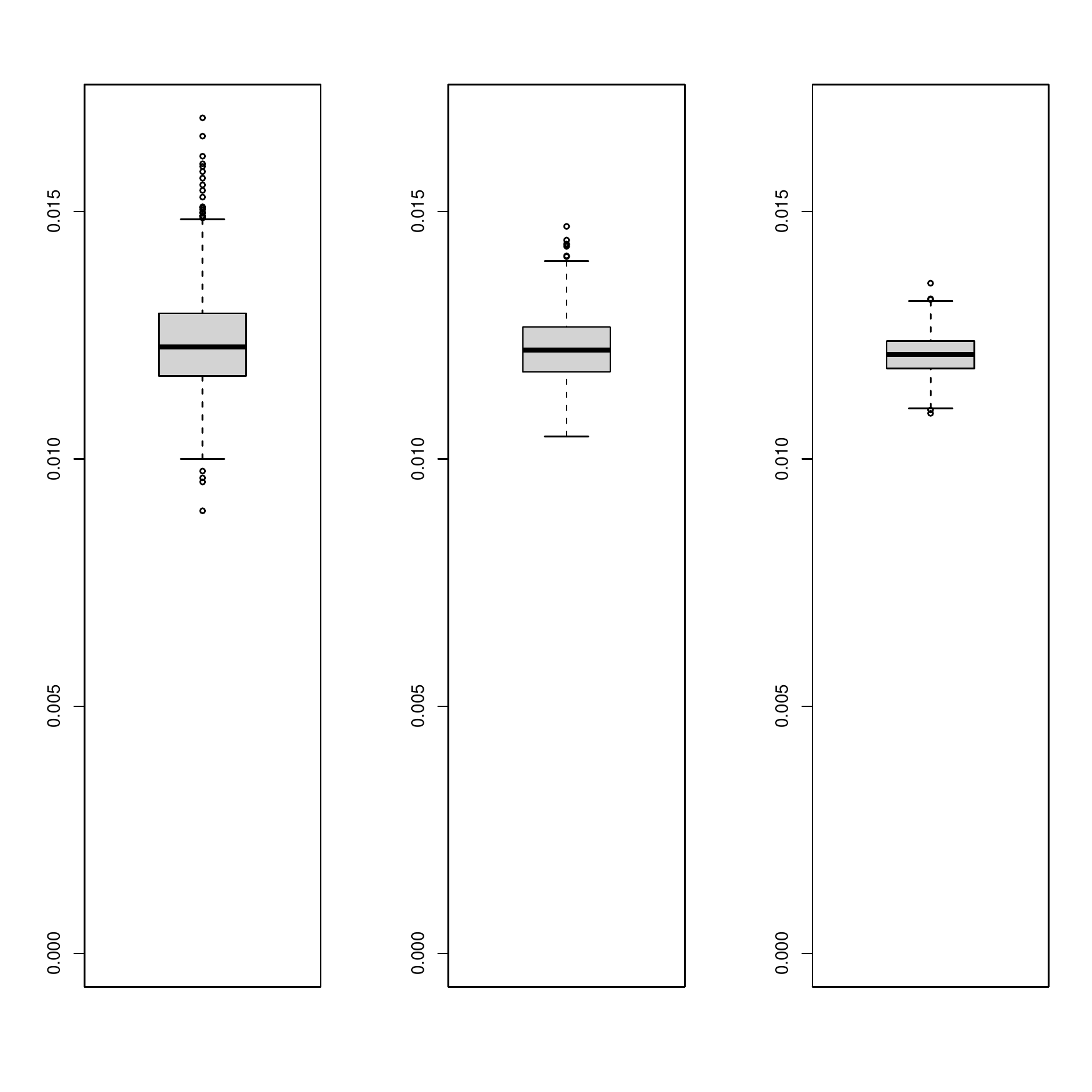}
	\centering
	\caption{Boxplots of the errors for the conditional shape function estimates $ISE_r^{\mathcal{F}}$ in the simulation setting for local Fr\'echet regression in section 5.2 using fixed $\alpha_n=1$, which does not grow with $n$, for $n=1000$ (left), $ n=2000$ (middle) and $n=5000$ (right).}\label{fig:boxplotISE_alphanFixed}
\end{figure}

Figure~\ref{fig:boxplotISE_alphanFixed} shows the integrated error metric for the shape part in the simulation settings for local Fr\'echet regression as outlined in section 5.2 when fixing $\alpha_n=1$, so that $\alpha_n$ does not grow with $n$, and thus violates a  basic assumption. We find that consistent recovery of the conditional intensity function is not possible if $\alpha_n$ is not allowed to increase with $n$ as the integrated error box plots for the shape component show no decline in bias  and stay well  bounded away from zero for increasing sample size.

\subsection*{\sf A.8 Comparison between standard Euclidean intensity regression function and local Fr\'echet regression}

In this section we compare through a simulation example the standard Euclidean intensity regression function and the local Fr\'echet regression counterpart. We consider the same Gaussian data generation mechanism as the one outlined in the global framework in section 5.3 but modifying the following population parameters: $a_2=b_2=1/3$, $a_3=0.05$, $b_3=0$, $e_1=-0.15$, $f_1=0.15$, $\sigma_1=3$, $\sigma_2=0$. 
Here we do not consider the error $\varepsilon_2$ on the standard deviation $\sigma(x)$ of the generated random densities, but rather keep it constant at $\sigma(x)=a_3$ for all $x$. This reflects a similar situation of horizontal translation of a Gaussian random variable as here only the mean increases with $x$ while the standard deviation remains small and constant. It is easy to show that the standard Euclidean intensity regression function $g_x(\cdot)=E(\Lambda(\cdot) \vert X=x)$ is given by $g_x(\cdot) =E(f(\cdot)\vert X=x) E(\tau\vert X=x)$. We approximate $E(f(\cdot) \vert X=x)$ through a Monte Carlo approach where we average across random densities $f_i$ generated at predictor level $x$. Similarly, for the local Fr\'echet regression we obtain $E(Q(\cdot) \vert X=x)$ by averaging the corresponding random quantiles $Q_i$. To compare both quantities, denote by $f_x(\cdot)$ the density function which corresponds to the truncated Gaussian model considered before but disregarding the error $\varepsilon_1$ affecting its mean. Thus $f_x(\cdot)$ is the true density (without noise) at level $X=x$. The intensity signal $\Lambda_x(\cdot)$ is then constructed as $\Lambda_x=f_x E(\tau\vert X=x)$ and corresponds to the underlying intensity function in dependence on $x$  after removing noise.

\begin{figure}
	\includegraphics[width=0.45\textwidth]{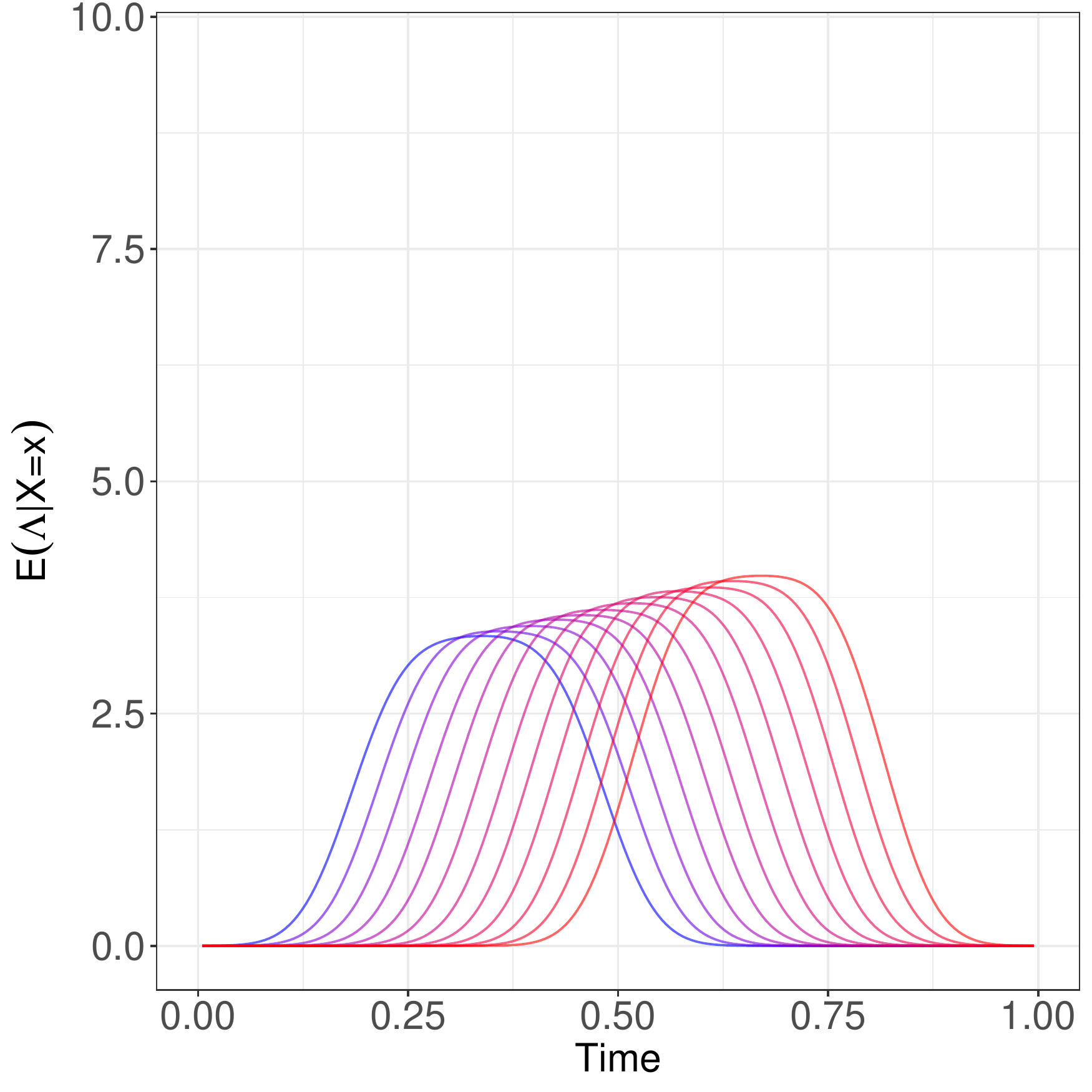}%
	\includegraphics[width=0.45\textwidth]{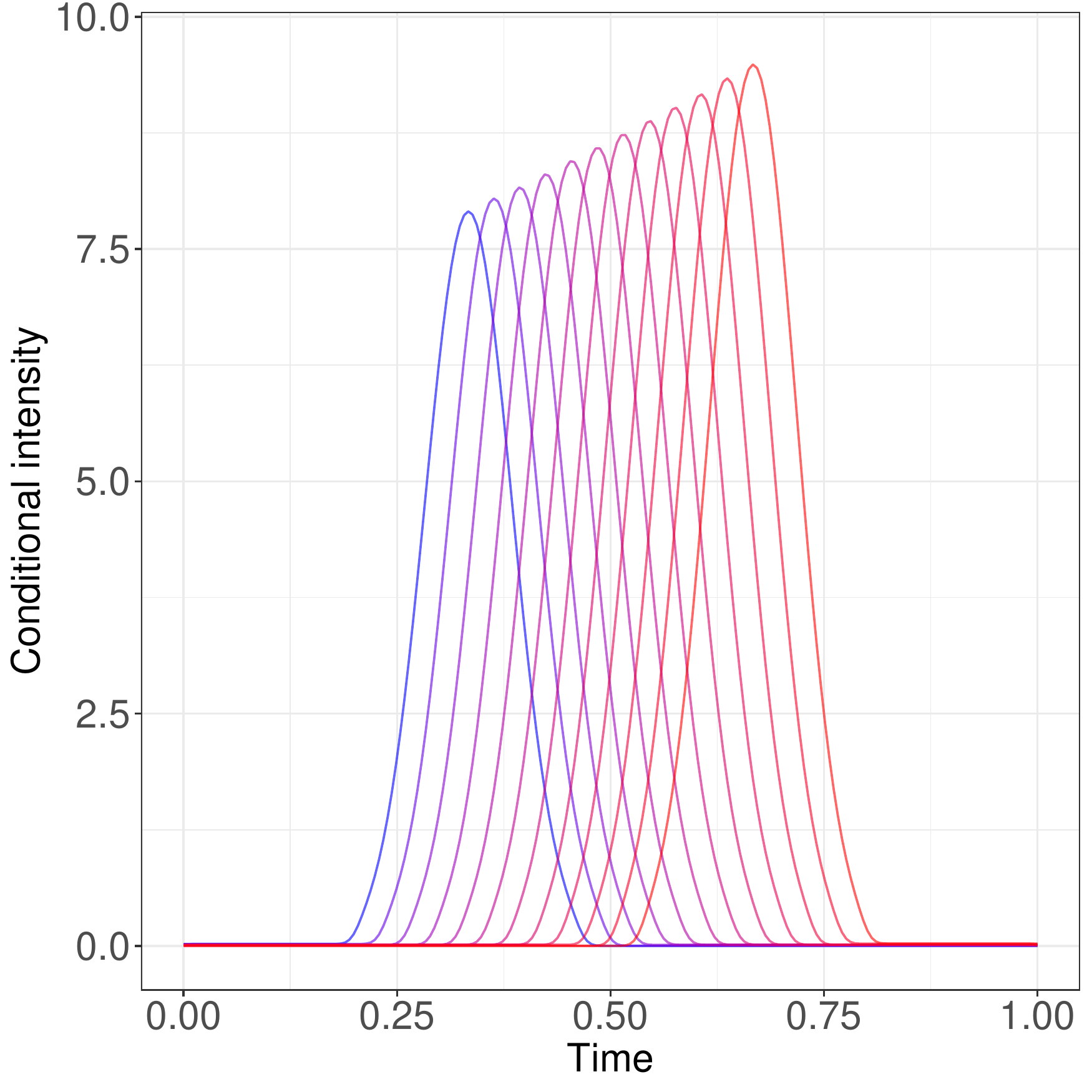}
	\includegraphics[width=0.45\textwidth]{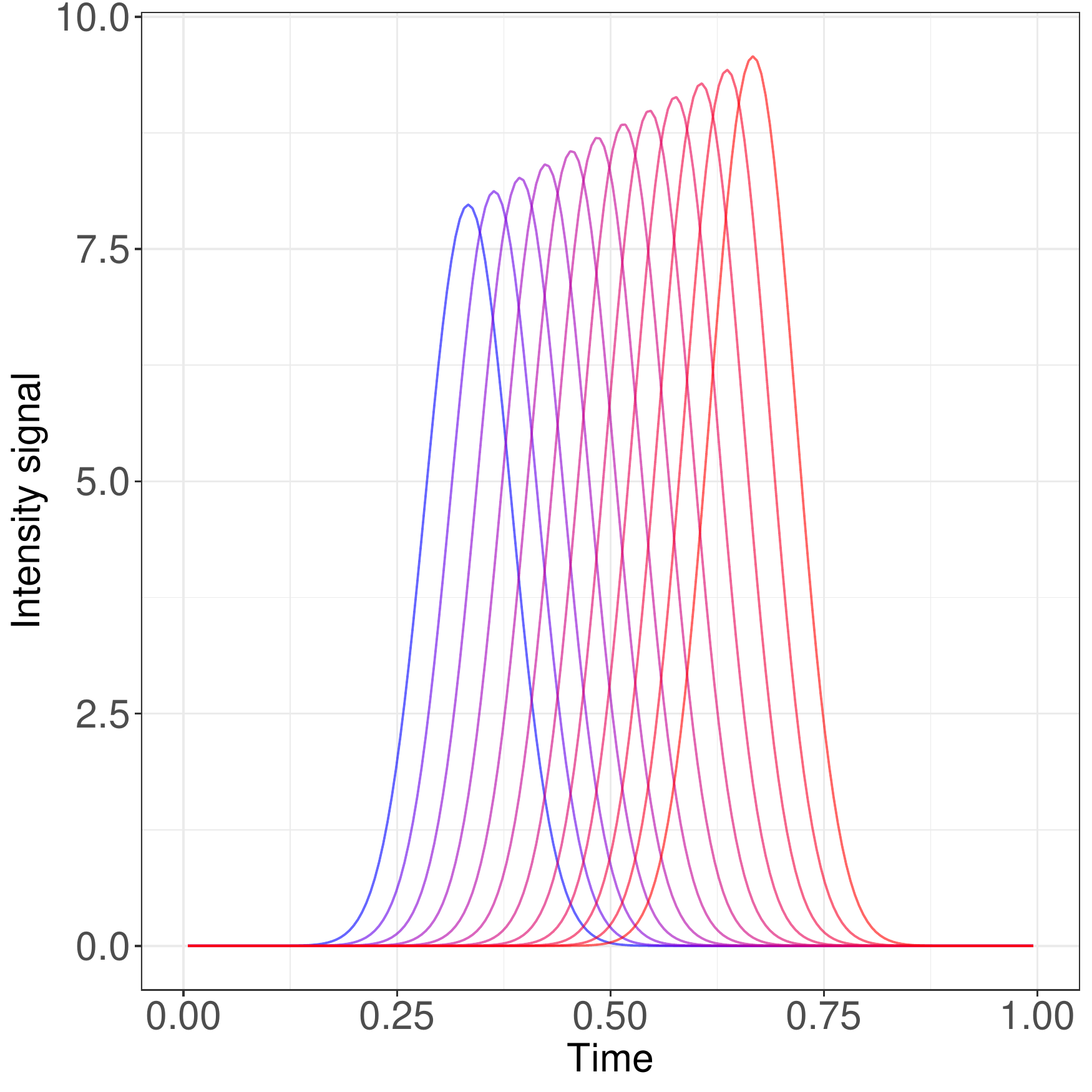}
	\centering
	\caption{Conditional intensity functions in the simulation setting over a dense grid of predictor levels $x$, displayed in blue when $x=0$ to red when $x=1$. The upper left panel illustrates the results for   classical Euclidean  intensity regression function $E(\Lambda\vert X=x)$ while the upper right panel illustrates the results for the conditional intensity $\Lambda_\oplus(x)$ that corresponds to the proposed point process regression. The bottom panel shows the true intensity signal $\Lambda_x$ when there is no noise in either the intensity factor or the mean of the simulated conditional shape component.}
\label{fig:comparisonClassicalIntensity}
\end{figure}

 Figure~\ref{fig:comparisonClassicalIntensity} presents  $g_x$ in the upper left panel over a grid of values for $x$ while the upper right panel displays  the local Fr\'echet regression function $\Lambda_{\oplus}(x)$. The true intensity signal $\Lambda_x$ is  in the bottom panel. One finds  that the shape component of the standard regression function $g_x$ is distorted and does not lie in the Gaussian class where the random densities are situated. This is due to the noise in the mean function $\mu(x)$ which produces Gaussian densities that are centered at the true signal $f_x$, and thus $E(f\vert X=x)$ corresponds to a mixture distribution which resides outside the Gaussian class. The conditional Fr\'echet regression function  defined through the $2$-Wasserstein barycenter is able to correctly capture the underlying geometry of the intensity space as its shape components remain in the Gaussian ensemble where the true signal was generated from, and thus provides a reasonable notion of center or mean in intensity space. If the variance of the noise $\varepsilon_1$ is very low, then both quantities are similar.

\section*{\sf Acknowledgments}\label{s:acknowledgement}
This research was supported in part by the National Science Foundation grant DMS-2014626.

\clearpage
\bibliography{2-1-19}
\bibliographystyle{rss}

\end{document}

%% file: commands0.tex
\usepackage{geometry,float}
\usepackage{float}
\usepackage{bm}
\usepackage{amssymb}
\usepackage{amstext}
\usepackage{graphics}
\usepackage{graphicx}
\usepackage{amsmath}
\usepackage{mathtools}
\usepackage{enumitem}
\usepackage{latexsym}
\usepackage{color}
\usepackage{rotating}
\usepackage{multirow}

\usepackage[colorlinks=true, allcolors=blue]{hyperref}
\usepackage{natbib}

\graphicspath{{figures/}}

\renewcommand{\baselinestretch}{1.6} %{2} %{1.4}
\AtBeginDocument{}
\makeatletter
\newsavebox{\mybox}\newsavebox{\mysim}
\newcommand{\distas}[1]{%
  \savebox{\mybox}{\hbox{\kern3pt$\scriptstyle#1$\kern3pt}}%
  \savebox{\mysim}{\hbox{$\sim$}}%
  \mathbin{\overset{\text{#1}}{\kern\z@\resizebox{\wd\mybox}{\ht\mysim}{$\sim$}}}%
}
\makeatother
\DeclarePairedDelimiterX{\inner}[2]{\langle}{\rangle}{#1, #2}
\DeclarePairedDelimiterX{\innerp}[2]{\langle}{\rangle_p}{#1, #2}
\newcommand{\Nin}{N_i^{(n)}(T)}

\newcommand{\sumin}{\sum_{i=1}^n}

\newcommand{\bbR}{\mathbb{R}}

\newcommand{\Var}{\,\text{Var}}

\newcommand{\hsigma}{\hat{\sigma}}

\newcommand{\hmu}{\hat{\mu}}

\newcommand{\hSigma}{\hat{\Sigma}}

\newcommand{\hLambda}{\hat{\Lambda}}

\newcommand*{\inv}{^{-1}}
\newcommand{\bea}{\begin{eqnarray*}}
\newcommand{\eea}{\end{eqnarray*}}
\newcommand{\be}{\begin{eqnarray}}
\newcommand{\ee}{\end{eqnarray}}
\newcommand{\ed}{\end{document}}
\newcommand{\no}{\noindent}

\newcommand{\btab}{\begin{tabular}}
\newcommand{\etab}{\end{tabular}}
\newcommand{\bc}{\begin{center}}
\newcommand{\ec}{\end{center}}

\newcommand{\bfi}{\begin{figure}}
\newcommand{\efi}{\end{figure}}
\newcommand{\ben}{\begin{enumerate}}
\newcommand{\een}{\end{enumerate}}
\newcommand{\bdes}{\begin{description}}
\newcommand{\edes}{\end{description}}
\newcommand{\bay}{\begin{array}}
\newcommand{\eay}{\end{array}}

\def\bco{\iffalse}

\def\cp{\citep}

\def\i1{_{i1}}

\def\s2{\sum_{k=1}^\infty\lambda_k\phi_k(t) \phi_k(t)}
\def\s1{\sum_{k=1}^\infty\lambda_k\phi^{(1)}_k(s) \phi_k(t)}

\def\pt{\phi_k(t)}
\def\ps{\phi_k(s)}
\def\pt1{\phi^{(1)}_k(t)}
\def\ps1{\phi^{(1)}_k(s)}

\def\singlespace{\def\baselinestretch{1}\@normalsize}

\newcommand{\single}{\renewcommand{\baselinestretch}{1.2}\normalsize}
\newcommand{\double}{\renewcommand{\baselinestretch}{1.5}\normalsize}

%% file: ms.bbl
\begin{thebibliography}{52}
\expandafter\ifx\csname natexlab\endcsname\relax\def\natexlab#1{#1}\fi
\expandafter\ifx\csname url\endcsname\relax
  \def\url#1{\texttt{#1}}\fi
\expandafter\ifx\csname urlprefix\endcsname\relax\def\urlprefix{URL }\fi

\bibitem[{Agueh and Carlier(2011)}]{ague:11}
Agueh, M. and Carlier, G. (2011) Barycenters in the {W}asserstein space.
\newblock \textit{SIAM Journal on Mathematical Analysis}, \textbf{43},
  904--924.

\bibitem[{Berthet \textit{et~al.}(2020)Berthet, Blondel, Teboul, Cuturi, Vert
  and Bach}]{berthe:20}
Berthet, Q., Blondel, M., Teboul, O., Cuturi, M., Vert, J.-P. and Bach, F.
  (2020) Learning with differentiable perturbed optimizers.
\newblock \textit{arXiv preprint arXiv:2002.08676}.

\bibitem[{Bigot \textit{et~al.}(2018)Bigot, Cazelles and Papadakis}]{bigo:18}
Bigot, J., Cazelles, E. and Papadakis, N. (2018) Data-driven regularization of
  wasserstein barycenters with an application to multivariate density
  registration.
\newblock \textit{arXiv preprint arXiv:1804.08962}.

\bibitem[{Bigot \textit{et~al.}(2013)Bigot, Gadat, Klein and
  Marteau}]{bigo:13:2}
Bigot, J., Gadat, S., Klein, T. and Marteau, C. (2013) Intensity estimation of
  non-homogeneous {P}oisson processes from shifted trajectories.
\newblock \textit{Electronic {J}ournal of {S}tatistics}, \textbf{7}, 881--931.

\bibitem[{Bolstad \textit{et~al.}(2003)Bolstad, Irizarry, {\AA}strand and
  Speed}]{bols:03}
Bolstad, B.~M., Irizarry, R., {\AA}strand, M. and Speed, T. (2003) A comparison
  of normalization methods for high density oligonucleotide array data based on
  variance and bias.
\newblock \textit{Bioinformatics}, \textbf{19}, 185--193.

\bibitem[{Borgnat \textit{et~al.}(2011)Borgnat, Abry, Flandrin, Robardet,
  Rouquier and Fleury}]{borg:11}
Borgnat, P., Abry, P., Flandrin, P., Robardet, C., Rouquier, J.-B. and Fleury,
  E. (2011) Shared bicycles in a city: A signal processing and data analysis
  perspective.
\newblock \textit{Advances in Complex Systems}, \textbf{14}, 415--438.

\bibitem[{Bouzas and Ruiz-Fuentes(2015)}]{bouz:15}
Bouzas, P. and Ruiz-Fuentes, N. (2015) A review on functional data analysis for
  {C}ox processes.
\newblock \textit{Boletin de Estadistica e Investigacion Operativa}, 215--230.

\bibitem[{Bouzas \textit{et~al.}(2006)Bouzas, Valderrama, Aguilera and
  Ruiz-Fuentes}]{bouz:06:2}
Bouzas, P.~R., Valderrama, M., Aguilera, A.~M. and Ruiz-Fuentes, N. (2006)
  Modeling the mean of a doubly stochastic {P}oisson process by functional data
  analysis.
\newblock \textit{Computational Statistics and Data Analysis}, \textbf{50},
  2655--2667.

\bibitem[{Burago \textit{et~al.}(2001)Burago, Burago and Ivanov}]{bura:01}
Burago, D., Burago, Y. and Ivanov, S. (2001) \textit{A {C}ourse in {M}etric
  {G}eometry}, vol.~33.
\newblock American Mathematical Society.

\bibitem[{Cazelles \textit{et~al.}(2018)Cazelles, Seguy, Bigot, Cuturi and
  Papadakis}]{caze:18}
Cazelles, E., Seguy, V., Bigot, J., Cuturi, M. and Papadakis, N. (2018)
  Geodesic {PCA} versus log-{PCA} of histograms in the {W}asserstein space.
\newblock \textit{SIAM Journal on Scientific Computing}, \textbf{40},
  B429--B456.

\bibitem[{Cowling \textit{et~al.}(1996)Cowling, Hall and Phillips}]{cowl:96:1}
Cowling, A., Hall, P. and Phillips, M.~J. (1996) Bootstrap confidence regions
  for the intensity of a {P}oisson point process.
\newblock \textit{Journal of the {A}merican {S}tatistical {A}ssociation},
  \textbf{91}, 1516--1524.

\bibitem[{Cox and Isham(1980)}]{cox:80}
Cox, D.~R. and Isham, V. (1980) \textit{Point {P}rocesses}.
\newblock London: Chapman \& Hall.
\newblock Monographs on Applied Probability and Statistics.

\bibitem[{Cucala(2008)}]{cuca:08}
Cucala, L. (2008) Intensity estimation for spatial point processes observed
  with noise.
\newblock \textit{Scandinavian Journal of Statistics}, \textbf{35}, 322--334.

\bibitem[{Daley and Vere-Jones(2003)}]{dale03}
Daley, D.~J. and Vere-Jones, D. (2003) \textit{An {I}ntroduction to the
  {T}heory of {P}oint {P}rocesses{:} volume I{:} {E}lementary {T}heory and
  {M}ethods, {S}econd {E}dition}.
\newblock Springer, New York.

\bibitem[{Diggle and Marron(1988)}]{digg:88}
Diggle, P. and Marron, J. (1988) Equivalence of smoothing parameter selections
  in density and intensity estimation.
\newblock \textit{Journal of the American Statistical Association},
  \textbf{83}, 793--800.

\bibitem[{Diggle(1985)}]{digg:85}
Diggle, P.~J. (1985) A kernel method for smoothing point process data.
\newblock \textit{Applied Statistics}, \textbf{34}, 138--147.

\bibitem[{Diggle \textit{et~al.}(2013)Diggle, Moraga, Rowlingson and
  Taylor}]{digg13}
Diggle, P.~J., Moraga, P., Rowlingson, B. and Taylor, B.~M. (2013) Spatial and
  spatio-temporal log-{G}aussian {C}ox processes: extending the geostatistical
  paradigm.
\newblock \textit{Statistical Science}, \textbf{28}, 542--563.

\bibitem[{Fan and Gijbels(1996)}]{fan:96}
Fan, J. and Gijbels, I. (1996) \textit{Local Polynomial Modelling and its
  Applications}.
\newblock London: Chapman \& Hall.

\bibitem[{Fan \textit{et~al.}(1996)Fan, Gijbels, Hu and Huang}]{fan:96:3}
Fan, J., Gijbels, I., Hu, T.-C. and Huang, L.-S. (1996) A study of variable
  bandwidth selection for local polynomial regression.
\newblock \textit{Statistica Sinica}, 113--127.

\bibitem[{Fr{\'e}chet(1948)}]{frec:48}
Fr{\'e}chet, M. (1948) Les {\'e}l{\'e}ments al{\'e}atoires de nature quelconque
  dans un espace distanci{\'e}.
\newblock In \textit{Annales de l'Institut Henri Poincar{\'e}}, vol.~10,
  215--310.

\bibitem[{Gervini(2017)}]{gerv:17}
Gervini, D. (2017) Multiplicative component models for replicated point
  processes.
\newblock \textit{arXiv preprint arXiv:1705.09693}.

\bibitem[{Gervini and Khanal(2019)}]{gerv:19}
Gervini, D. and Khanal, M. (2019) Exploring patterns of demand in bike sharing
  systems via replicated point process models.
\newblock \textit{Journal of the Royal Statistical Society Series C: Applied
  Statistics}, \textbf{68}, 585--602.

\bibitem[{Grenander(1950)}]{gren:50}
Grenander, U. (1950) Stochastic processes and statistical inference.
\newblock \textit{Arkiv f\"or Matematik}, \textbf{1}, 195--277.

\bibitem[{Henderson(2003)}]{hend:03}
Henderson, S.~G. (2003) Estimation for nonhomogeneous {P}oisson processes from
  aggregated data.
\newblock \textit{Operations Research Letters}, \textbf{31}, 375--382.

\bibitem[{Hsing and Eubank(2015)}]{hsin:15}
Hsing, T. and Eubank, R. (2015) \textit{Theoretical Foundations of Functional
  Data Analysis, with an Introduction to Linear Operators}.
\newblock John Wiley \& Sons.

\bibitem[{Kao and Chang(1988)}]{Kao:88}
Kao, E. P.~C. and Chang, S.-L. (1988) Modeling time-dependent arrivals to
  service systems: A case in using a piecewise-polynomial rate function in a
  nonhomogeneous {P}oisson process.
\newblock \textit{Management Science}, \textbf{34}, 1367--1379.

\bibitem[{Kleffe(1973)}]{klef:73}
Kleffe, J. (1973) Principal components of random variables with values in a
  separable {H}ilbert space.
\newblock \textit{Statistics: A Journal of Theoretical and Applied Statistics},
  \textbf{4}, 391--406.

\bibitem[{Kuhl and Bhairgond(2000)}]{kuhl:00:02}
Kuhl, M. and Bhairgond, P. (2000) Nonparametric estimation of nonhomogeneous
  {P}oisson processes using wavelets.
\newblock In \textit{2000 Winter Simulation Conference Proceedings (Cat.
  No.00CH37165)}, vol.~1, 562--571 vol.1.

\bibitem[{Kuhl and Wilson(2000)}]{kuhl:00}
Kuhl, M.~E. and Wilson, J.~R. (2000) Least squares estimation of nonhomogeneous
  {P}oisson processes.
\newblock \textit{Journal of Statistical Computation and Simulation},
  \textbf{67}, 699--712.

\bibitem[{Kuhl and Wilson(2001)}]{kuhl:01}
--- (2001) Modeling and simulating {P}oisson processes having trends or
  nontrigonometric cyclic effects.
\newblock \textit{European Journal of Operational Research}, \textbf{133},
  566--582.

\bibitem[{Kuhl \textit{et~al.}(1997)Kuhl, Wilson and Johnson}]{kuhl:97}
Kuhl, M.~E., Wilson, J.~R. and Johnson, M.~A. (1997) Estimating and simulating
  {P}oisson processes having trends or multiple periodicities.
\newblock \textit{IIE Transactions}, \textbf{29}, 201--211.

\bibitem[{Lawless(1987)}]{lawl87}
Lawless, J.~F. (1987) Regression methods for {P}oisson process data.
\newblock \textit{Journal of the American Statistical Association},
  \textbf{82}, 808--815.

\bibitem[{Lee \textit{et~al.}(1991)Lee, Wilson and Crawford}]{Lee:91}
Lee, S., Wilson, J.~R. and Crawford, M.~M. (1991) Modeling and simulation of a
  nonhomogeneous {P}oisson process having cyclic behavior.
\newblock \textit{Communications in Statistics - Simulation and Computation},
  \textbf{20}, 777--809.

\bibitem[{Leemis(1991)}]{Leem:91}
Leemis, L.~M. (1991) Nonparametric estimation of the cumulative intensity
  function for a nonhomogeneous {P}oisson process.
\newblock \textit{Management Science}, \textbf{37}, 886--900.

\bibitem[{Lewis and Shedler(1976)}]{Lewi:76}
Lewis, P. A.~W. and Shedler, G.~S. (1976) Statistical analysis of
  non-stationary series of events in a data base system.
\newblock \textit{IBM Journal of Research and Development}, \textbf{20},
  465--482.

\bibitem[{Li and Guan(2014)}]{li:14}
Li, Y. and Guan, Y. (2014) Functional principal component analysis of
  spatiotemporal point processes with applications in disease surveillance.
\newblock \textit{Journal of the American Statistical Association},
  \textbf{109}, 1205--1215.

\bibitem[{Mikosch(2009)}]{miko:09}
Mikosch, T. (2009) \textit{Non-{L}ife {I}nsurance {M}athematics: {A}n
  {I}ntroduction with the {P}oisson {P}rocess}.
\newblock Springer.

\bibitem[{Panaretos and Zemel(2016)}]{pana:16}
Panaretos, V.~M. and Zemel, Y. (2016) Amplitude and phase variation of point
  processes.
\newblock \textit{Annals of Statistics}, \textbf{44}, 771--812.

\bibitem[{{Panaretos} and {Zemel}(2019)}]{pana:19}
{Panaretos}, V.~M. and {Zemel}, Y. (2019) {Statistical Aspects of Wasserstein
  Distances}.
\newblock \textit{Annual Review of Statistics and Its Application}, \textbf{6},
  405--431.

\bibitem[{Petersen and M\"{u}ller(2016)}]{mull:16:1}
Petersen, A. and M\"{u}ller, H.-G. (2016) Functional data analysis for density
  functions by transformation to a {H}ilbert space.
\newblock \textit{Annals of Statistics}, \textbf{44}, 183--218.

\bibitem[{Petersen and M{\"u}ller(2019)}]{mull:18:3}
Petersen, A. and M{\"u}ller, H.-G. (2019) Fr{\'e}chet regression for random
  objects with {E}uclidean predictors.
\newblock \textit{The Annals of Statistics}, \textbf{47}, 691--719.

\bibitem[{Peyr{\'e} and Cuturi(2019)}]{peyr:19}
Peyr{\'e}, G. and Cuturi, M. (2019) Computational optimal transport: With
  applications to data science.
\newblock \textit{Foundations and Trends{\textregistered} in Machine Learning},
  \textbf{11}, 355--607.

\bibitem[{Reddy and Dass(2006)}]{redd:06}
Reddy, S.~K. and Dass, M. (2006) Modeling on-line art auction dynamics using
  functional data analysis.
\newblock \textit{Statistical Science}, \textbf{21}, 179--193.

\bibitem[{Reynaud-Bouret(2003)}]{reyn:03}
Reynaud-Bouret, P. (2003) Adaptive estimation of the intensity of inhomogeneous
  {P}oisson processes via concentration inequalities.
\newblock \textit{Probability Theory and Related Fields}, \textbf{126},
  103--153.

\bibitem[{Reynaud-Bouret and Rivoirard(2010)}]{reyn:10}
Reynaud-Bouret, P. and Rivoirard, V. (2010) Near optimal thresholding
  estimation of a {P}oisson intensity on the real line.
\newblock \textit{Electronic Journal of Statistics}, \textbf{4}, 172--238.

\bibitem[{Sayarshad and Chow(2016)}]{Saya:16}
Sayarshad, H.~R. and Chow, J. Y.~J. (2016) Survey and empirical evaluation of
  nonhomogeneous arrival process models with taxi data.
\newblock \textit{Journal of Advanced Transportation}, \textbf{50}, 1275--1294.

\bibitem[{Shmueli \textit{et~al.}(2007)Shmueli, Russo and Jank}]{shmu:07}
Shmueli, G., Russo, R.~P. and Jank, W. (2007) The barista: A model for bid
  arrivals in online auctions.
\newblock \textit{The Annals of Applied Statistics}, \textbf{1}, 412--441.

\bibitem[{Utsu \textit{et~al.}(1995)Utsu, Ogata, S and Matsu'ura}]{utsu:95}
Utsu, T., Ogata, Y., S, R. and Matsu'ura (1995) The centenary of the {O}mori
  formula for a decay law of aftershock activity.
\newblock \textit{Journal of Physics of the Earth}, \textbf{43}, 1--33.

\bibitem[{Villani(2003)}]{vill:03}
Villani, C. (2003) \textit{Topics in Optimal Transportation}.
\newblock American Mathematical Society.

\bibitem[{{Willett} and {Nowak}(2007)}]{will:07}
{Willett}, R.~M. and {Nowak}, R.~D. (2007) Multiscale {P}oisson intensity and
  density estimation.
\newblock \textit{{IEEE} {T}ransactions on {I}nformation {T}heory},
  \textbf{53}, 3171--3187.

\bibitem[{Wu \textit{et~al.}(2013)Wu, M{\"u}ller and Zhang}]{mull:13:1}
Wu, S., M{\"u}ller, H.-G. and Zhang, Z. (2013) Functional data analysis for
  point processes with rare events.
\newblock \textit{Statistica Sinica}, \textbf{23}, 1--23.

\bibitem[{Zhang and Kou(2010)}]{zhan:10}
Zhang, T. and Kou, S.~C. (2010) Nonparametric inference of doubly stochastic
  {P}oisson process data via the kernel method.
\newblock \textit{The Annals of Applied Statistics}, \textbf{4}, 1913--1941.

\end{thebibliography}
